\documentclass[twocolumn]{aastex62}

\received{August 30, 2018}
\revised{November 7, 2018}
\accepted{November 14, 2018}
\submitjournal{ApJ}

\shorttitle{Fast-to-\alfven mode conversion}
\shortauthors{P.A.~Gonz\'alez-Morales et al.}

\usepackage{mathtools}
\usepackage{fixmath}
\newcommand{\dt}[1]{\frac{\partial{#1}}{\partial{t}}}

\newcommand{\coma}{\hspace{3pt},}
\newcommand{\punto}{\hspace{3pt}.}

\newcommand\mancha{\textsc{Mancha3D~}}

\newcommand\alfven{Alfv\'en~}
\newcommand\alfvenic{Alfv\'enic~}
\begin{document}

\title{Fast-to-\alfven mode conversion mediated by Hall current. II Application to the solar atmosphere.}

\correspondingauthor{P. A.~Gonz\'alez-Morales}
\email{pagm@iac.es}

\author[0000-0003-0600-2210]{P. A.~Gonz\'alez-Morales}
\affiliation{Instituto de Astrof\'isica de Canarias (IAC), E-38205 La Laguna, Tenerife, Spain.}
\affiliation{ Universidad de La Laguna (ULL), Dpto. Astrof\'isica, E-38206 La Laguna, Tenerife, Spain.}
\nocollaboration

\author{E.~Khomenko}
\affiliation{Instituto de Astrof\'isica de Canarias (IAC), E-38205 La Laguna, Tenerife, Spain.}
\affiliation{ Universidad de La Laguna (ULL), Dpto. Astrof\'isica, E-38206 La Laguna, Tenerife, Spain.}
\nocollaboration

\author[0000-0001-5794-8810]{P.~S.~Cally}
\affiliation{School of Mathematical Sciences, Monash University, Clayton, Australia. }
\affiliation{Monash Centre for Astrophysics, Monash University, Clayton, Australia.}
\nocollaboration



\begin{abstract}
\noindent Coupling between fast magneto-acoustic and \alfven waves can be observe in fully ionized plasmas mediated by stratification and 3D geometrical effects. In Paper I, \citet{2015ApJ...814..106C} have shown that in a weakly ionized plasma, such as the solar photosphere and chromosphere, the Hall current introduces a new coupling mechanism. The present study extends the results from Paper I to the case of warm plasma. We report on numerical experiments where mode transformation is studied using quasi-realistic stratification in thermodynamic parameters resembling the solar atmosphere. This redresses the limitation of the cold plasma approximation assumed in Paper I, in particular allowing the complete process of coupling between fast and slow magneto-acoustic modes and subsequent coupling of the fast mode to the \alfven mode through the Hall current. Our results confirm the efficacy of the mechanism proposed in Paper I for the solar case. We observe that the efficiency of the transformation is a sensitive function of the angle between the wave propagation direction and the magnetic field, and of the wave frequency. The efficiency increases when the field direction and the wave direction are aligned for increasing wave frequencies. After scaling our results to typical solar values, the maximum amplitude of the transformed \alfven waves, for a frequency of 1 Hz, corresponds to an energy flux (measured above the height of peak Hall coupling) of $\sim10^3$ $\rm W\,m^{-2}$, based on an amplitude of 500 $\rm m\,s^{-1}$ at $\beta=1$, which is sufficient to play a major role in both quiet and active region coronal heating.
\end{abstract}



\section{Introduction}\label{sec:intro}

\alfven waves \citep{1942Natur.150..405A} are plasma perturbations whose restoring force is magnetic tension instead of gas or magnetic pressure. They carry energy along the magnetic field lines but the motion of the charged particles, which provide the inertia, and the magnetic field perturbation are transverse to the field. 

In the solar atmosphere, the presence of gradients and strong vertical stratification allow for the process of mode transformation. Fast/slow magnetoacoustic coupling takes place where the acoustic and \alfven speeds match ($c_s = v_A$). In the solar atmosphere the stratification-induced mode transformation usually occurs somewhere in the upper photosphere and low chromosphere, depending on the strength of the magnetic structures. In sunspot umbra it is typically sub-surface. The mathematical formalism of mode conversion was developed by \citet{2006MNRAS.372..551S} and \citet{2006RSPTA.364..333C, Cally:2007cn} based on the generalized ray theory of \citet{2003PhPl...10.2147T}. In the solar literature, the equipartition layer is either defined as where the plasma $\beta$ or the ratio of acoustic and \alfven speeds squared, $c_s^2/v_A^2$, reach unity. These measures differ by the factor $\gamma/2$. The wave speed equipartition is the more physically relevant criterion, but in practice there is little difference. 
 
Ideal-MHD fast-to-\alfven transformation happens around where the fast wave reflects from the density stratification. This is typically well above the equipartition level, but depends on wave frequency and horizontal wavenumber. In a low $\beta$ plasma, the reflection point is near where the horizontal phase speed matches the Alfv\'en speed, and so is higher for high frequency and for near-vertical propagation. This process could be an important mechanism to provide energy to the solar corona. This is so because unlike fast waves, which reflect in the transition region or chromosphere due to the rapidly increasing \alfven speed, \alfven waves generated through mode transformation close to the transition region are more able to overcome this barrier, and to reach the corona. The efficiency of geometrical mode conversion depends on the local relative inclination between the wave vector and the background magnetic field \citep{2008SoPh..251..251C, 2009MNRAS.395.1309C, 2011ApJ...738..119C, Khomenko:2011fx, Khomenko:2012gw, Felipe:2012kq}. Observations realized by \citet{DePontieu:2007eh}, \citet{2007Sci...317.1192T} or \citet {Jess-2009} have shown that \alfven waves are everywhere in the corona. Later on, observations from \citet{McIntosh:2011cy} and \cite{ 2012ApJ...744L...5J} have shown that these `Alfv\'enic' waves are of sufficient amplitude to heat some regions and contribute to the acceleration of the solar wind. Recently, \citet{Srivastava:2017kw} have reported the first observation of high-frequency torsional \alfven waves  ($\sim$$12$--$42$ mHz) in the solar chromosphere.

In general, astrophysical plasmas are formed by electrons, ions, neutrals, and dust grains and all these particles interact with the magnetic field, either directly or via collisional coupling between charged particles and neutrals. When differing inertia and collisional interactions produce a drift between electrons and ions, ideal magnetohydrodynamic (MHD) theory has to be modified to include the Hall effect. This introduces a new Hall electric field proportional to the cross product of the current density and the magnetic field, which thereby contributes to a  generalized Ohm's law. In order to treat the Hall effect, one has to apply so-called Hall-MHD theory.

In the atmosphere of the Sun or solar-like stars, the plasma can be weakly ionized, reaching, for example, an ionization fraction as low as $f\sim10^{-4}$ around the Sun's temperature minimum. Under these conditions, \cite{2012ApJ...750....6C} investigated the effects of ambipolar diffusion and Hall currents on the formation of structures in photospheric magneto-convection, showing how the Hall current can couple magneto-acoustic and \alfven waves. Later on, \citet{2015ApJ...814..106C} provided a corrected theory of Hall-current mediated coupling between the fast and \alfven waves in cold (i.e., high beta) plasmas. They demonstrated that coupling efficiency is proportional to the dimensionless Hall parameter $\epsilon_\mathrm{Hall}=\omega/\Omega_i f$, where $\Omega_i$ is the mean ion gyro-frequency and $\omega$ is the wave frequency. Due to the smallness of the ionization fraction $f$, coupling can be produced even for relatively low-frequency waves. They show that the Hall effect produces an oscillation between the \alfven and magneto-sonic states and the precession would be the beating between those modes. They found that this coupling occurs in places where the wave vector is nearly aligned with the magnetic field vector. It is also more efficient for relatively low magnetic field strengths, and for higher frequency waves.  \citet{2015ApJ...814..106C} speculated that such processes as reconnection, where high-frequency waves are produced, can be affected.

Unlike the geometrical mode conversion studied by \citet{2008SoPh..251..251C}, \citet{ Khomenko:2012gw}, or \citet{Felipe:2012kq} among others, Hall current induced mode conversion can happen even if initially the wave vector and the magnetic field vector are in the same two-dimensional plane.  Moreover, once the transformation happens, the waves keep travelling long distance nearly aligned with the magnetic field so they could transfer energy to the surrounding plasma during the precession process.

The purpose of the current paper is to advance the initial work by \citet{2015ApJ...814..106C}, relaxing the approximation of cold plasmas. That approximation excludes acoustic modes. Nevertheless, the current schematic picture of mode generation, propagation and transformation in a stratified solar atmosphere suggests that acoustic waves play an important role in the chain of wave energy transmission to the corona. Acoustic $p$-modes propagating below the surface (and being fast modes there since $c_s > v_A$) reach the equipartition layer where they are partially transformed into fast magnetic waves. These subsequently refract and reflect back to the solar surface. As mentioned above, \alfven waves are produced through a secondary transformation while the fast waves refract. By considering cold plasma, the first process (fast acoustic to fast magnetic transformation) is not considered, and, therefore, the efficiency of the production of \alfven waves through the Hall current mechanism cannot be evaluated correctly for the solar case.

Here we perform simulations of the complete process beginning with imposed acoustic wave generation below the surface, and all subsequent transformations, including the Hall effect, are addressed on the way to the corona. We take account of realistic plasma stratification for the Sun, and realistic values of the magnetic field. In Section \ref{sec:eq} we show the set of equations corresponding to a single fluid description with a generalized Ohm's law including the Hall term. In Section \ref{sec:num_setup} we present our numerical setup. In Section \ref{sec:numexp} we show the results of our numerical experiments, while in Section \ref{sec:sum} a brief summary is presented.

\section{Equations}\label{sec:eq}

For this work we adopt the Hall-MHD formulation for a partially ionized solar plasma. We neglect all other non-ideal effects (such as ambipolar diffusion and the battery effect) and consider only the non-ideal Hall term under the single-fluid approach \citep{2014PhPl...21i2901K, 2018SSRv..214...58B}. In this approximation, the equations to be solved are the continuity equation,
\begin{equation}\label{eq:cont}
\dt{\rho}+\nabla\cdot(\rho \mathbf{v})=0\coma 
\end{equation}
the momentum conservation equation,
\begin{equation}\label{eq:mom}
\frac{\partial(\rho\mathbf{v})}{\partial t} + \nabla \cdot \Bigg[\rho\mathbf{vv}+\Bigg(p+\frac{\mathbf{B}^2}{2\mu_0}\Bigg)\mathbf{I}-\frac{\mathbf{BB}}{\mu_0}\Bigg]= \rho \mathbf{g} \coma
\end{equation}
the induction equation,
\begin{equation}\label{eq:induc}
\begin{aligned}
\frac{\partial \mathbf{B}}{\partial t} = {} & \nabla \times \Bigg[(\mathbf{v}\times\mathbf{B})-\eta_H\frac{\mu_0}{|B|}(\mathbf{J}\times\mathbf{B}) \Bigg] \coma
\end{aligned}
\end{equation}
where we retained the convective and Hall terms on the right hand side. Here $\eta_H$ is the Hall coefficient and is written in units of a diffusivity coefficient ($l^2t^{-1}$, i.e. $m^2s^{-1}$ in SI), 
\begin{equation}\label{eq:hall}
\eta_H=\frac{|B|}{en_e\mu_0} \punto
\end{equation}
The total energy conservation equation,
\begin{equation}\label{eq:ener}
\begin{aligned}
\dt{e_\mathrm{tot}}{} & + \nabla\cdot\Bigg[ \frac{\rho \mathbf{v}^2}{2}\mathbf{v}+\frac{\gamma p}{\gamma-1}\mathbf{v} + \\ & + \frac{1}{\mu_0} \nabla\cdot\left[ \mathbf{B}\times\left( \mathbf{v}\times\mathbf{B}\right) \right] \Bigg] = \rho\mathbf{v}\cramped{\cdot}\mathbf{g} \coma
\end{aligned}
\end{equation}
is written in terms of the total energy density per volume unit $e_\mathrm{tot}$, which is the sum of the kinetic, internal and magnetic energies,
\begin{equation}\label{eq:totener}
e_\mathrm{tot}=\frac{1}{2}\rho \mathbf{v}^2+\frac{p}{\gamma-1}+\frac{\mathbf{B}^2}{2\mu_0} \punto
\end{equation}
\noindent The electric current density $\mathbf{J}$ is defined through
\begin{equation}
\mu_0\mathbf{J}=\nabla\times\mathbf{B} \coma
\end{equation}
and Gauss's law for magnetism is
\begin{equation}\label{eq:gauss}
\nabla\cdot\mathbf{B}=0\punto
\end{equation}
To close the system, the equation of state for an ideal gas is used. The equations above assume charge neutrality ($n_e=n_i$) and negligible electron mass ($m_e\ll m_i$). 

\section{Numerical setup}\label{sec:num_setup}
%
\begin{figure}
 \includegraphics[width=1\columnwidth]{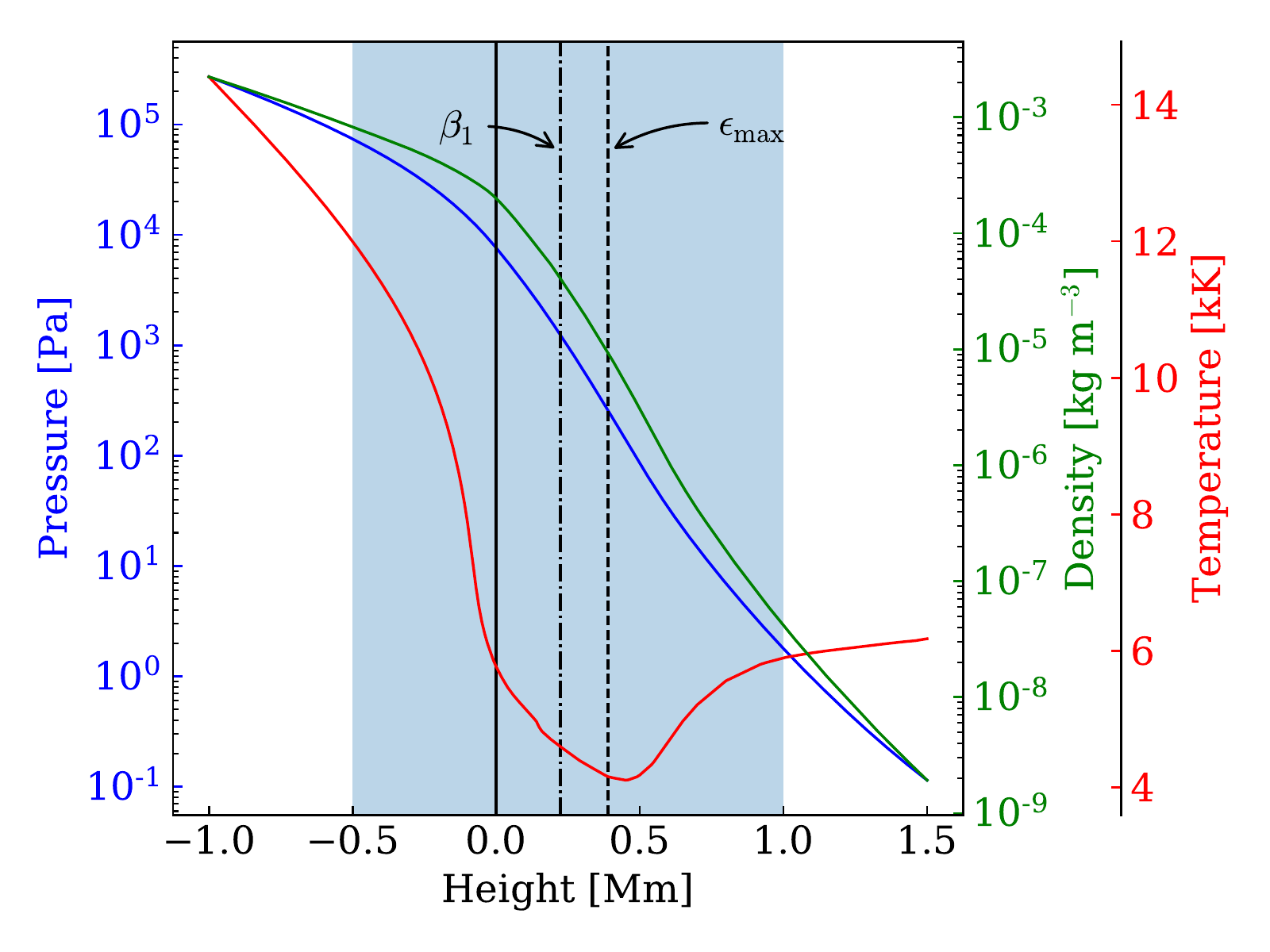}
 \caption{\footnotesize Pressure (blue), density (green), and temperature (red) as a function of vertical coordinate in the model atmosphere assumed in this study. The dot-dashed vertical line indicates the location of $\beta=1$ layer. The dashed line corresponds to the location of the maximum of the Hall parameter ($\epsilon_\mathrm{max}$). The blue rectangle indicates the location of our numerical box.}
 \label{fig:atmosfera}
\end{figure}
\begin{figure}
\includegraphics[width=1\columnwidth]{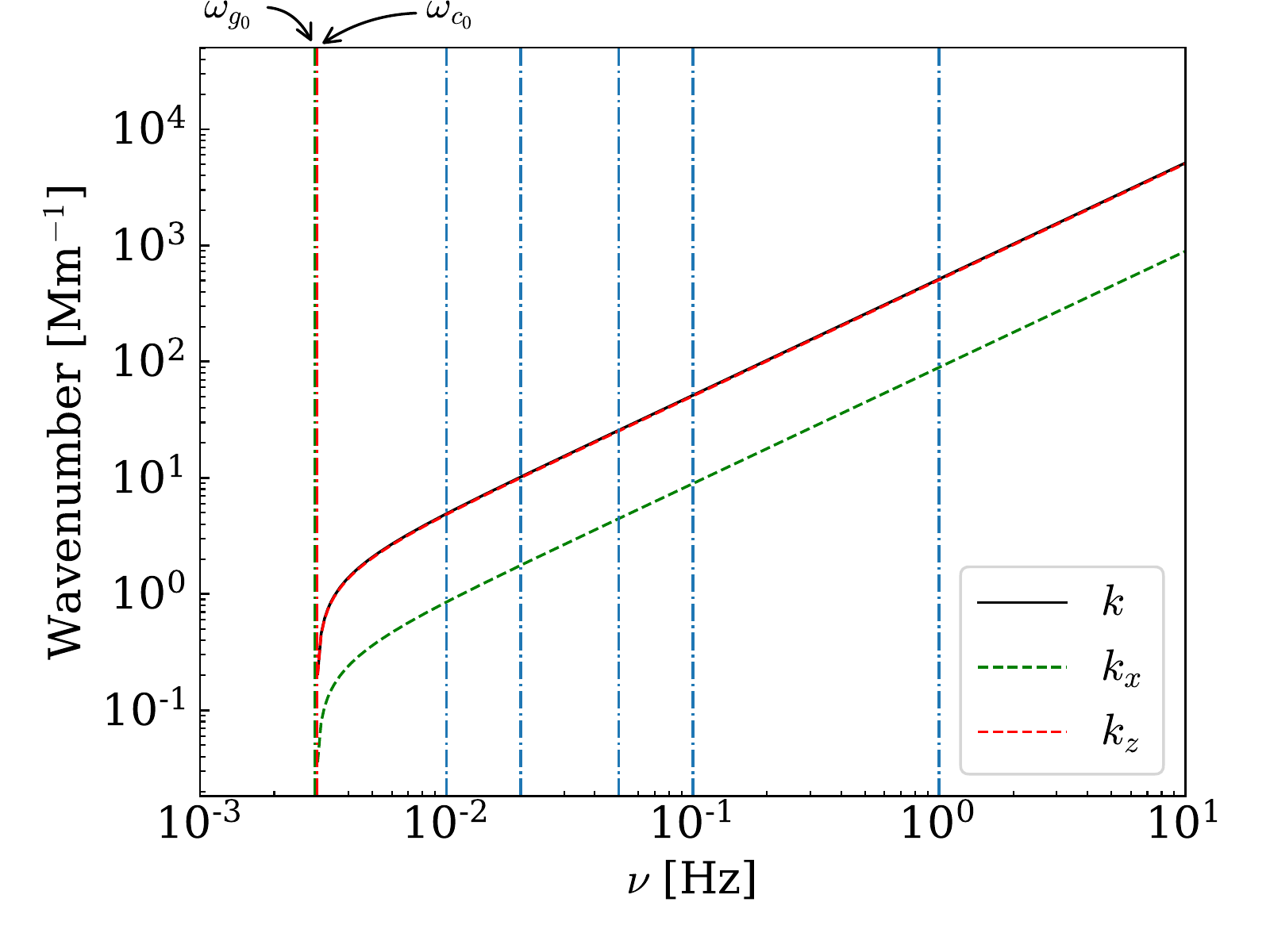}{\caption{\footnotesize Modulus of the wave number $k$, as a function of the wave frequency at the bottom boundary of the numerical box, calculated according to Eq. \ref{eq:kmodule}. The blue vertical dot-dashed lines are the selected frequencies for the experiments. The red dot-dashed vertical line is acoustic cut-off frequency $\omega_{c_0}$ and the green one $\omega_{g_0}$, both calculated at the bottom of the numerical domain.}\label{fig:frec}}
\end{figure}
\begin{figure}
\includegraphics[width=1\columnwidth]{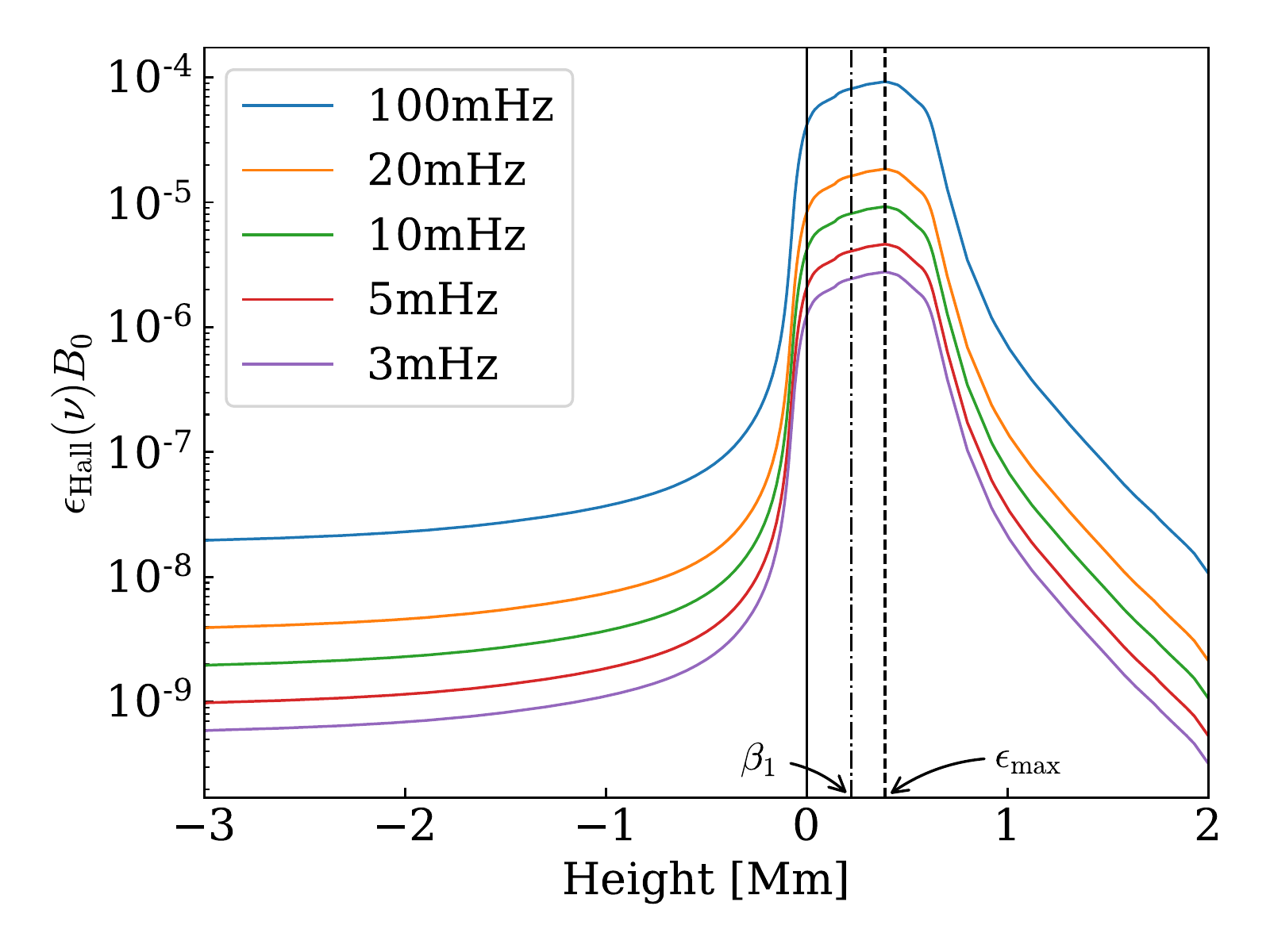}{\caption{\footnotesize Hall parameter, in magnetic field units, as a function of height for different wave frequencies $\nu$. The vertical line corresponds to the photospheric level, $z=0$ km. The vertical dashed line shows where the Hall parameter has its maximum value, $z\approx390$ km. The vertical dot-dashed line indicates the location of the equipartition layer ($\beta=1$), $z\approx228$ km}\label{fig:ehall}}
\end{figure}
\begin{figure}
 \centering
 \includegraphics[width=0.48\columnwidth]{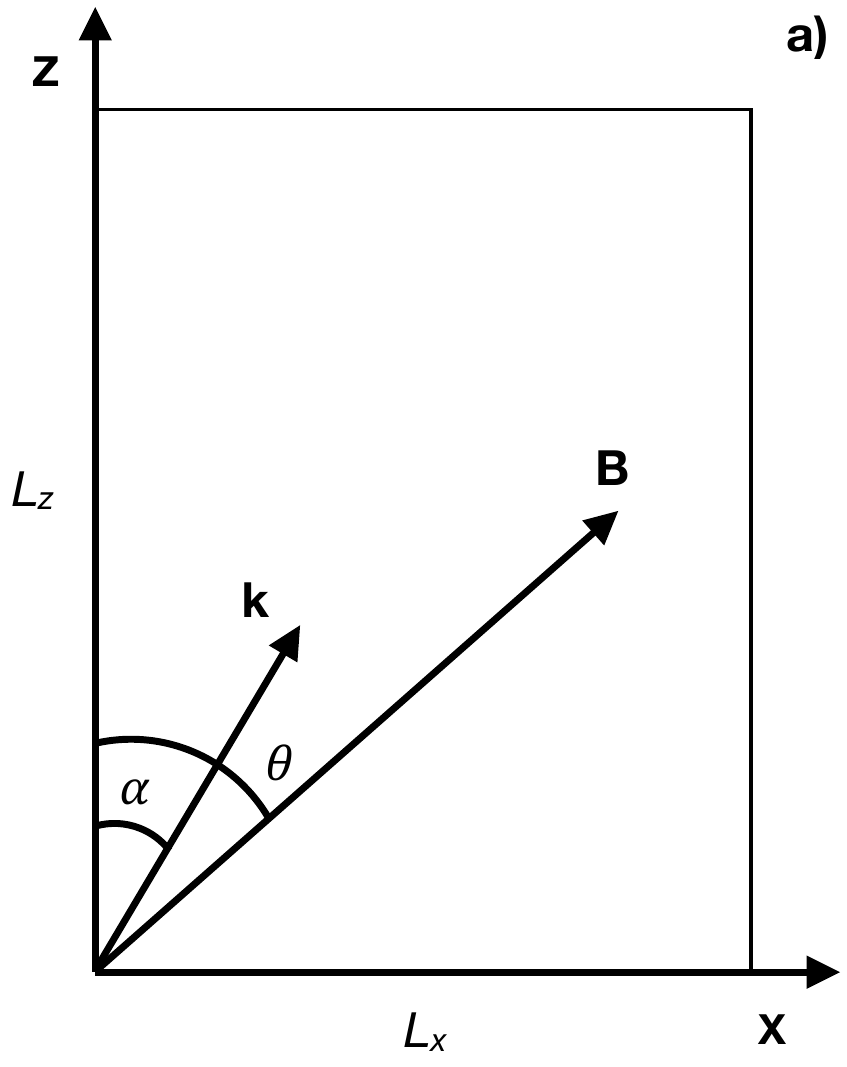}
 \includegraphics[width=0.48\columnwidth]{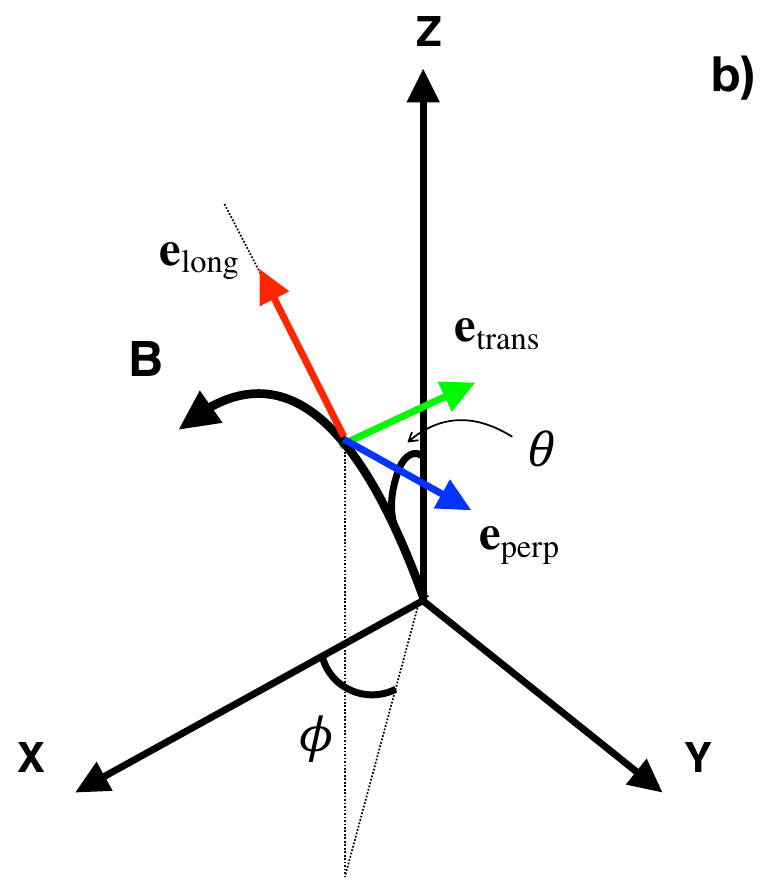}
 \caption{\footnotesize{\emph{Panel a)}: Schematic diagram illustrating the geometrical configuration of the numerical scenario. \emph{Panel b)}: Schematic diagram to show the three characteristic directions.}}
 \label{fig:esquemas}    
\end{figure}
The numerical experiments described in this work are produced by the \mancha code \citep{2006ApJ...653..739K} which solves the non-linear non-ideal 3D single-fluid MHD equations for the perturbations written in conservative form \citep{2010ApJ...719..357F}. Our recently implemented numerical treatment of the Hall term is described in \citet{{2018arXiv180304891G}}.

The modelled solar stratification is build using a 2.5D approximation, that is, vectors are three dimensional objects but the derivatives apply only in two directions, one horizontal and one vertical, taking those to be $x$ and $z$. The model is horizontally homogeneous. 

The vertical stratification is build by merging the standard solar Model-S of \citet{1996Sci...272.1286C} with the chromospheric model VAL-C of \citet{1981ApJS...45..635V} smoothly at height $z=140$ km. Because the upper layers of the solar convection zone are super-adiabatic and unstable against convection, we modify the stratification in thermodynamic parameters to make it convectively stable and avoid the generation of modes that grow exponentially. To do this, we apply the procedure described by \cite{2011SoPh..271....1S} solving the system iteratively and considering the hydrostatic equilibrium condition to the stratification with a constant value for the gravity $g_0=273.98$ m s$^{-2}$ and a constant adiabatic coefficient $\gamma = 5/3$. The resulting stratification is illustrated in Figure \ref{fig:atmosfera}. The numerical box is $L_z= 1.5$ Mm tall, extending from $z=-0.5$ Mm to $z=1$ Mm. 

Finally, a constant and uniform magnetic field $\mathbf{B}_0$ with adjustable inclination $\theta$ with respect to the vertical direction is included. The strength of the magnetic field is chosen based on the arguments explained below in this section.

To excite the waves in the simulation domain we impose an acoustic-gravity wave with a given frequency and horizontal wave number as a bottom boundary condition. The analytic solution at the bottom boundary is calculated according to \cite{1984oup..book.....M}, and provides a self-consistent perturbation in pressure, density, temperature and velocity. The temperature gradient and magnetic field are neglected in this analytical solution. The neglect of magnetic field is justified since the bottom boundary is located in the region where $v_A \ll c_s$, and we are interested in exciting acoustic fast waves. The analytical solutions applied are identical to  \citet{Khomenko:2012gw} and are provided in Appendix \ref{sec:source}.

For the top boundary we use a perfectly matched layer (PML; \citealt{1994JCoPh.114..185B, Berenger1996363, 1996JCoPh.129..201H, Hesthaven1998129, 2001JCoPh.173..455H, 2009ApJ...694..573P}) but slightly modified. We apply the absorption coefficients over all the split variables to obtain a better attenuation and stability for certain magnetic field angles. The horizontal boundaries are set to periodic. The horizontal size of the domain, $L_x$, is determined by the value of $k_x$. Table \ref{tab:experiments} provides the details of the simulation runs. Our goal is to study how the efficiency of the transformation depends on the wave frequency. Therefore, the wave frequency is chosen to be a free parameter varying from simulation to simulation. We used frequencies from 0.01 to 1 Hz. 

To initiate the simulation one has to specify the horizontal wave number, $k_x$, for the given frequency. Because we are interested in keeping the wave vector in the same direction for different source frequencies we calculate the components $(k_x, k_z)$ for a given angle $\alpha$ by writing the dispersion relation in terms of the wave vector modulus:
\begin{equation}\label{eq:kmodule}
k=\frac{\omega}{c_s}\sqrt{\frac{\omega_c^2-\omega^2}{\omega_g^2\sin^2{\alpha}-\omega^2}}\coma
\end{equation}
where $\omega_c = \gamma g / 2c_s$ is the acoustic cutoff frequency and $\omega_g = 2\omega_c \sqrt{\gamma-1} / \gamma$ is the Brunt-V\"ai\"sal\"a frequency. The wave vector components are then calculated as $k_x=|k|\sin{\alpha}$ and $k_z=|k|\cos{\alpha}$, being $\alpha$ the angle between the vertical and the wave number vector, see Fig. \ref{fig:esquemas}a. We choose $\alpha=10$ degrees.

Figure \ref{fig:frec} shows the behaviour of $k$ according to Eq. (\ref{eq:kmodule}) and the wave number vector components $k_x$, $k_z$ as a function of the source frequency for a fixed angle $\alpha=10^\circ$.
\begin{table*}[ht]
    \caption{\footnotesize Selected source frequency and relevant parameters for the numerical setups.}
    \label{tab:experiments}
    \centering
    \begin{tabular}{ccccccc}
    \hline
    $\nu$ [Hz] & $k_x$ [Mm$^{-1}$] & $k_z$ [Mm$^{-1}$] & $dx$ [km] & $dz$ [km] & $n_x$ & $n_z$ \\
    \hline
    $0.01$ & 0.85  & 4.84   & 147.21 & 10   & 50 & 172 \\
    $0.02$ & 1.77  & 10.02  & 71.13  & 10   & 50 & 172 \\
    $0.05$ & 4.46  & 25.28  & 28.19  & 5    & 50 & 332 \\
    $0.1$  & 8.92  & 50.62  & 14.08  & 1    & 50 & 1612 \\
    $1$    & 89.29 & 506.40 & 1.41   & 0.05 & 50 & 32012 \\
    \hline
    \end{tabular}
\end{table*}

In order to select the background magnetic field strength $B_0$ for our experiments we consider the behaviour of the Hall parameter $\epsilon_\mathrm{Hall}$, defined according to:
\begin{equation}\label{eq:ehall}
    \epsilon_\mathrm{Hall}(\nu,B)=\frac{\omega}{f\Omega_i}=\frac{2\pi\rho_0}{q_e n_e}\frac{\nu}{B_0} \coma
\end{equation}
where $\rho_0$ is the background density, $q_e$ the electron charge, $n_e$ the electron number density, $\nu$ is the wave source frequency, and $B_0$ is the background magnetic field. Figure \ref{fig:ehall} shows the Hall parameter, written in units of the background magnetic field, calculated for our stratification of thermodynamic parameters, as a function of height, for several wave frequencies. Independently of the frequency and the value of $B_0$, $\epsilon_\mathrm{Hall}$ has a maximum around $z\approx390$ km. Therefore, we choose $B_0$ to place the equipartition layer $c_s=v_A$ below the height where $\epsilon_\mathrm{Hall}$ is maximum. This is because we wish to study the process of conversion between fast magnetic and \alfven waves, and the fast magnetic waves are to be produced first through the primarily geometrical transformation at the $c_s=v_A$ layer. To affect this compromise, we choose a magnetic field $B_0=500$ G, which locates the equipartition layer at $z\approx228$ km, just below the maximum of the Hall parameter and inside the region where mode transformation can take place. 

\begin{figure*}
 \centering
 \includegraphics[width=0.32\textwidth]{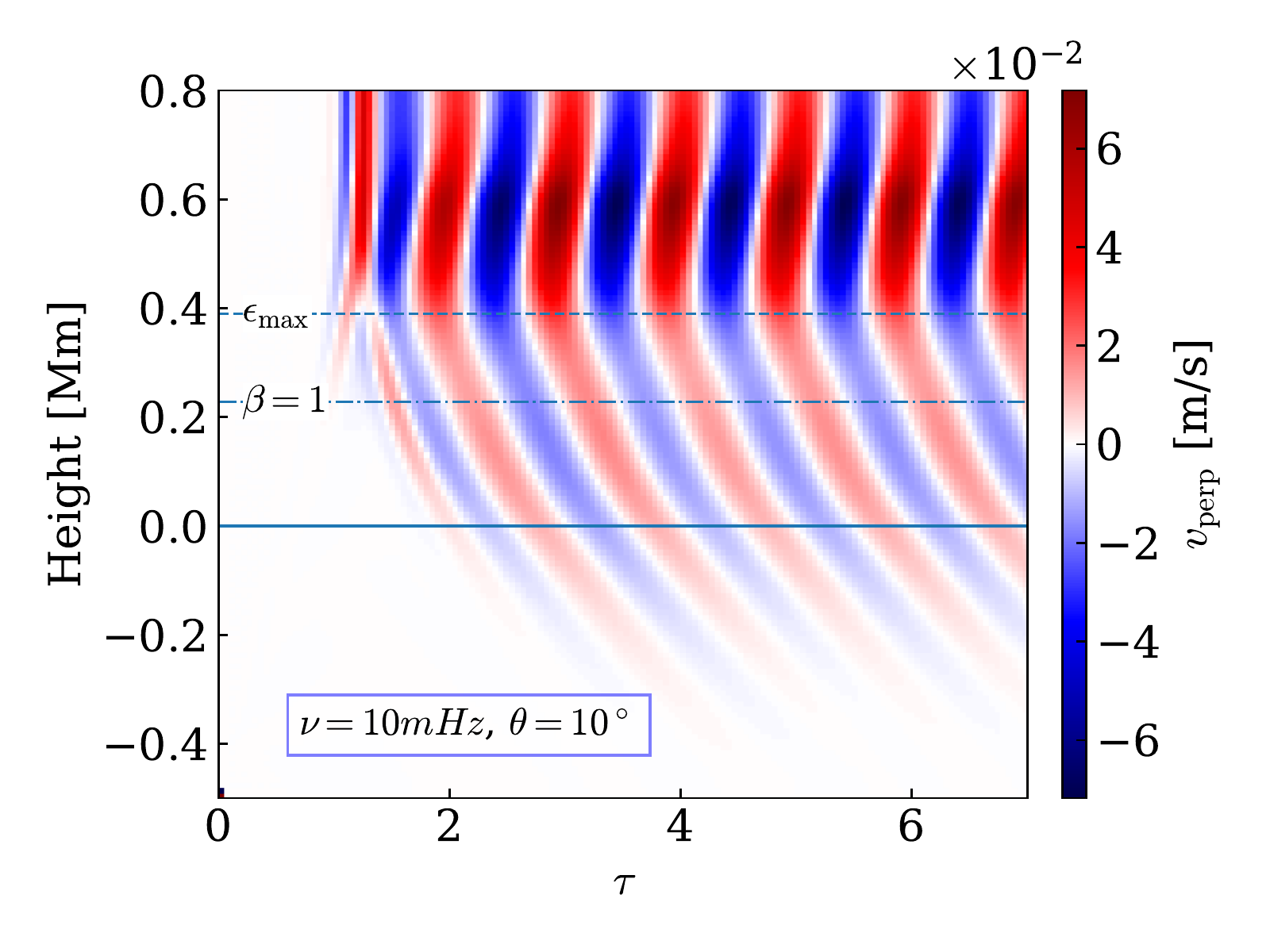}
 \includegraphics[width=0.32\textwidth]{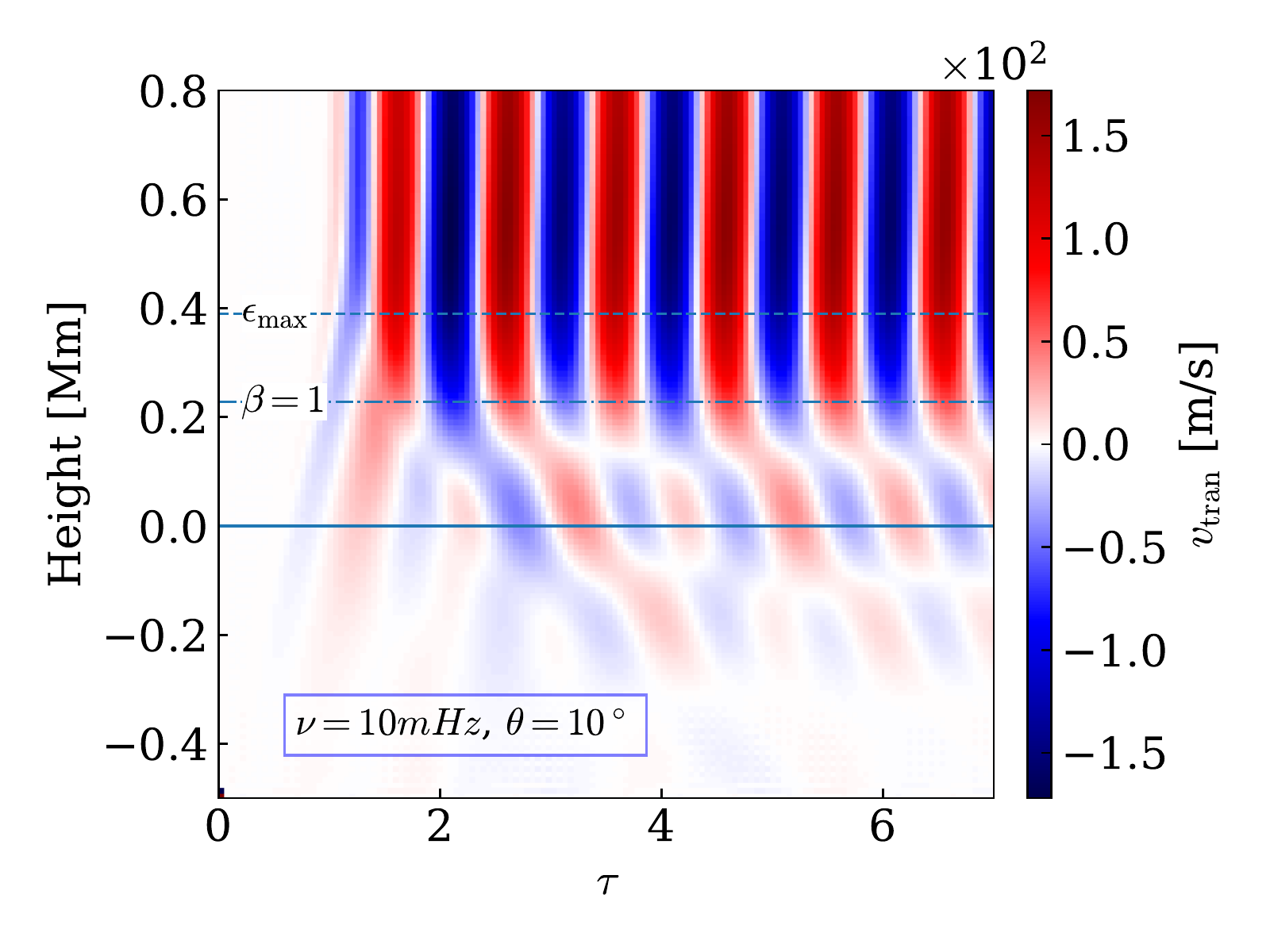}
 \includegraphics[width=0.32\textwidth]{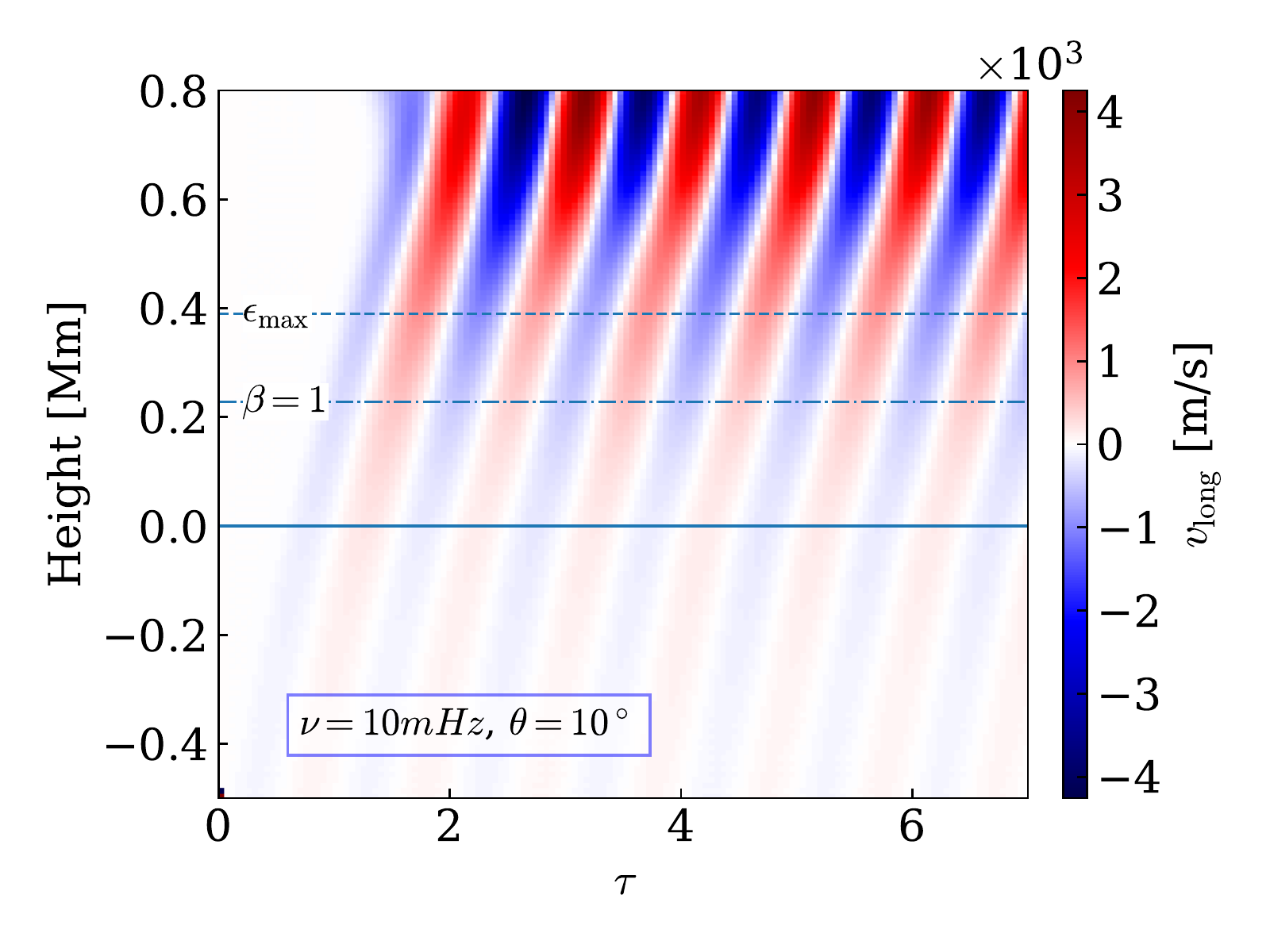}
 \caption{\footnotesize{Time-height diagrams showing the velocity projections according to Eqs. \ref{eq:pro} in the simulation with Hall term switched on. The background magnetic field was inclined by $\theta=10^\circ$, and a frequency was of $\nu=10$ mHz. From left to right, velocity projections into $\mathbf{e}_\mathrm{perp}$ (\alfven wave), $\mathbf{e}_\mathrm{tran}$ (fast wave), and $\mathbf{e}_\mathrm{long}$ (slow wave) directions. The units of the velocity colour scale are m/s. The horizontal axis is dimensionless ($\tau=\nu t$). Horizontal lines indicate the photospheric level ($z=0$), the equipartition layer ($\beta =1$), and the maximum for the Hall parameter ($\epsilon_\mathrm{max}$). Notice that the $v_y$ component is non-zero indicating the presence of the \alfven wave, generated after the mode transformation. }}
 \label{fig:projection}    
\end{figure*}
%

\section{Alfv\'en wave production in the warm plasma}\label{sec:numexp}
%

\begin{figure*}
 \centering
 \includegraphics[width=0.32\textwidth]{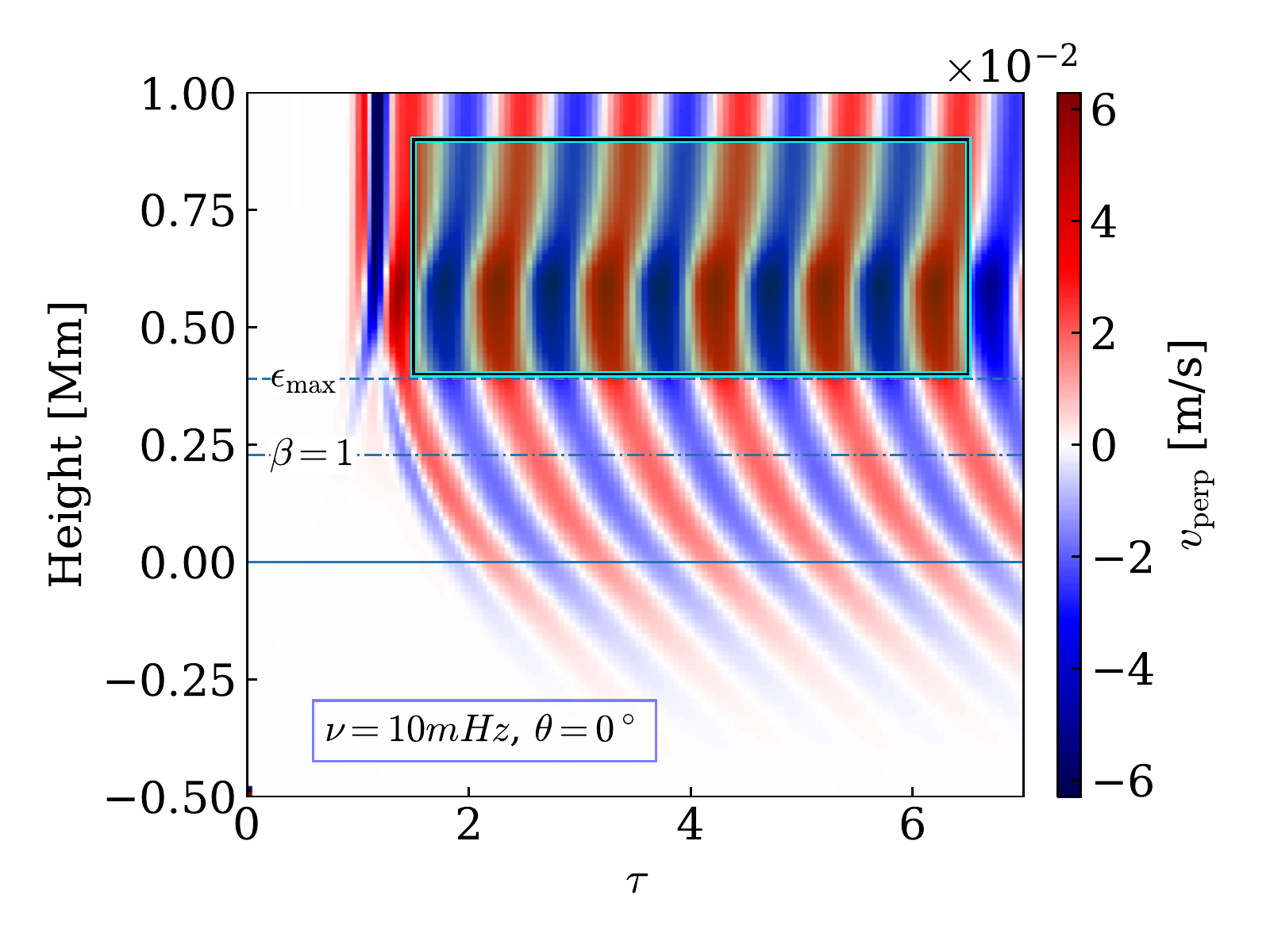}
 \includegraphics[width=0.32\textwidth]{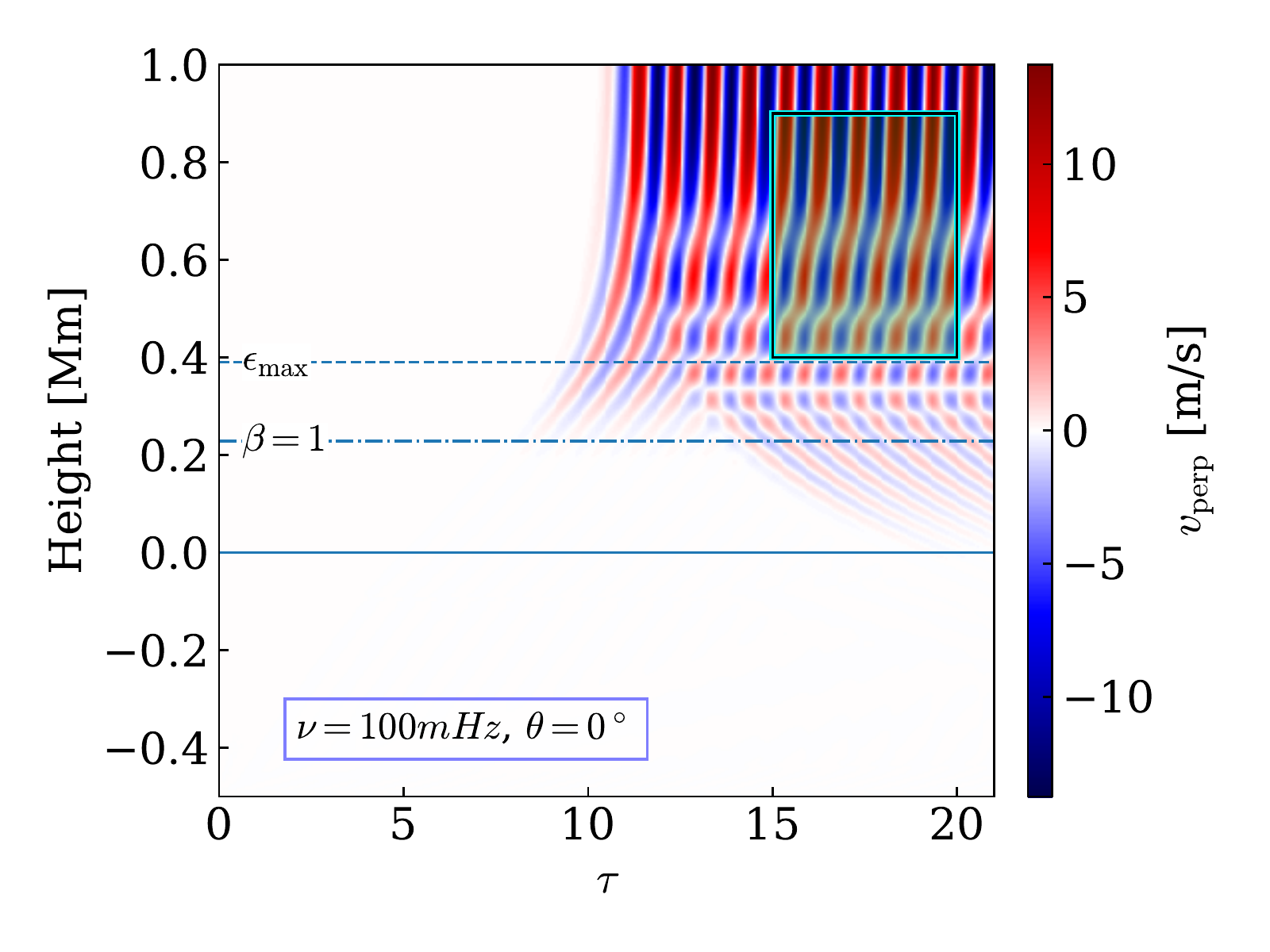}
 \includegraphics[width=0.32\textwidth]{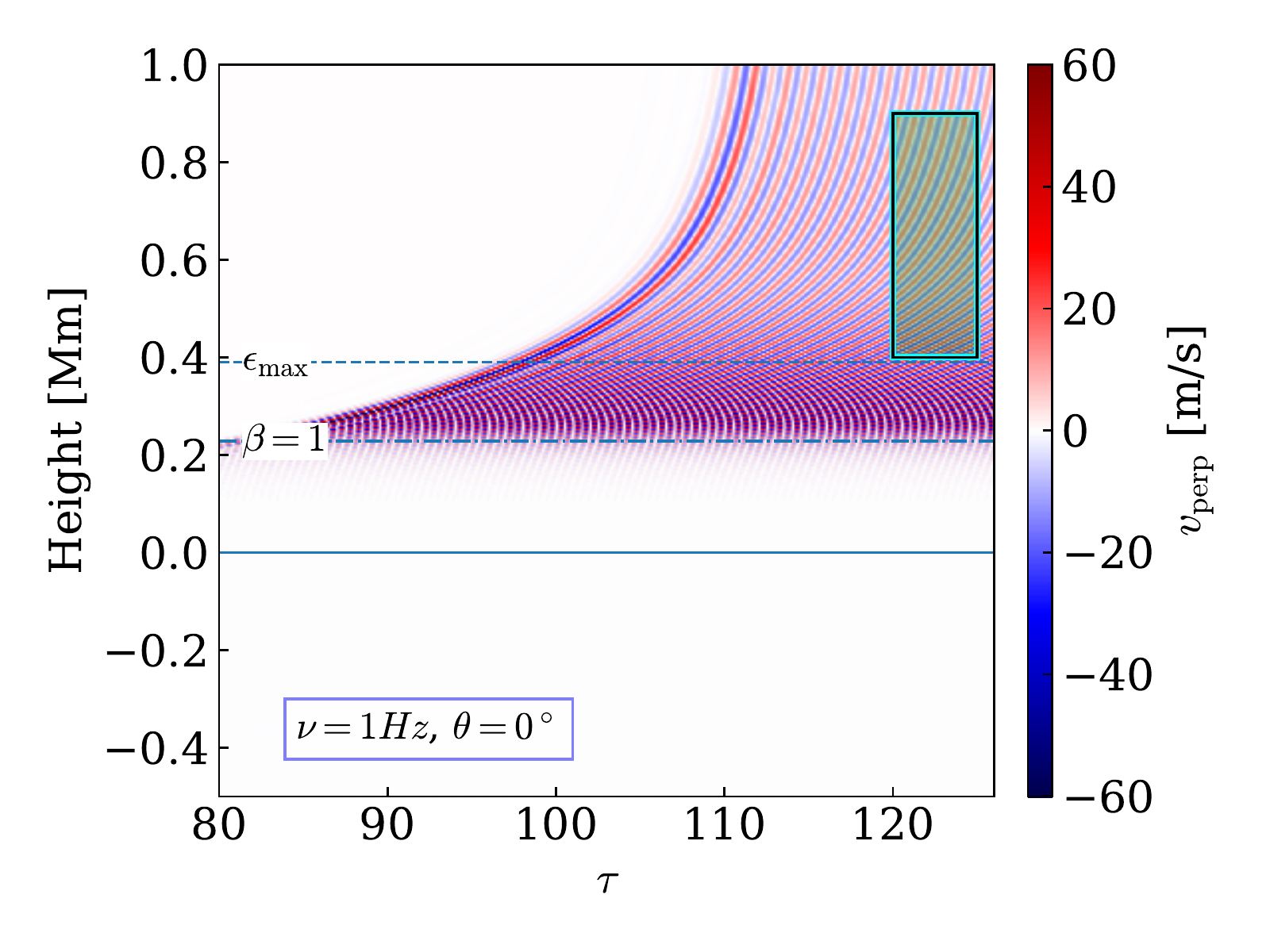}
  
 \includegraphics[width=0.32\textwidth]{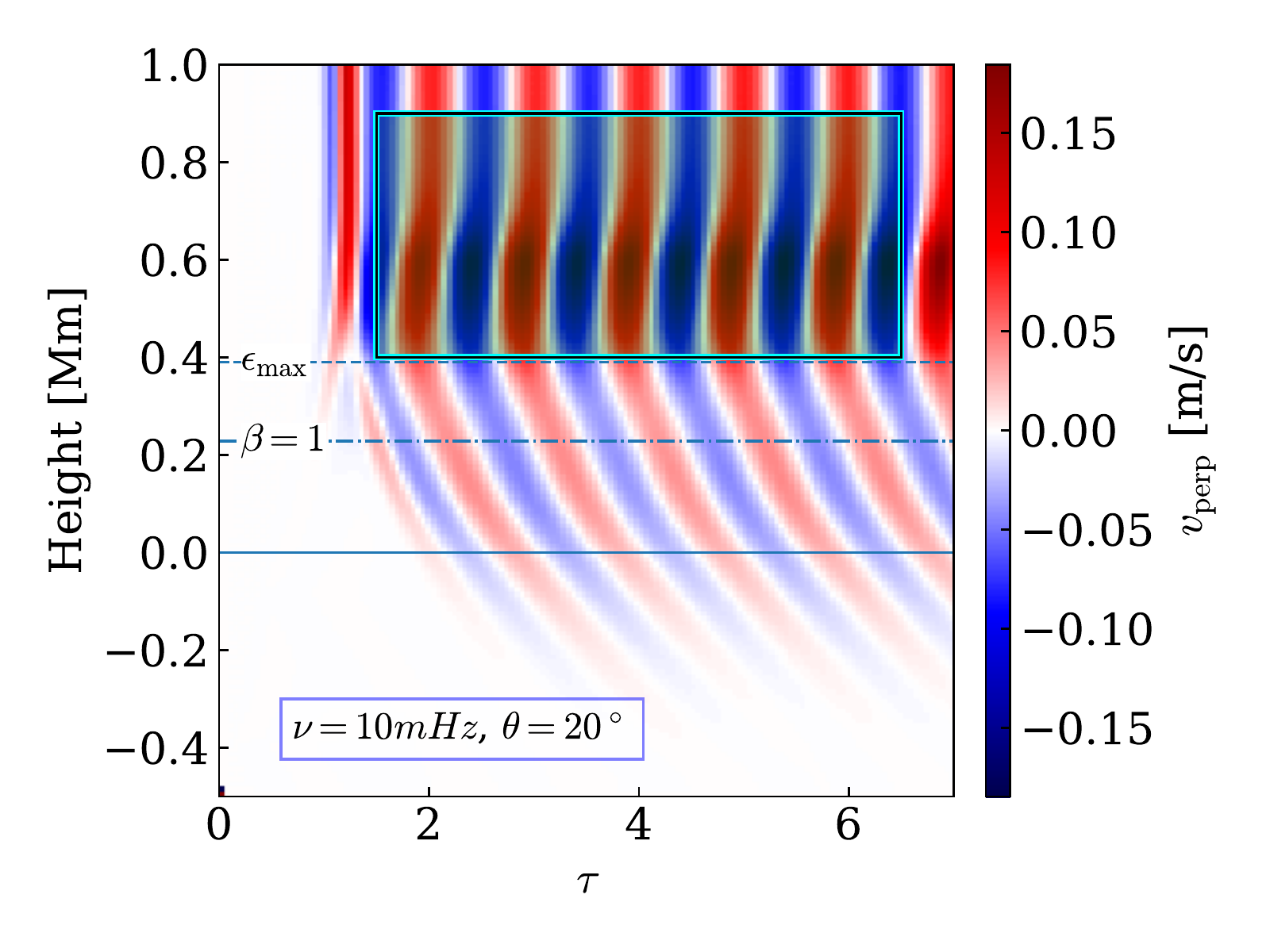}
 \includegraphics[width=0.32\textwidth]{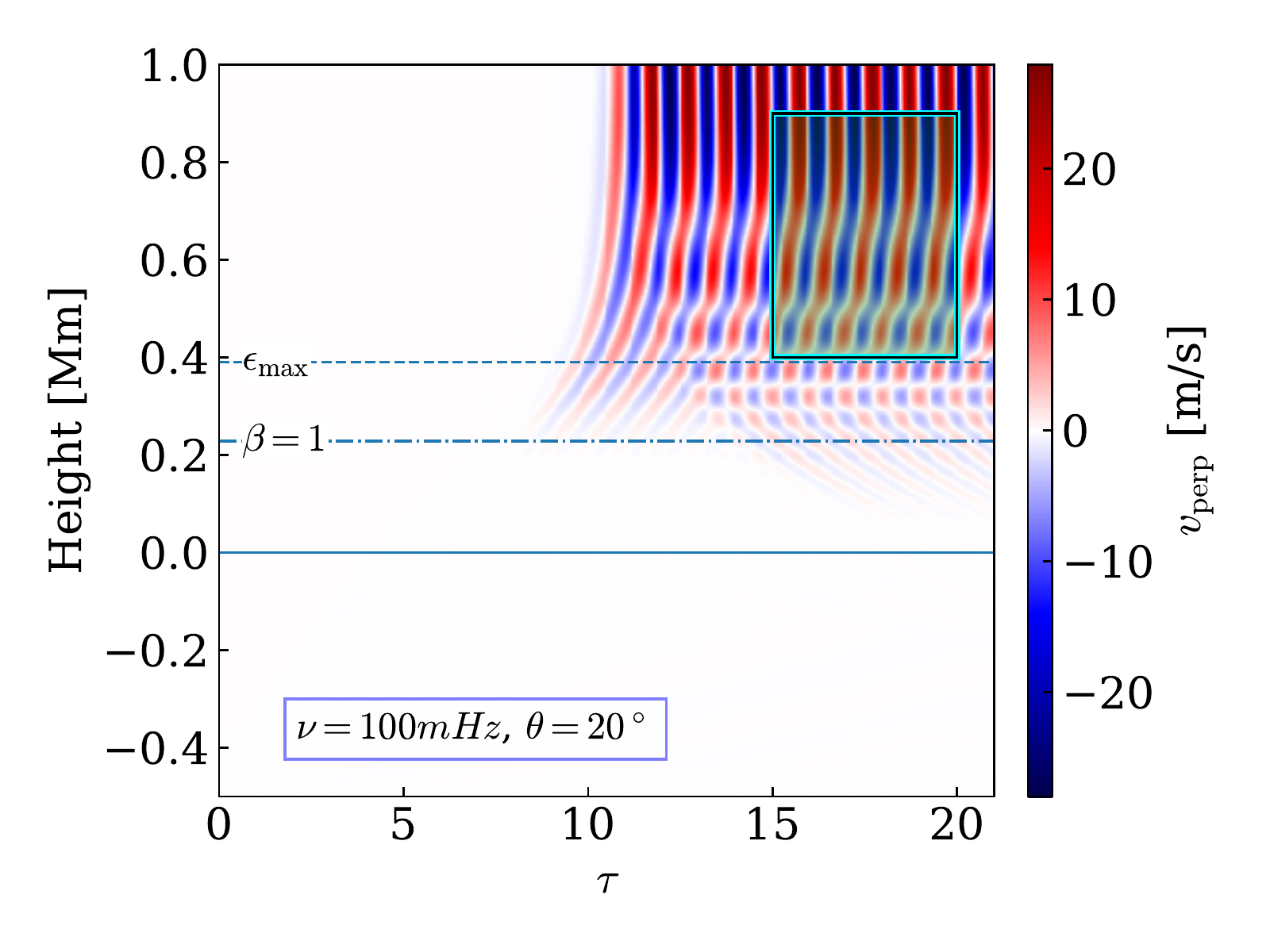}
 \includegraphics[width=0.32\textwidth]{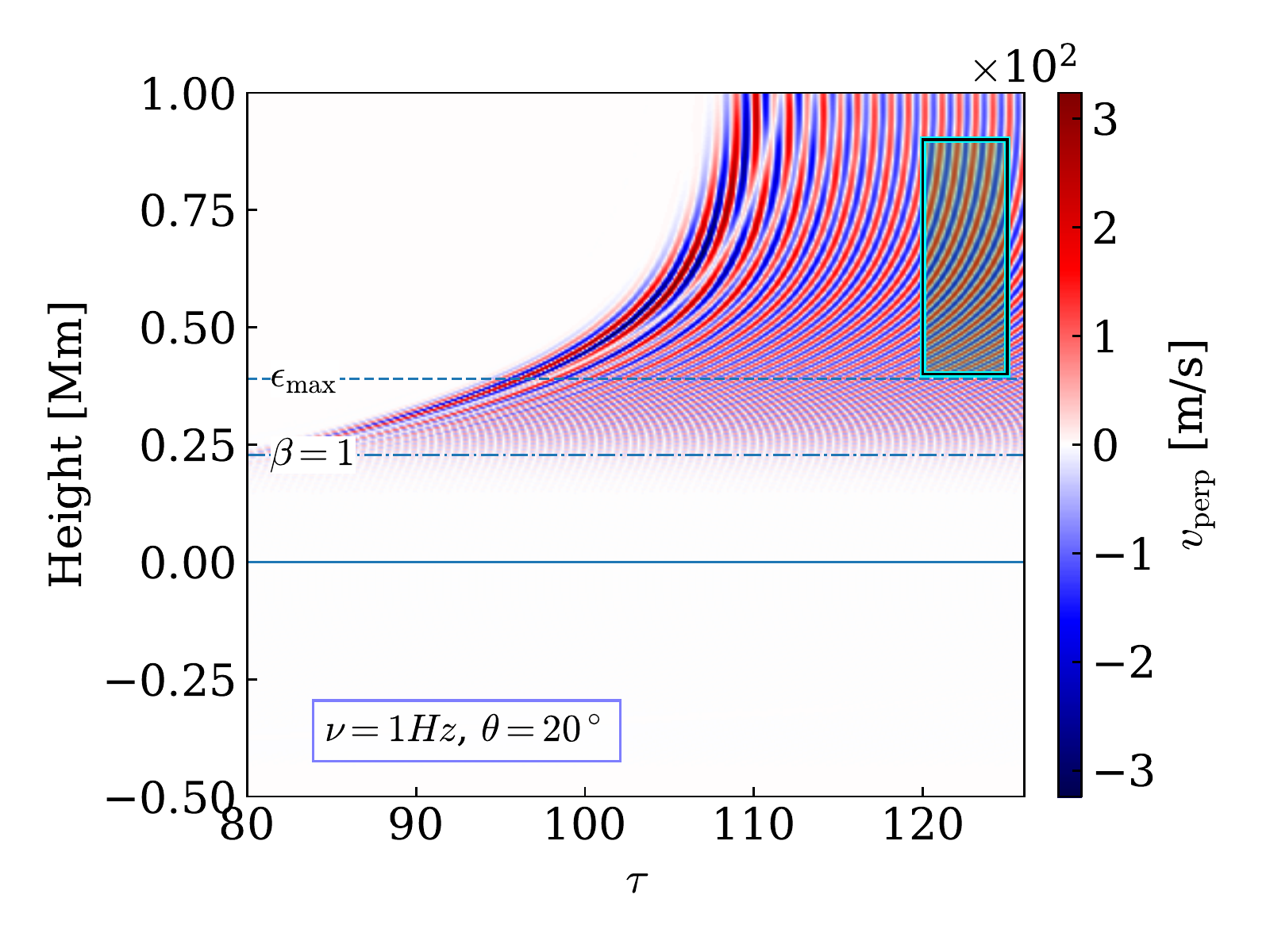}
  
 \includegraphics[width=0.32\textwidth]{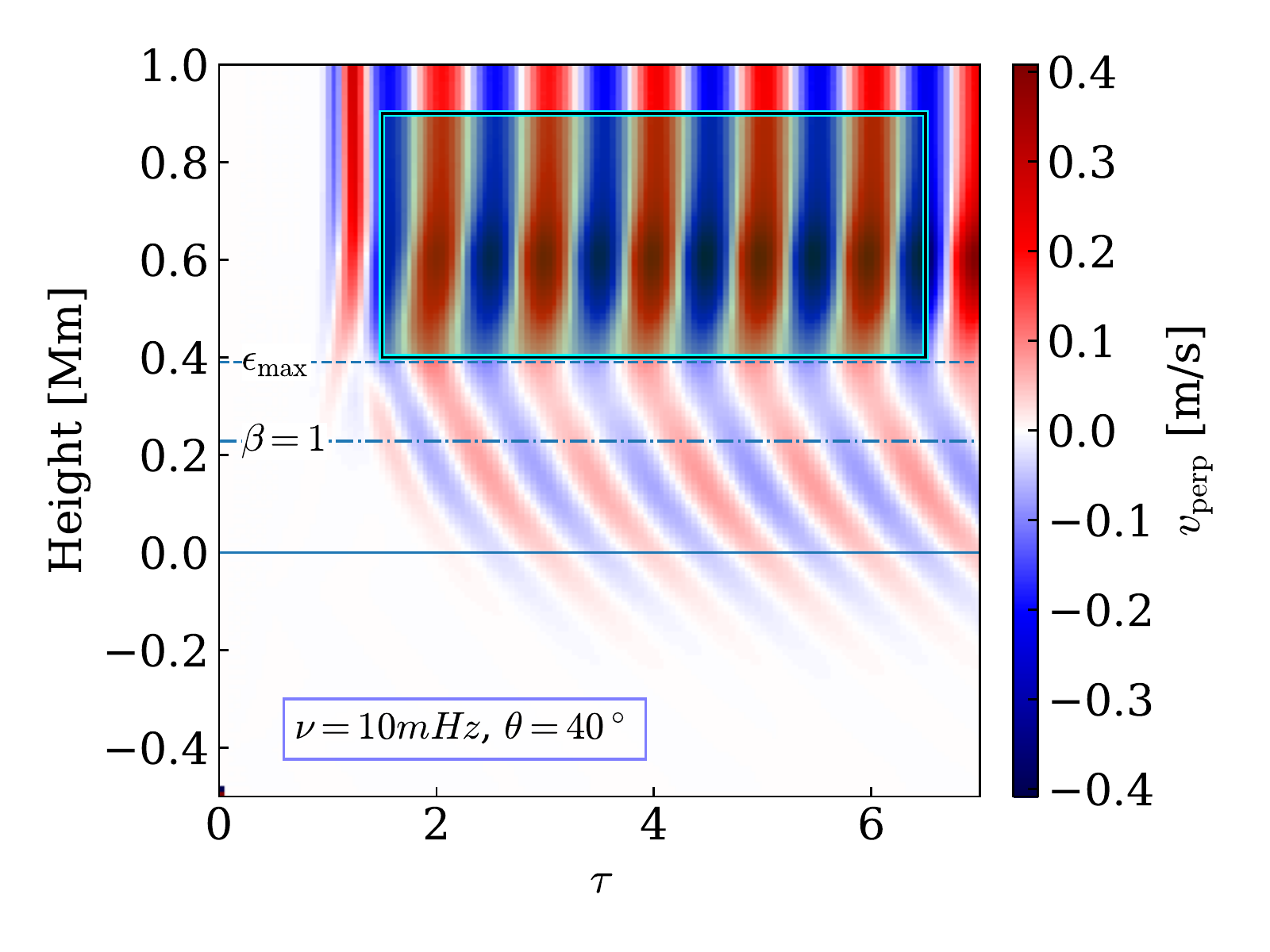}
 \includegraphics[width=0.32\textwidth]{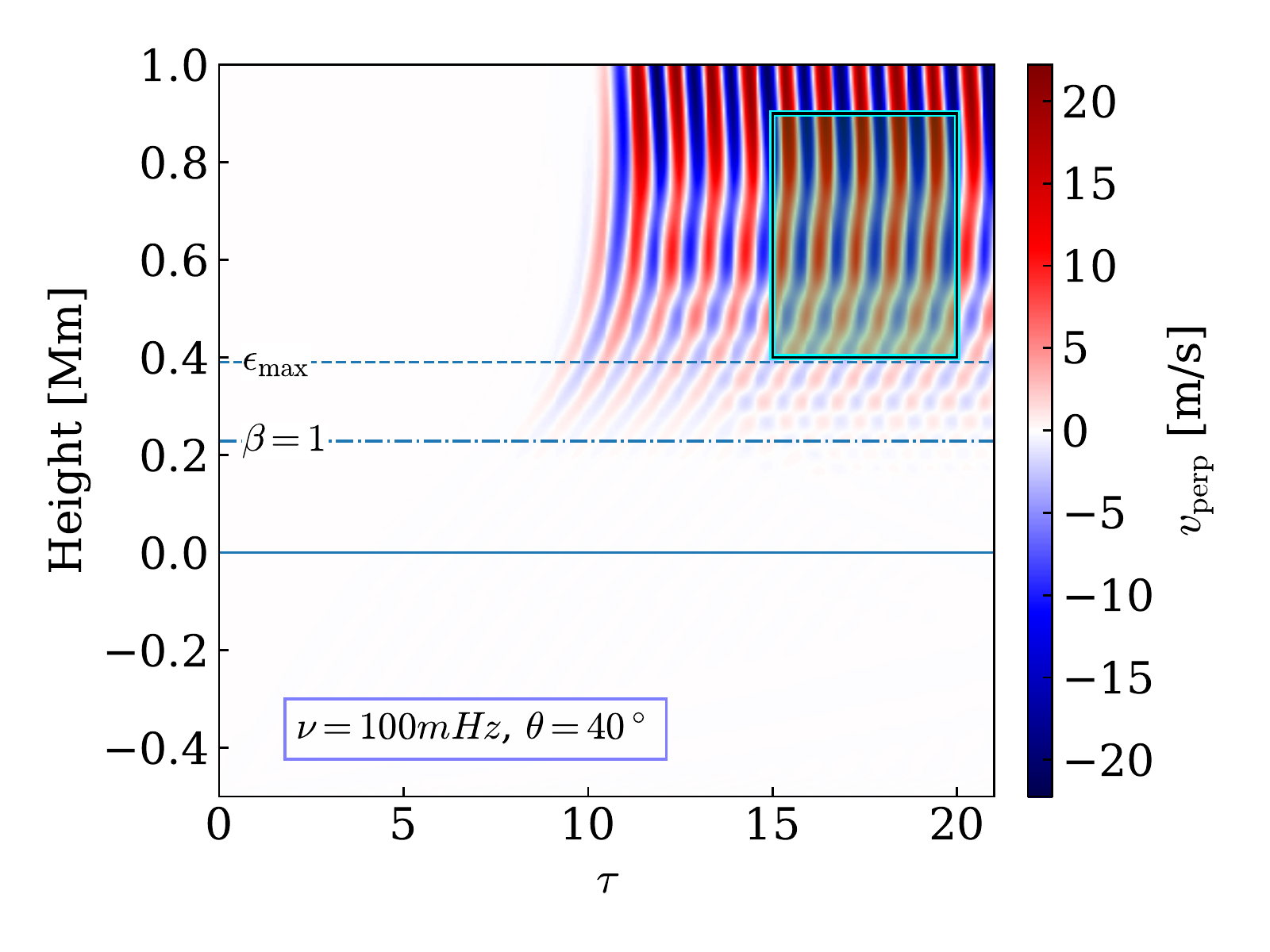}
 \includegraphics[width=0.32\textwidth]{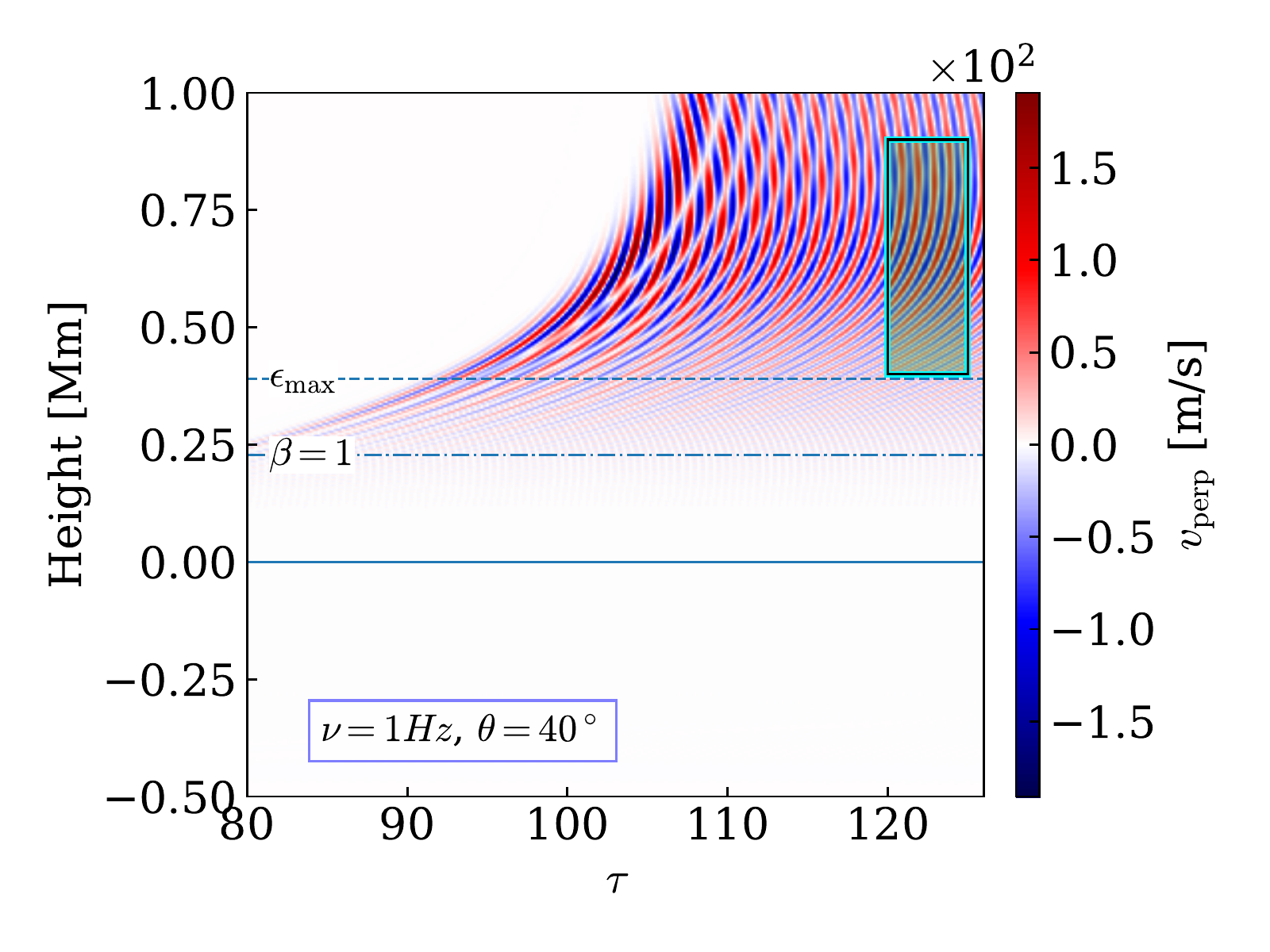} 
 
 \includegraphics[width=0.32\textwidth]{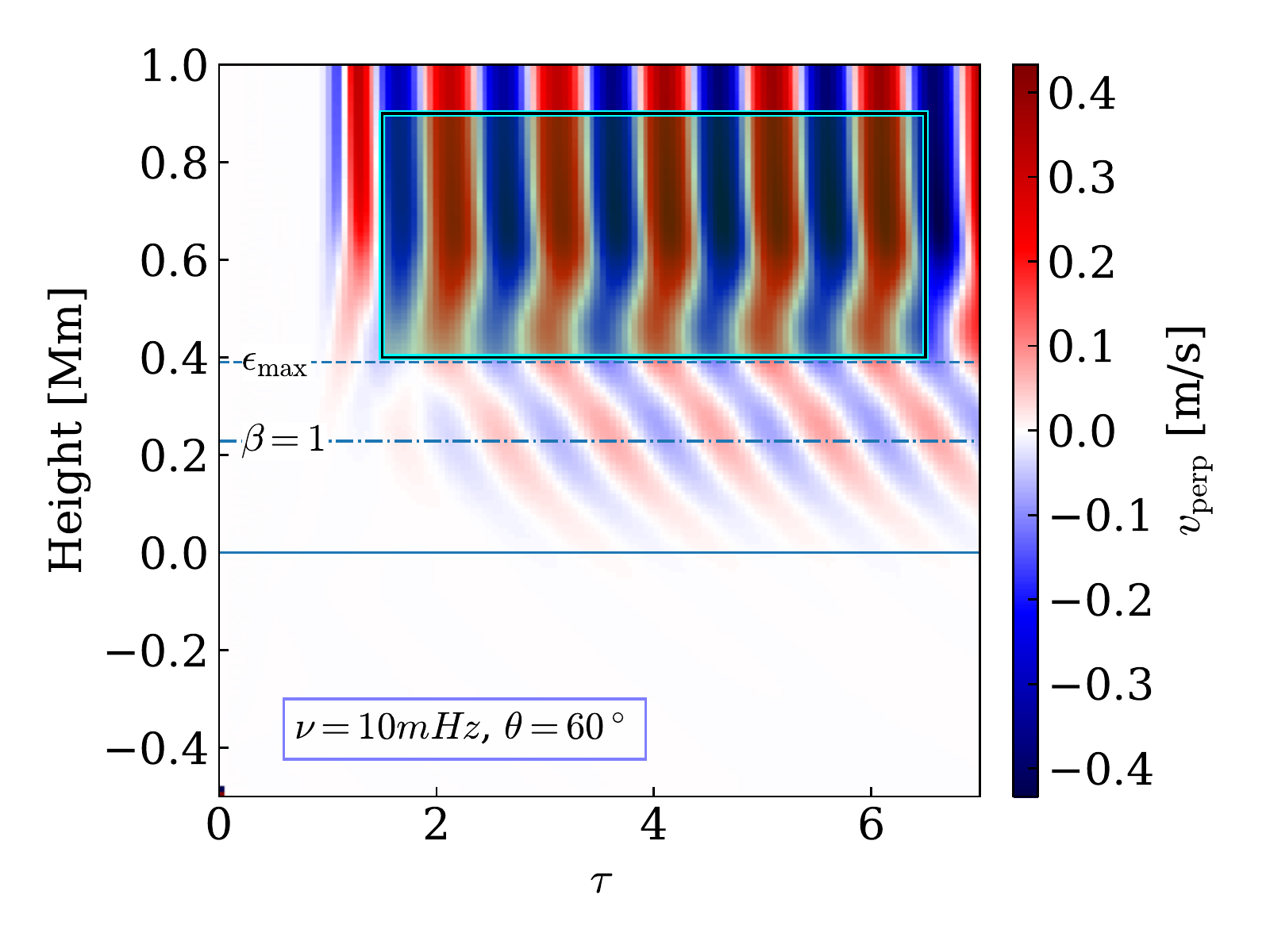}
 \includegraphics[width=0.32\textwidth]{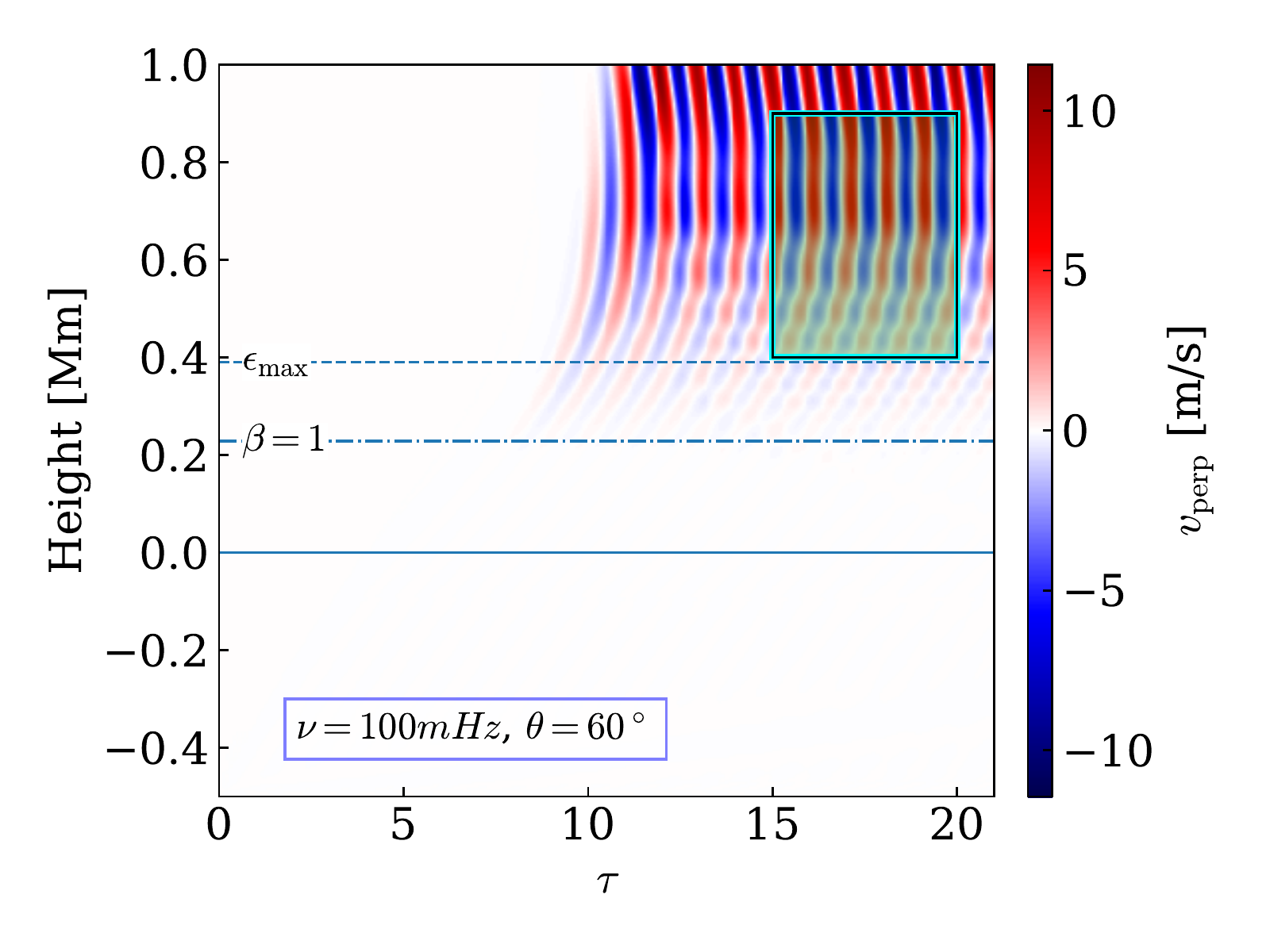}
 \includegraphics[width=0.32\textwidth]{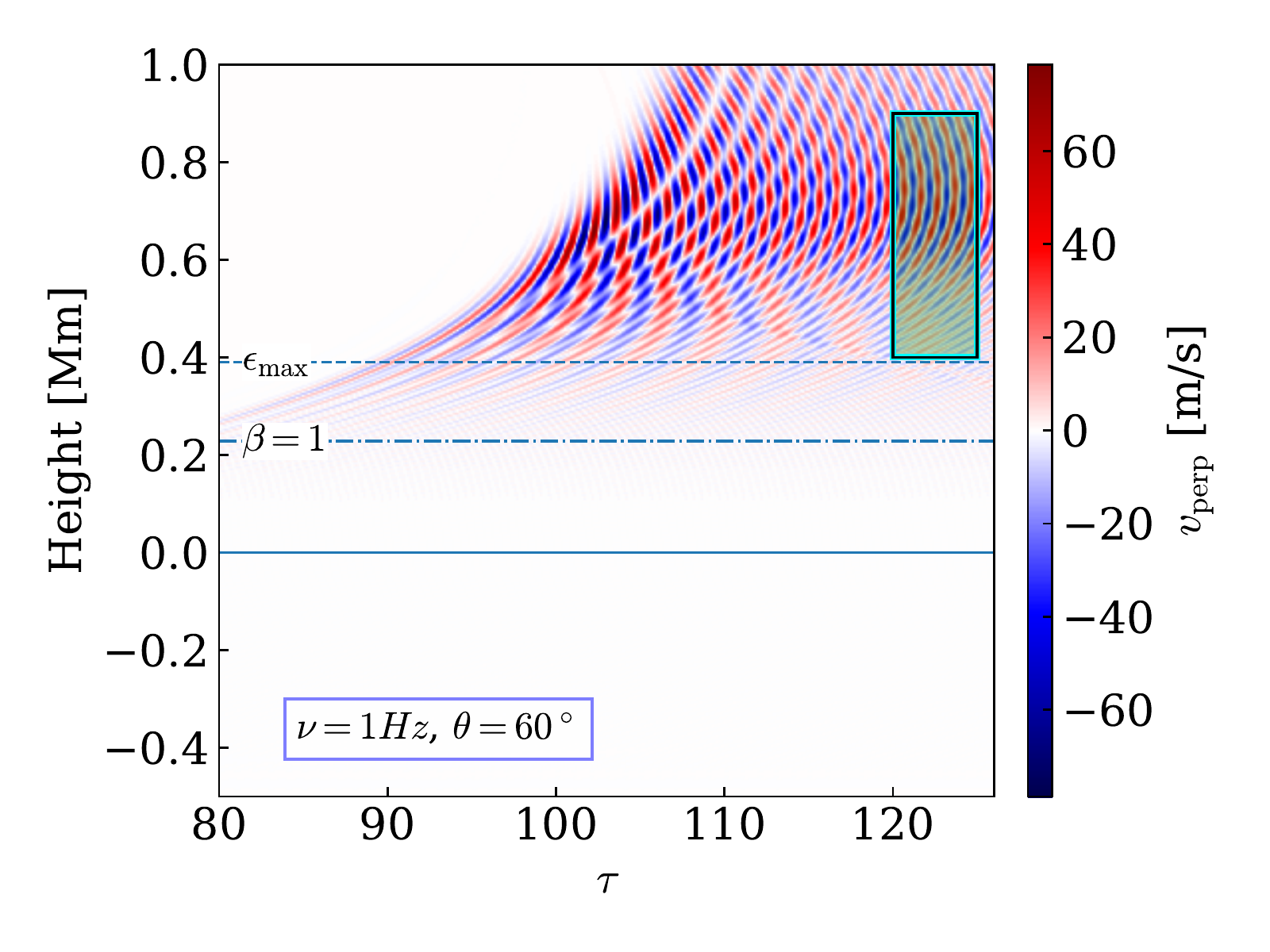} 
 
 \includegraphics[width=0.32\textwidth]{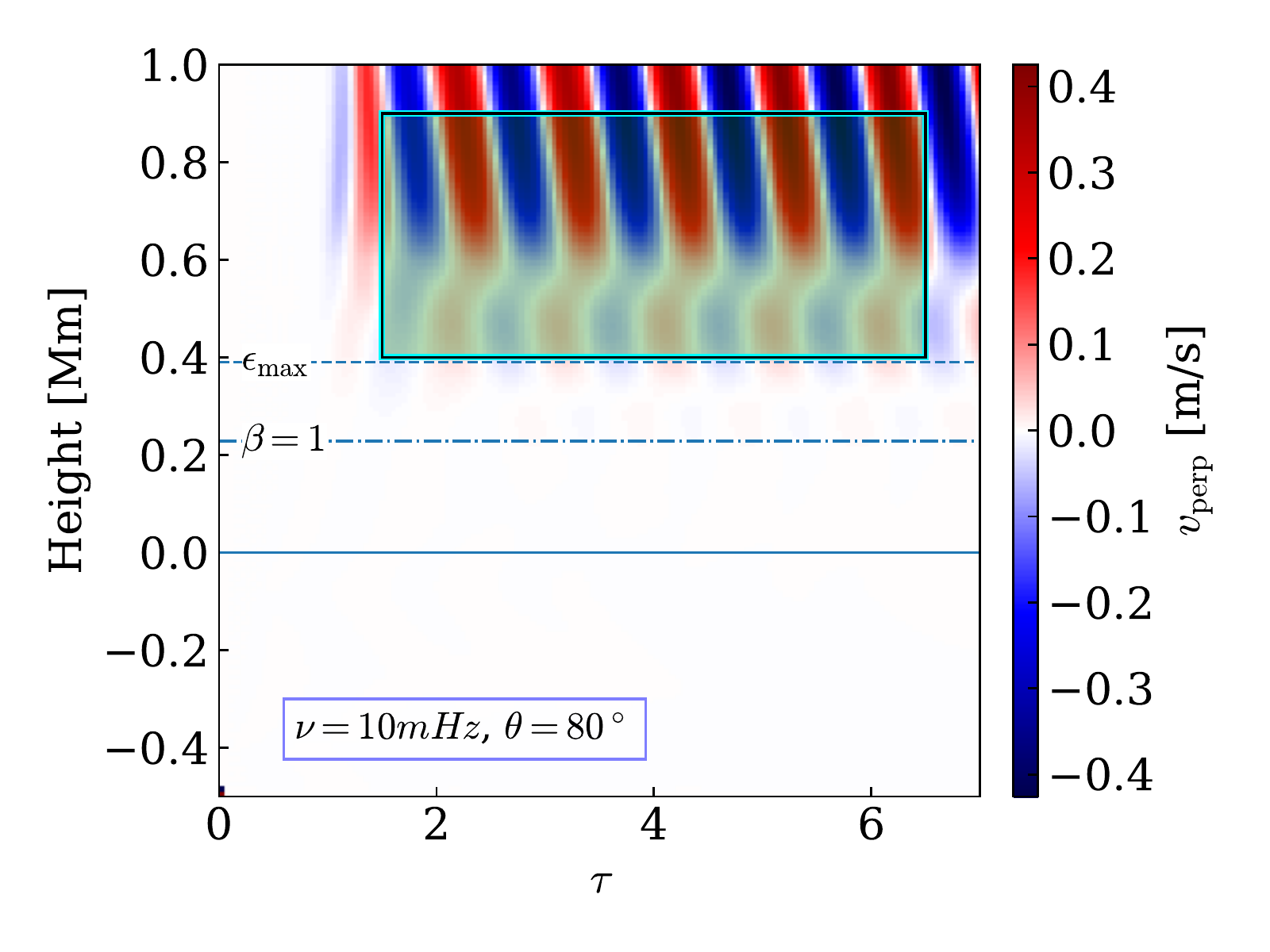}
 \includegraphics[width=0.32\textwidth]{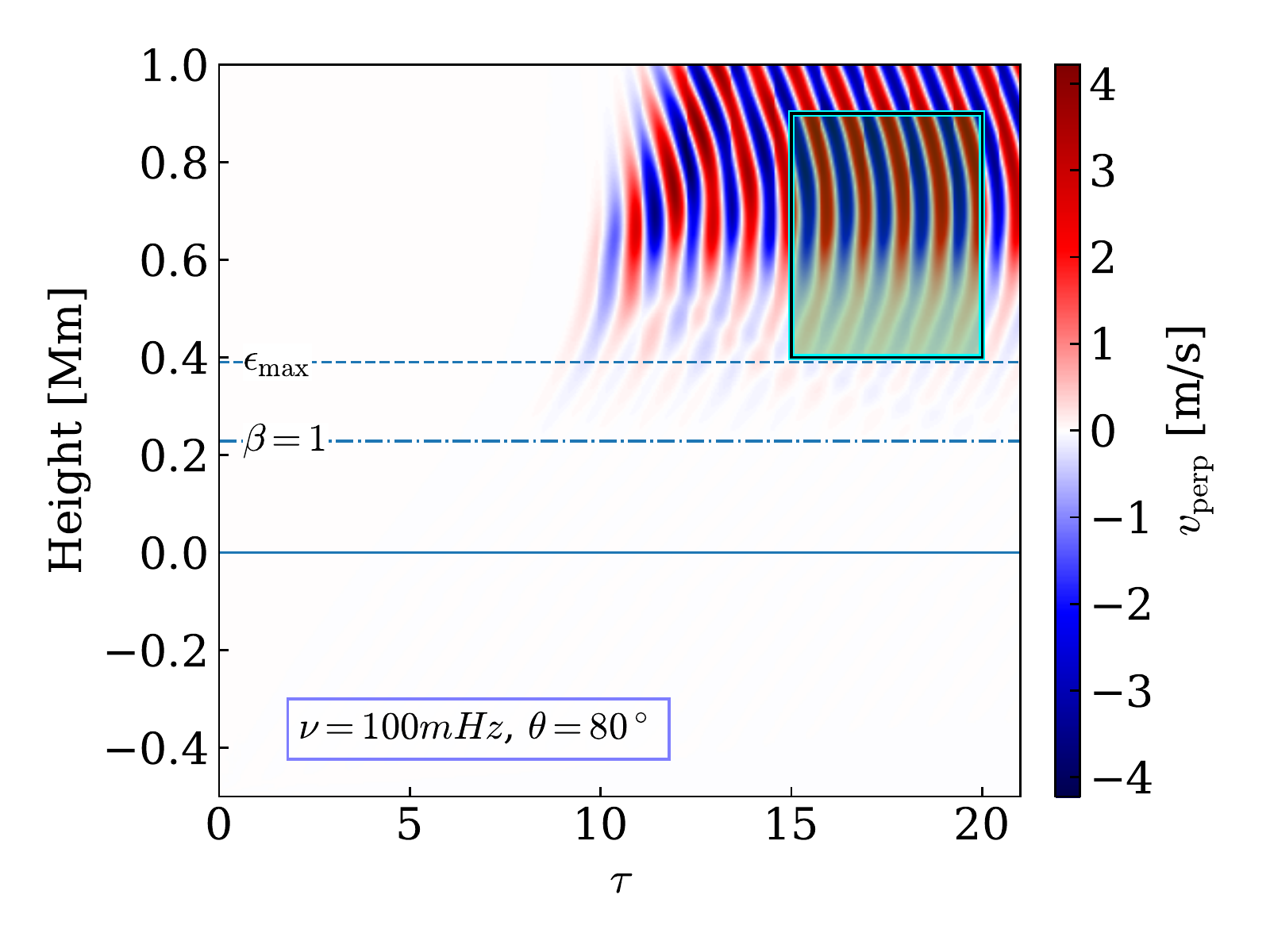}
 \includegraphics[width=0.32\textwidth]{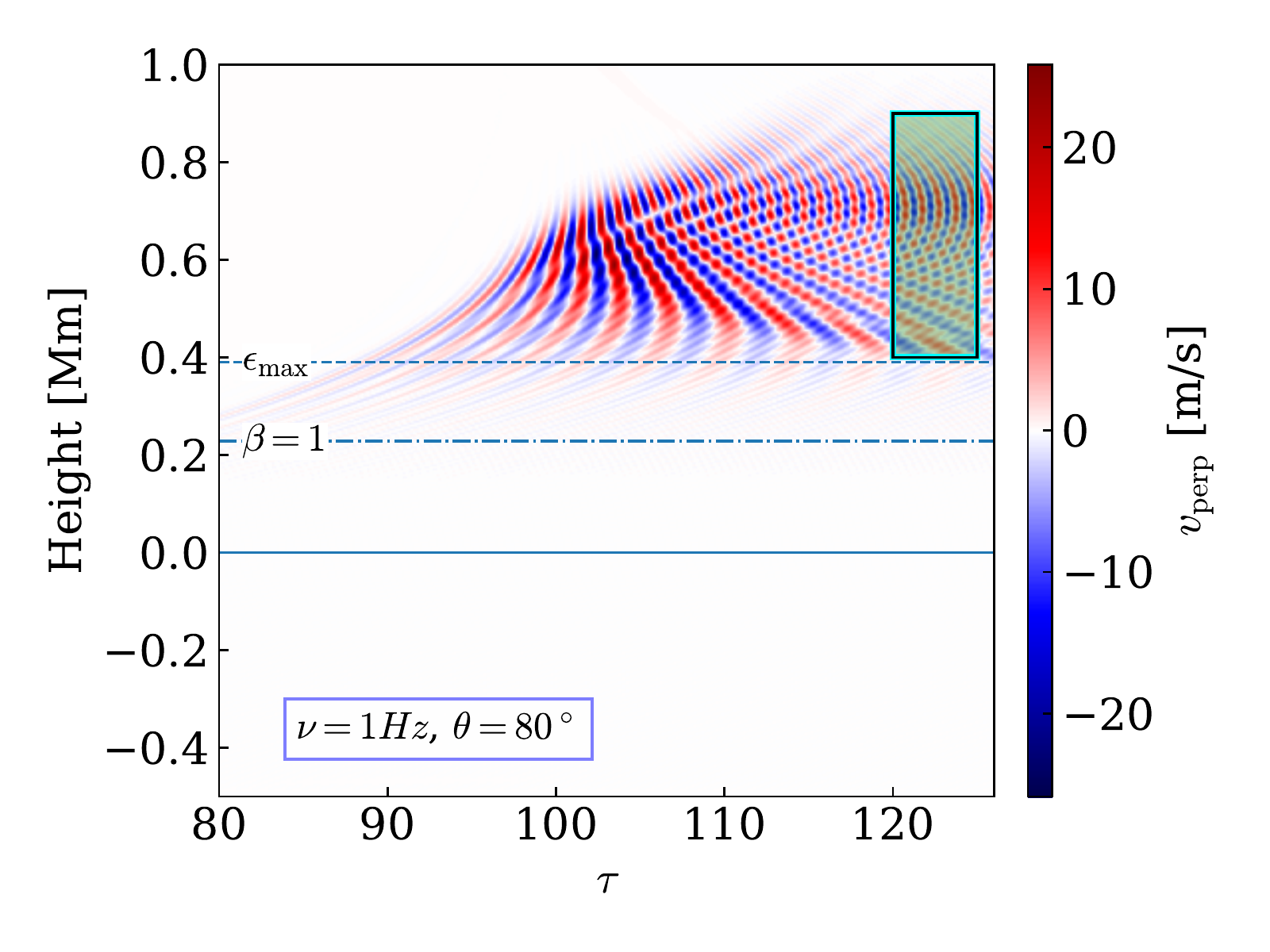}
 \vspace{-0.5cm}
 \caption{\footnotesize Time-height diagrams of the $v_\mathrm{perp}=v_y$ (\alfven wave) velocity component in simulations with varying frequency (panels from left to right) and varying inclination angle of the magnetic field (panels from top to bottom). The \alfven waves are generated through Hall mediated mode transformation. The values of the frequencies and inclination angles are indicated in each panel. The square box marks the region used to calculate the wave amplitudes in the stationary regime in the simulations summarized in Fig.~\ref{fig:maxvperp}.}
 \label{fig:vperp} 
\end{figure*}
%

First we confirm that Hall-current mediated transformation is indeed taking place in our model. For that we separate the contribution of the three wave modes. This can be done relatively easily in our numerical experiment, given the 2.5D setup and the knowledge of the direction of the wave vector and the magnetic field vector direction. We calculated the projection of the velocity vector following \citet{2008SoPh..251..251C}. This decomposition has been successfully used previously to identify the three wave modes propagating into a magnetized medium in similar models \citep[e.g.][]{2010ApJ...719..357F, Khomenko:2011fx, Khomenko:2012gw, Felipe:2012kq}. Figure \ref{fig:esquemas}b shows a schematic diagram for the decomposition. The longitudinal component of the velocity given by $\mathbf{e}_\mathrm{long}$, the one parallel to the magnetic field, selects the slow magneto-acoustic wave in a low-$\beta$ plasma. The other two components are perpendicular to the magnetic field. The component perpendicular to both $\mathbf{B}$ and $\nabla p$ (direction of the background pressure gradient, coinciding with the direction of gravitational stratification), given by $\mathbf{e}_\mathrm{perp}$, selects the \alfven wave. Finally, the component perpendicular to the previous two, $\mathbf{e}_\mathrm{tran}$, selects the fast magneto-acoustic wave in the low-$\beta$ plasma. Mathematically, this new basis can be written as: 
\begin{eqnarray}\label{eq:pro}
 \mathbf{e}_\mathrm{long} &=& (\cos{\phi}\sin{\theta}, \sin{\phi}\sin{\theta}, \cos{\theta}), \\ \nonumber
\mathbf{e}_\mathrm{perp} &=& (-\cos{\phi}\sin^2{\theta}\sin{\phi}, 1-\sin^2{\phi}\sin^2{\theta}, \\ \nonumber
& &-\cos{\theta}\sin{\theta}\sin{\phi}), \\ \nonumber
 \mathbf{e}_\mathrm{tran} &=& (-\cos{\theta}, 0, \cos{\phi}\sin{\theta}),
\end{eqnarray}
where $\theta$ is the magnetic field angle with the vertical $z$ axis, and the $\phi$ the azimuth angle measured from the $XZ$-plane. 

In a 2.5D case, we set the azimuth $\phi=0$, so the expressions for the components simplifies to:
\begin{eqnarray}\label{eq:pro2}
\mathbf{e}_\mathrm{long} &=& (\sin{\theta}, 0, \cos{\theta}), \\ \nonumber
\mathbf{e}_\mathrm{perp} &=& (0, 1, 0), \\ \nonumber
\mathbf{e}_\mathrm{tran} &=& (-\cos{\theta}, 0, \sin{\theta}),
\end{eqnarray}
this means that $v_\mathrm{perp}=v_y$, so the $v_y$ component of the velocity field is the one which selects \alfven waves.

A note of caution must be taken regarding the above decomposition. Following the book of  \cite{2003ASSL..294.....G}, chapter 5, the component $\mathbf{e}_\mathrm{perp}$ chooses the asymptotic polarization direction for the \alfven mode in a plasma with any value of $\beta$. However, two other directions, $\mathbf{e}_\mathrm{long}$ and $\mathbf{e}_\mathrm{tran}$ generally provide a mixture of the fast and slow magneto-acoustic modes. The particular contribution of each of the modes into $\mathbf{e}_\mathrm{long}$ and $\mathbf{e}_\mathrm{tran}$ depends on plasma $\beta$, see Appendix. In the limit of low-$\beta$, as mentioned above, most contribution to $\mathbf{e}_\mathrm{long}$ comes from longitudinal slow magneto-acoustic mode, while the direction $\mathbf{e}_\mathrm{tran}$ selects fast magneto-acoustic mode in this case.

As mentioned in Sec. \ref{sec:num_setup}, our source produces acoustic-gravity waves at the bottom of the numerical box. Since the plasma $\beta$ is large there, the waves generated by the source are fast, essentially acoustic, waves. These waves propagate upwards and suffer a first mode transformation at the equipartition layer where the acoustic and \alfven speeds coincide, splitting into the slow (essentially acoustic) wave component and fast (essentially magnetic) wave component in $v_A>c_s$. Due to the geometry of our numerical experiment setup, the wave vector $\mathbf{k}$ and the magnetic field vector lie in the same vertical plane (the $x$--$z$ plane in Figure \ref{fig:esquemas}b). Therefore, in ideal MHD with the Hall term switched off, only fast and slow MHD waves can exist in our system, and no transformation to the \alfven wave can take place. Mathematically, the velocity component $v_y$ is exactly zero, since there is no coupling out of the $x$--$z$ plane. 

However, when the Hall term is switched on, a secondary mode transformation can take place. This transformation happens when the fast magnetic mode, generated by the primary mode transformation at $v_A=c_s$, enters into the region where the Hall parameter becomes important, see Figure \ref{fig:ehall}. This way, \alfven waves are produced from fast magnetic waves. This double mode transformation can be seen in Figure \ref{fig:projection} \footnote{Fast-\alfven coupling is also produced by 3D geometric effects, though these are absent in the 2.5D system.}.

In the first example, we see in the right panel of Figure \ref{fig:projection} how the fast (acoustic) wave propagates with the acoustic speed upwards and how the fast and slow waves are generated after the primary mode transformation around 200 km height in the photosphere. The existence of both fast and slow components can be appreciated from different inclination of ridges in the middle and right panels. The fast (magnetic) mode ridges are much steeper above 200 km, indicating faster propagation speed. The fast magnetic mode can also be observed reflecting back to the sub-photosphere, which is seen as downward-inclined ridges and from the interference pattern below 200 km. The ridges become vertical above the reflection height, where the wave is evanescent. Such behaviour is well known and has been observed before in many simulations \citep{Khomenko:2009cx,Khomenko:2011fx, Felipe:2012kq, Khomenko:2012gw, Santamaria:2015boa}. The new feature one can observe in Figure \ref{fig:projection} is the generation of the \alfven mode, seen as a non-zero $v_y$ component in the left panel. This component starts to appear at heights around 400 km in the photosphere, coinciding with heights where the Hall parameter is maximum; see the left panel Fig.~\ref{fig:projection}. Nevertheless, the Hall-mediated transformation occurs throughout the height range. The amplitude of the \alfven waves produced by this mechanism is highest immediately above the height where the Hall parameter is maximum.

Although the height of the numerical box is insufficient to completely encompass the region where the fast wave reflection takes place, part of the fast mode that is reflected downward again travels through the  region with high Hall parameter values. This way, downward propagating \alfven waves are produced (ridges below $\epsilon_\mathrm{max}$ in the left panel of Fig. \ref{fig:projection}). A similar behaviour was observed by \citet{Khomenko:2011fx}, \citet{Khomenko:2012gw} and \citet{Felipe:2012kq}. However, in their simulations the \alfven waves were produced through geometrical mode transformation, and not through Hall-mediated transformation. 

One may notice that the amplitude of the \alfven waves produced through the Hall-mediated transformation in Figure \ref{fig:projection} is rather small, being about four orders of magnitude smaller than the amplitude of the slow acoustic mode at the equipartition layer. Nevertheless, the results of the theoretical investigation in cold plasma \citep{2015ApJ...814..106C} suggest that the amplitude of the \alfven waves is a sensitive function of their frequency and of the inclination between the magnetic field and the wave vector. In order to study these dependencies in warm plasmas, we have repeated the simulations of Figure \ref{fig:projection}, but with different inclinations of magnetic field (from 0 to 90 degrees) and different wave frequencies (as indicated in Table 1).


\begin{figure*}
 \centering
 \includegraphics[width=0.32\textwidth]{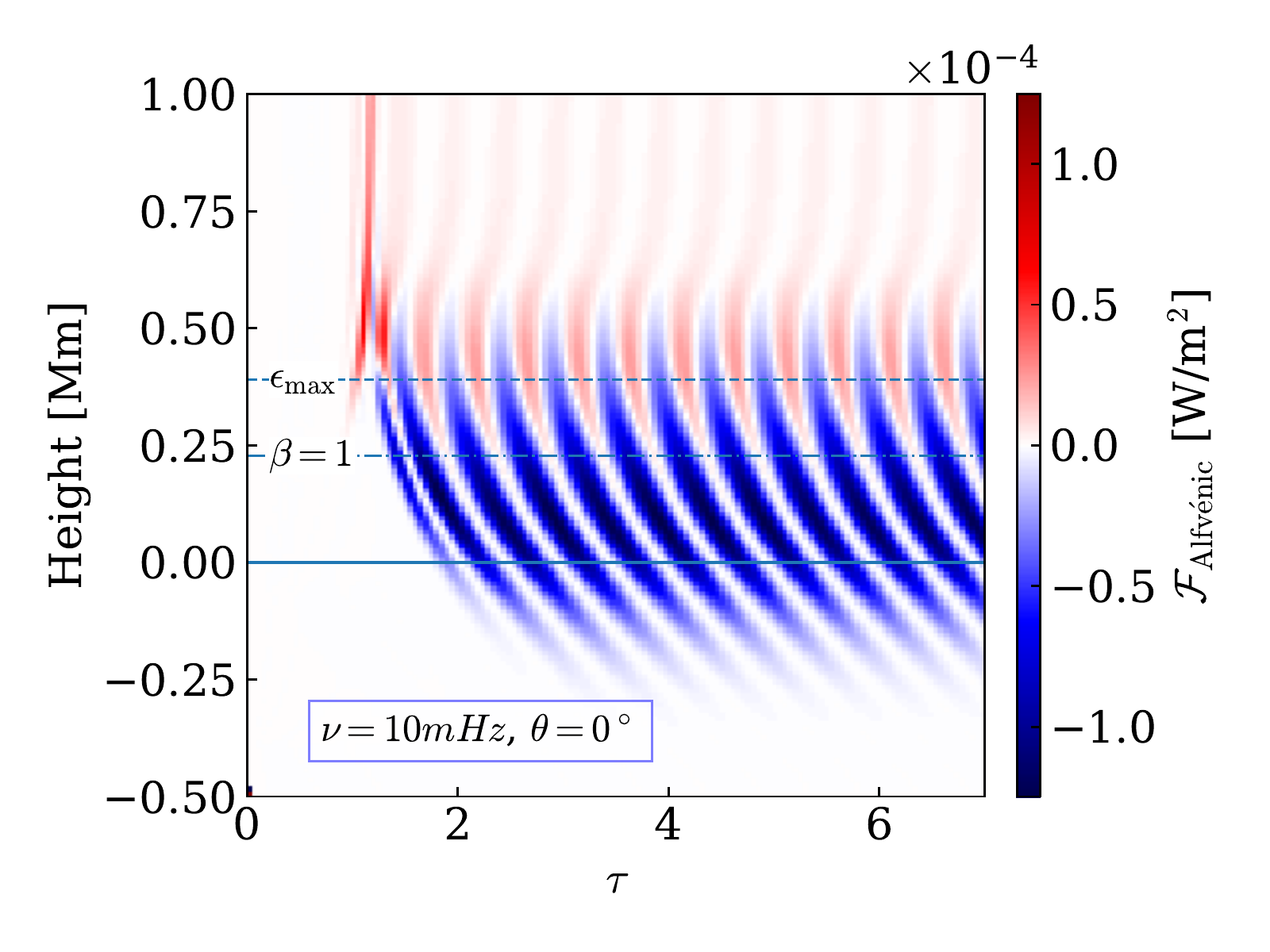}
 \includegraphics[width=0.32\textwidth]{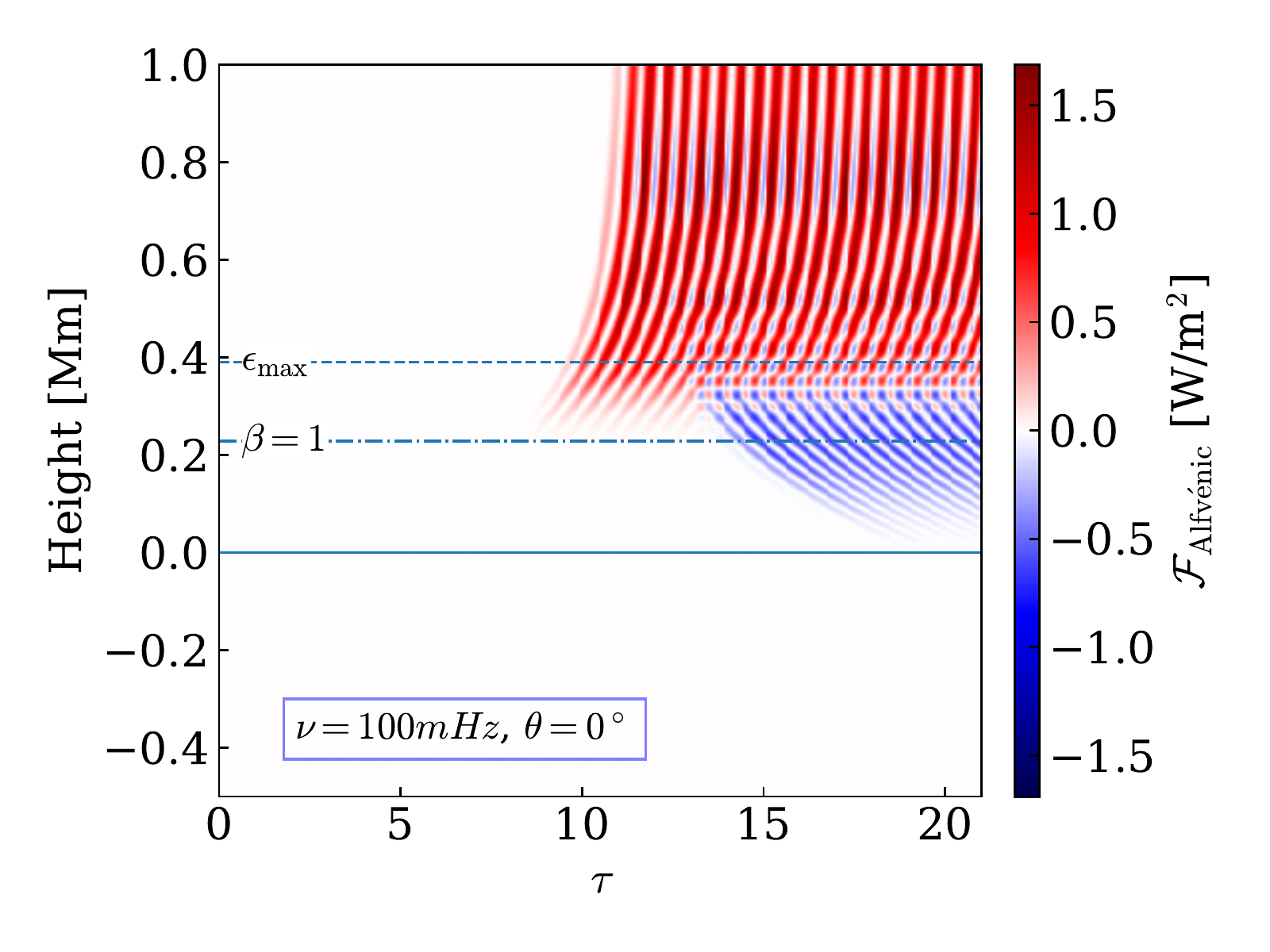}
 \includegraphics[width=0.32\textwidth]{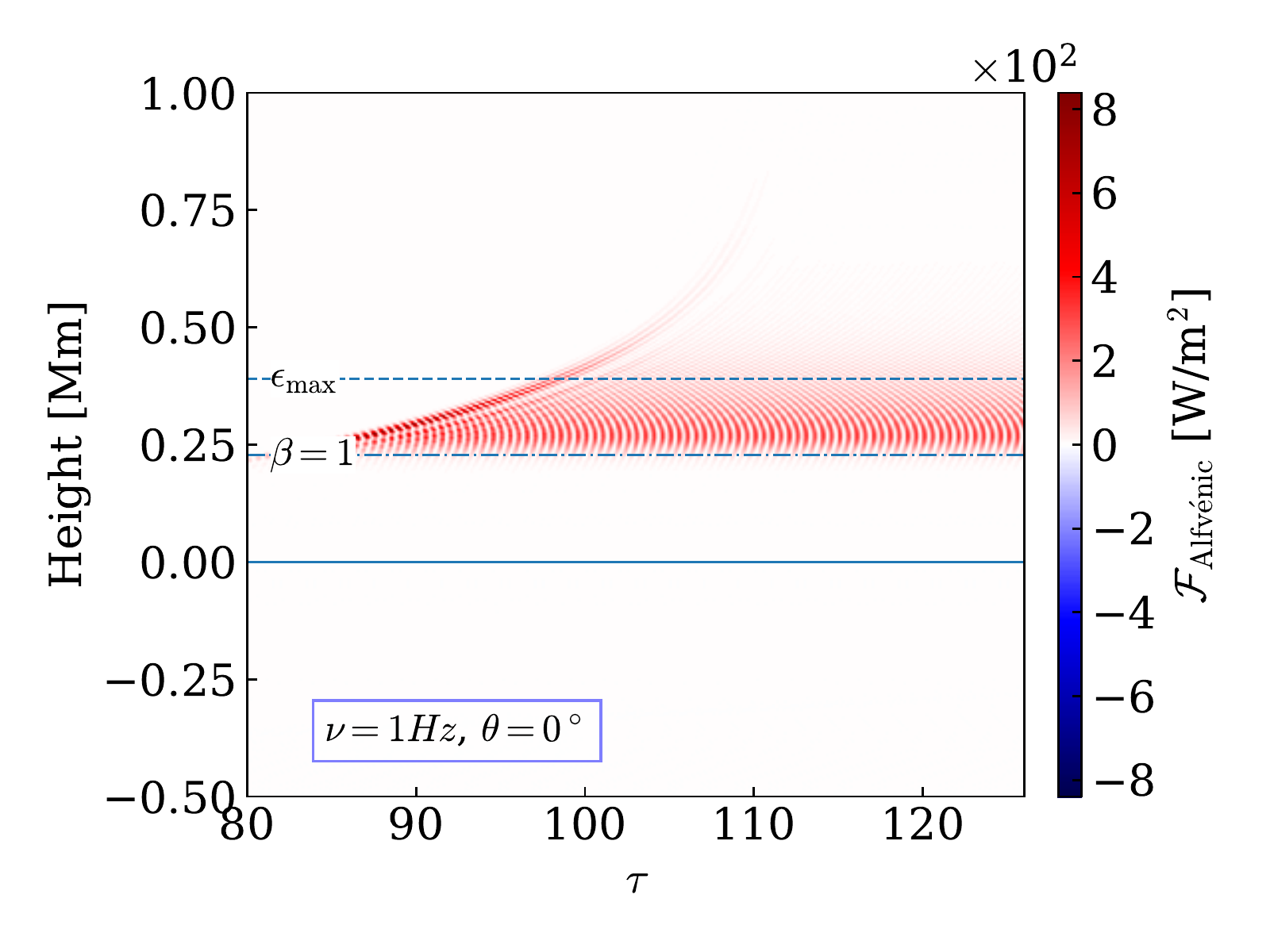}
  
 \includegraphics[width=0.32\textwidth]{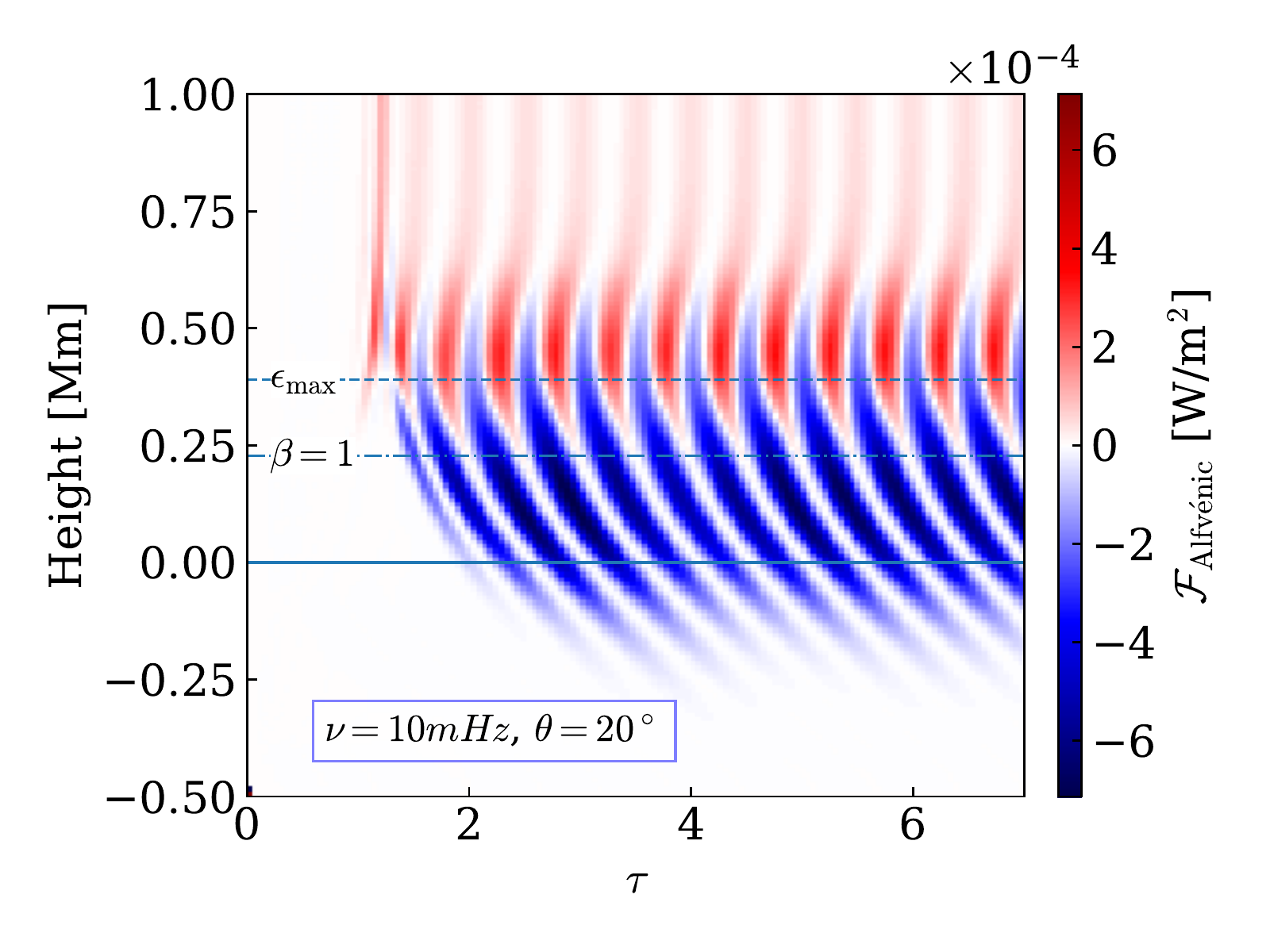}
 \includegraphics[width=0.32\textwidth]{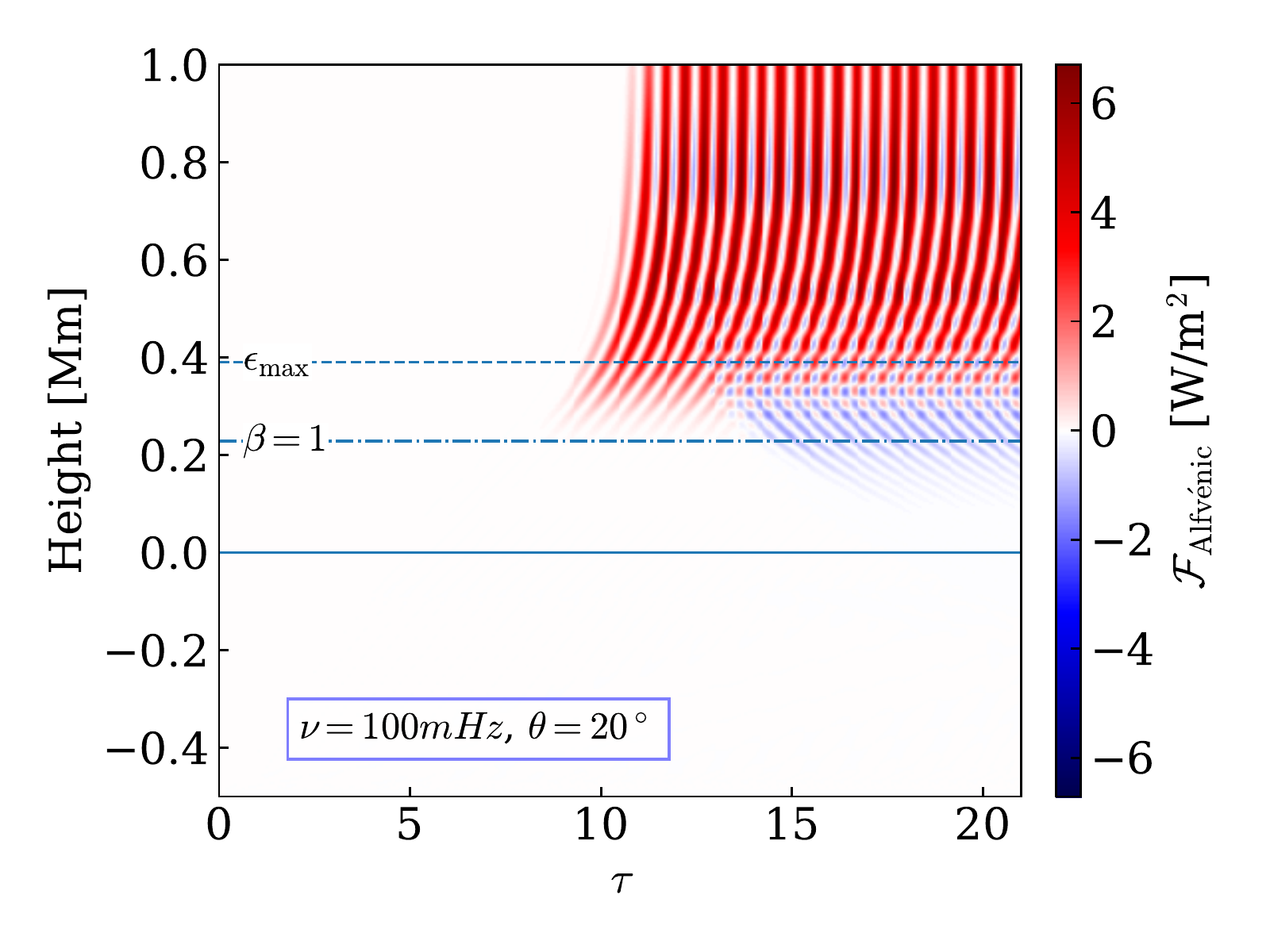}
 \includegraphics[width=0.32\textwidth]{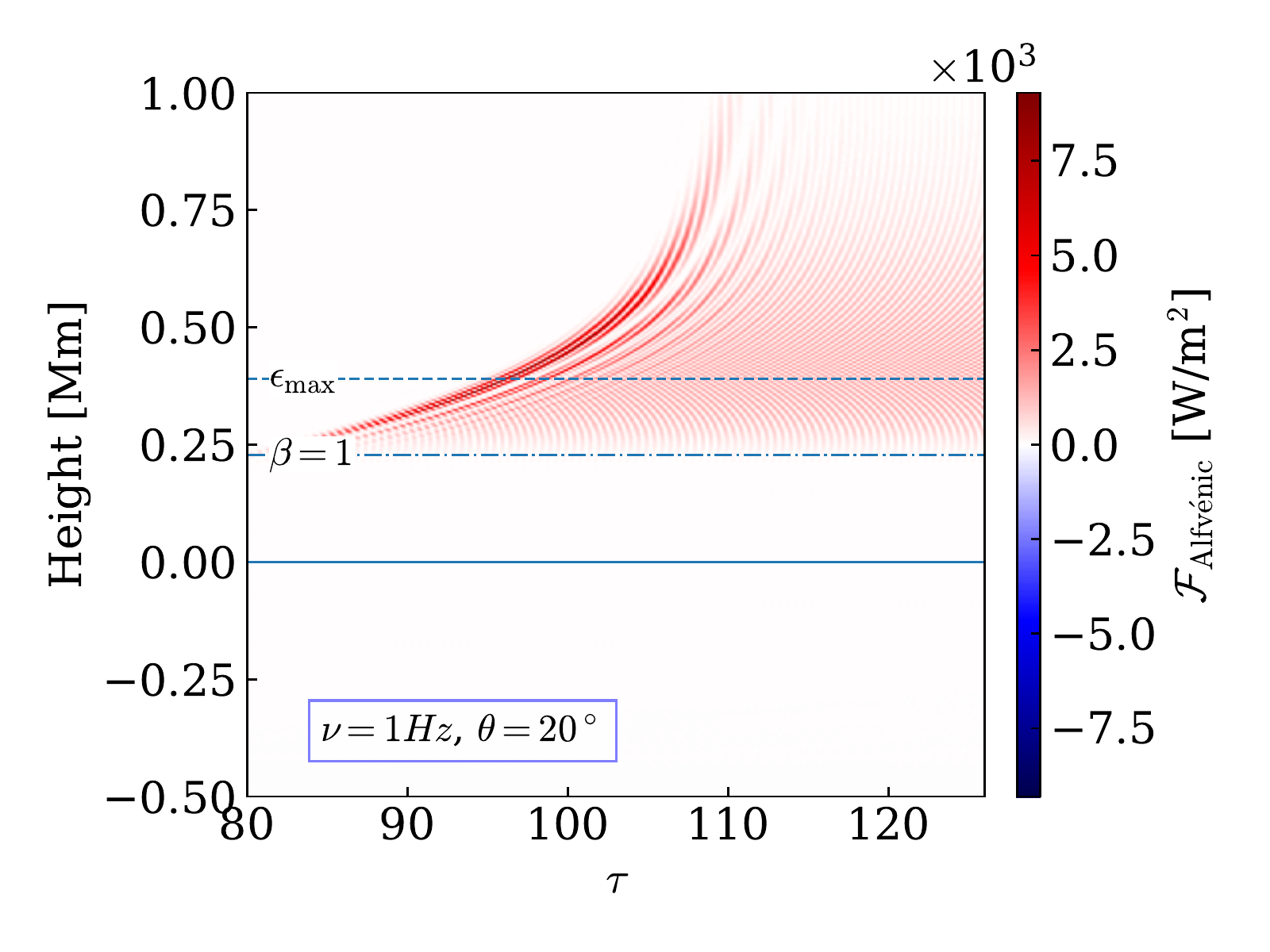}
  
 \includegraphics[width=0.32\textwidth]{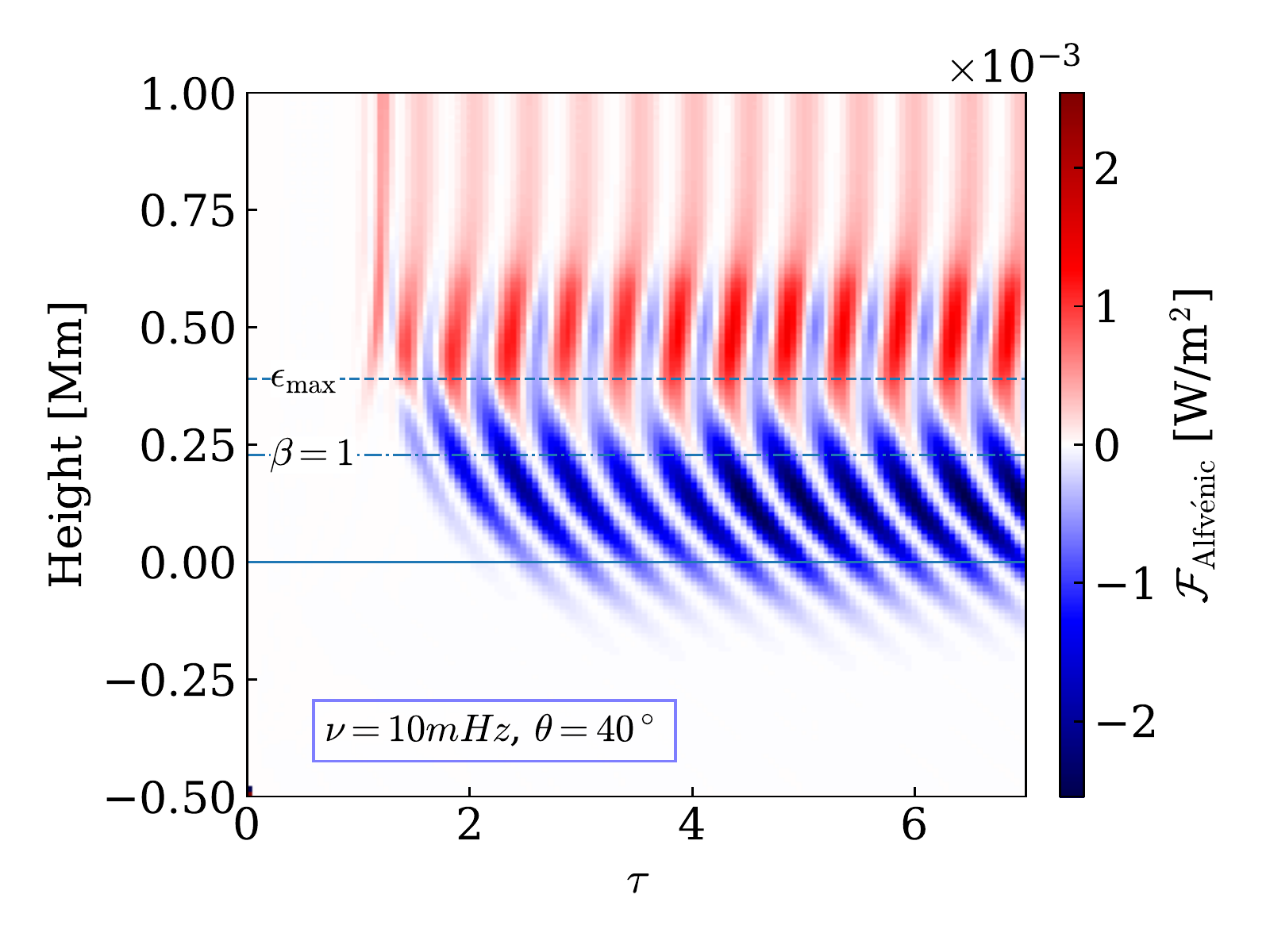}
 \includegraphics[width=0.32\textwidth]{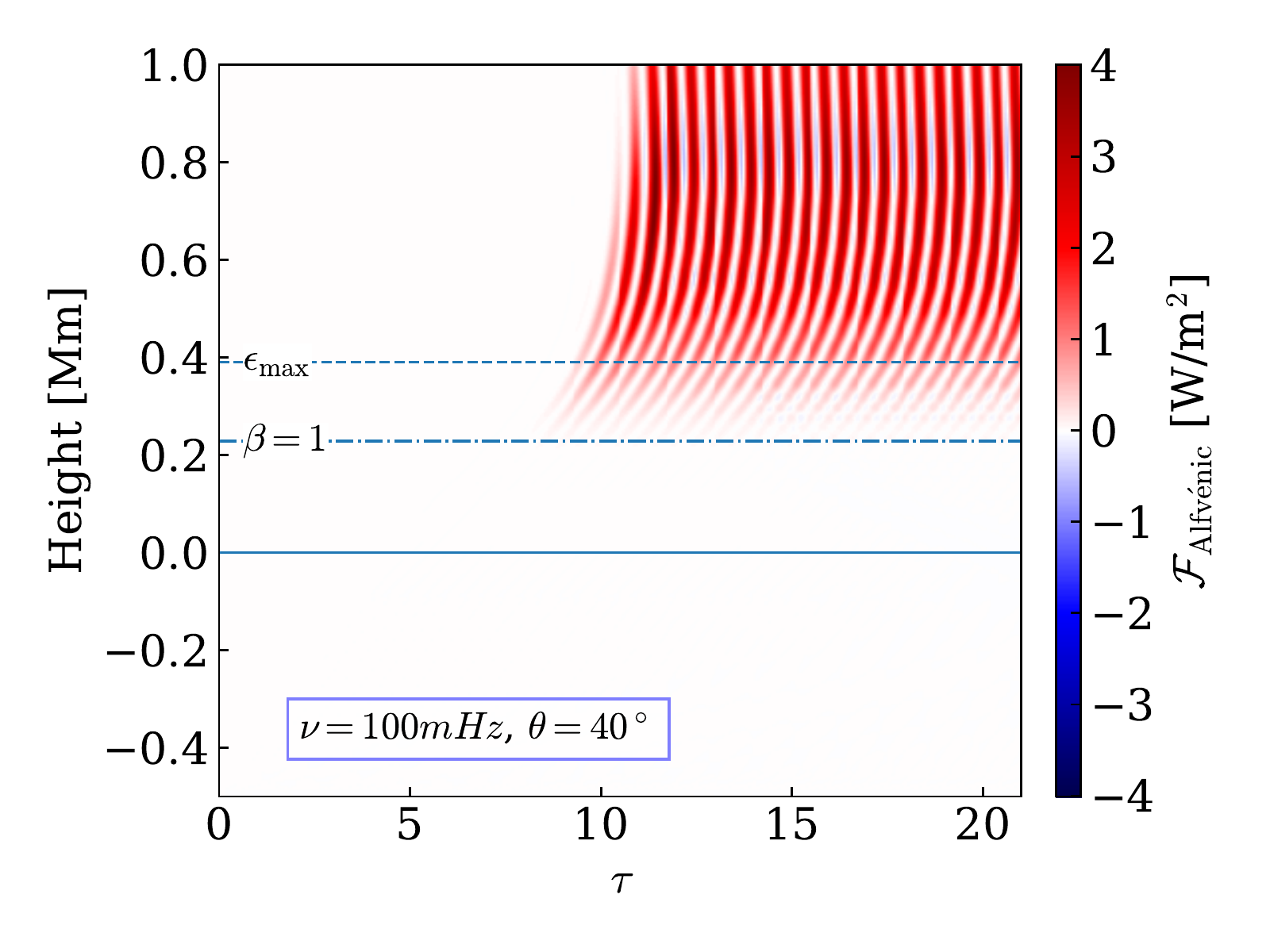}
 \includegraphics[width=0.32\textwidth]{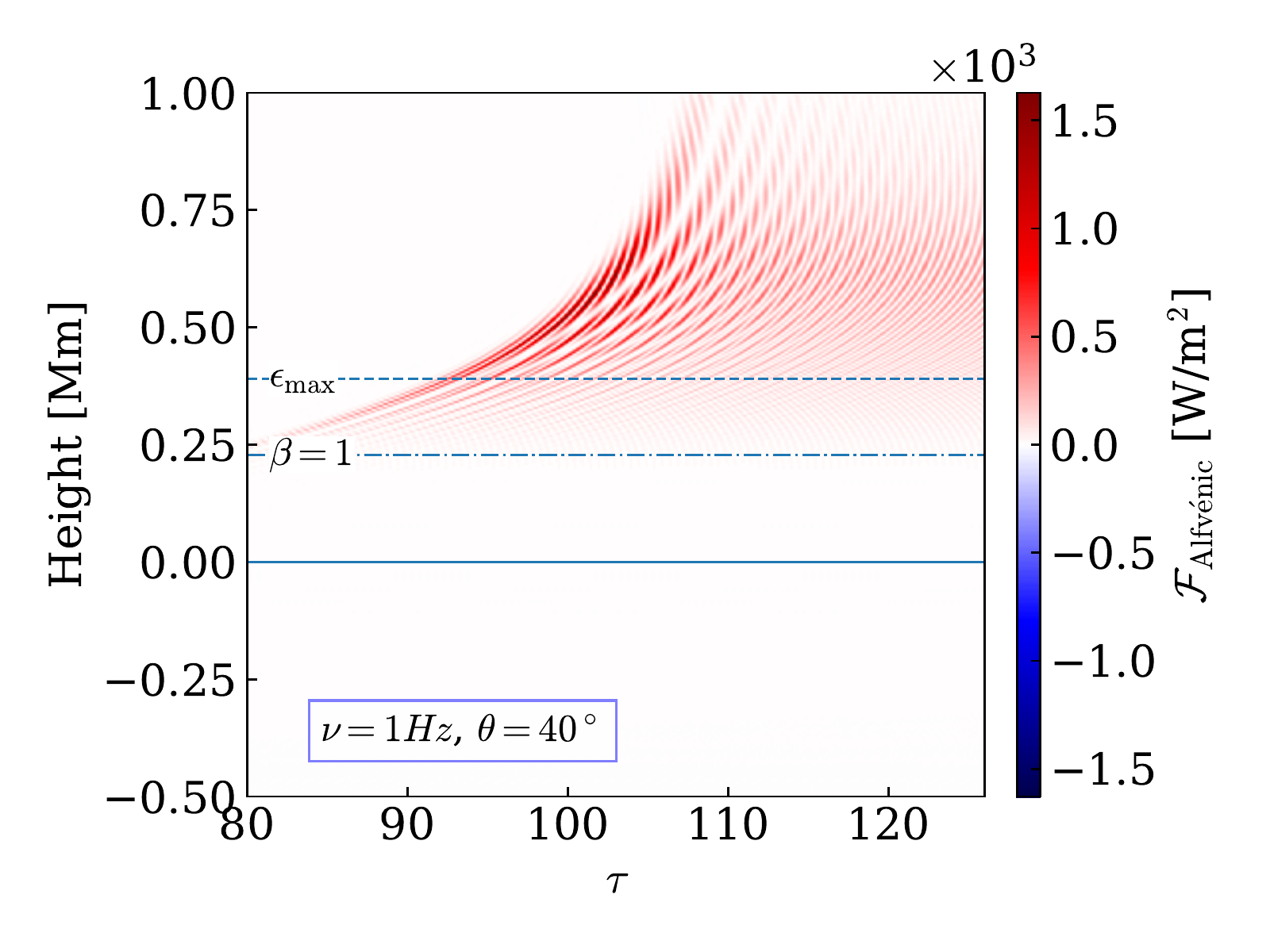} 
 
 \includegraphics[width=0.32\textwidth]{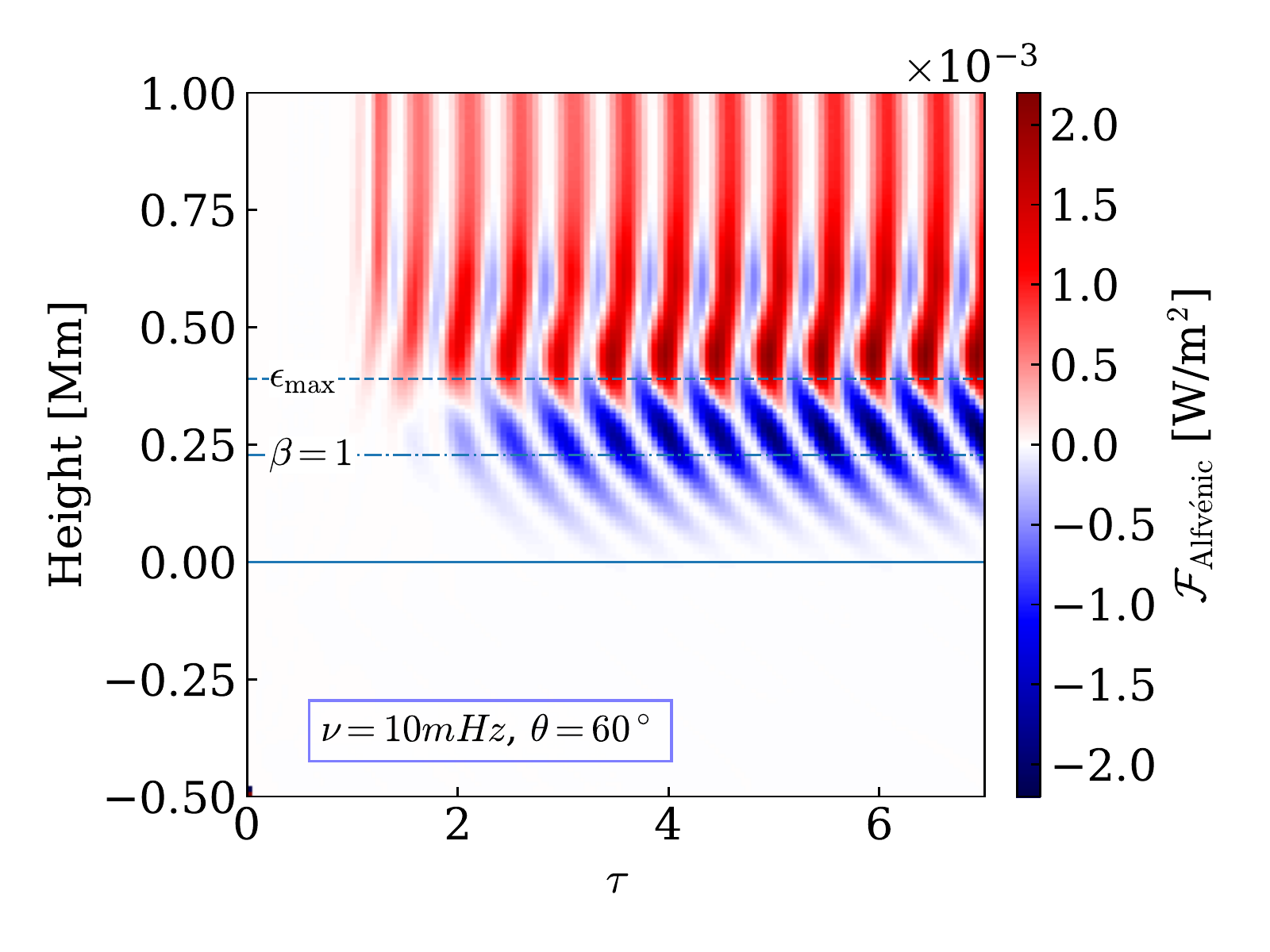}
 \includegraphics[width=0.32\textwidth]{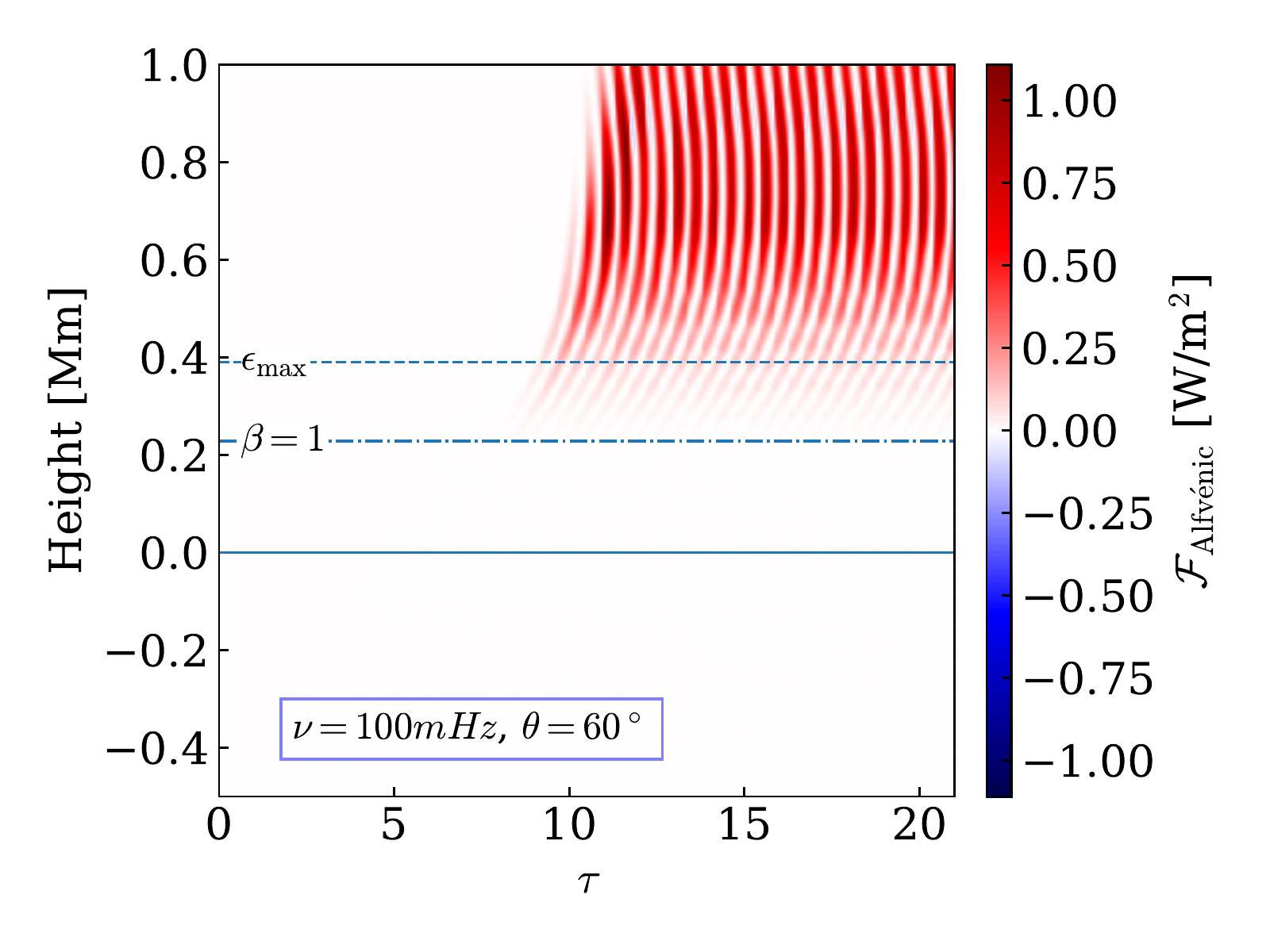}
 \includegraphics[width=0.32\textwidth]{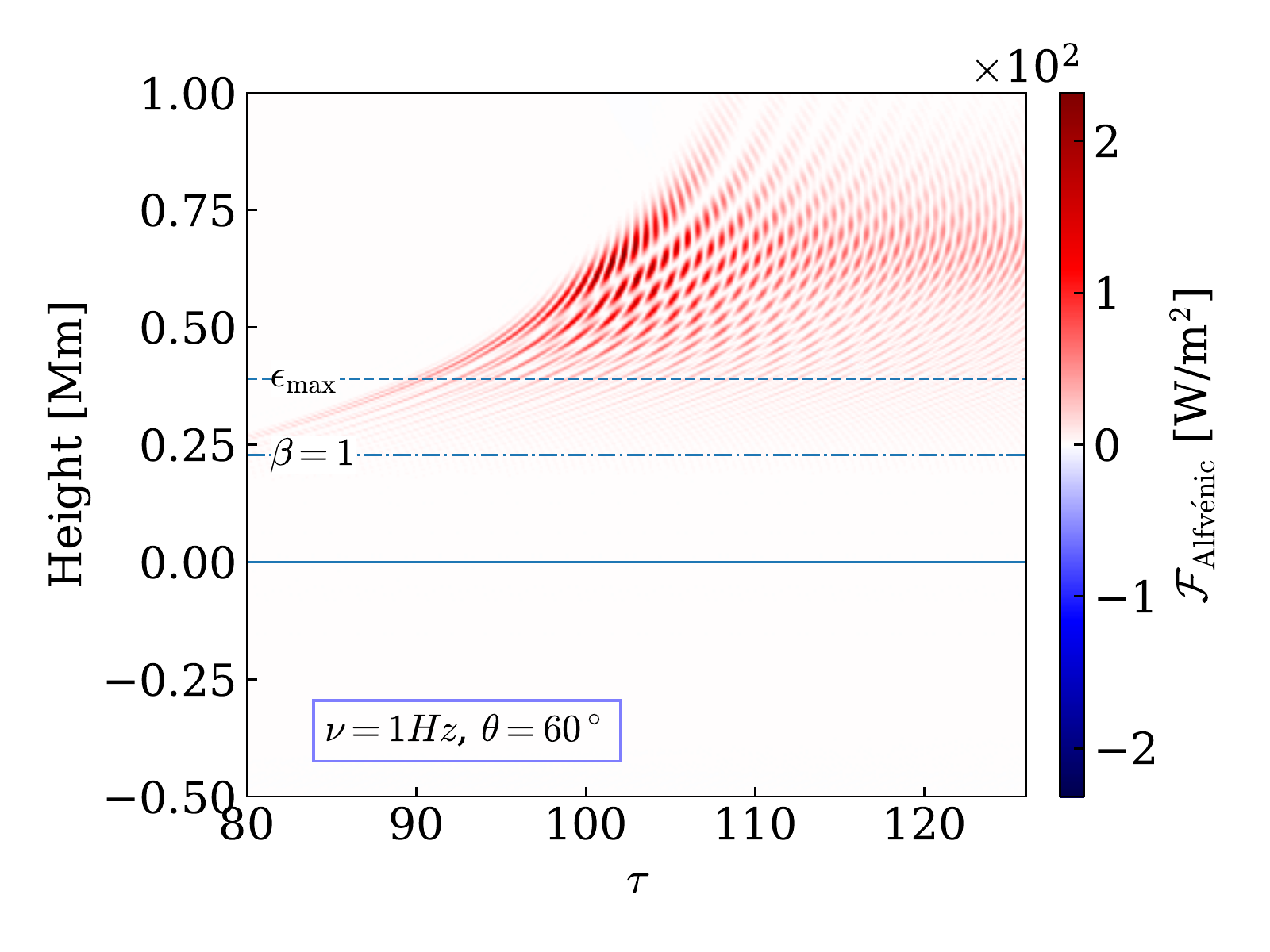} 
 
 \includegraphics[width=0.32\textwidth]{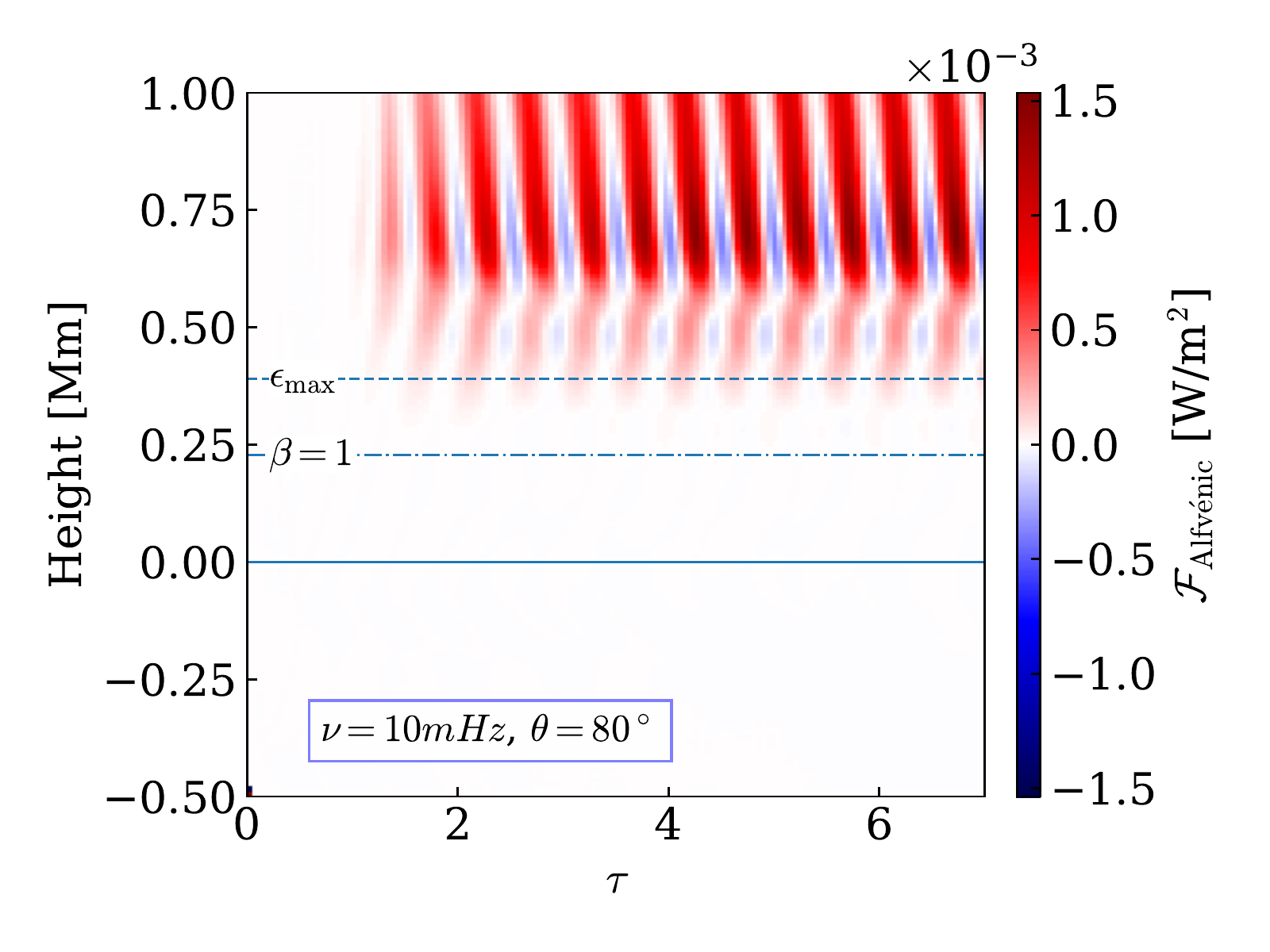}
 \includegraphics[width=0.32\textwidth]{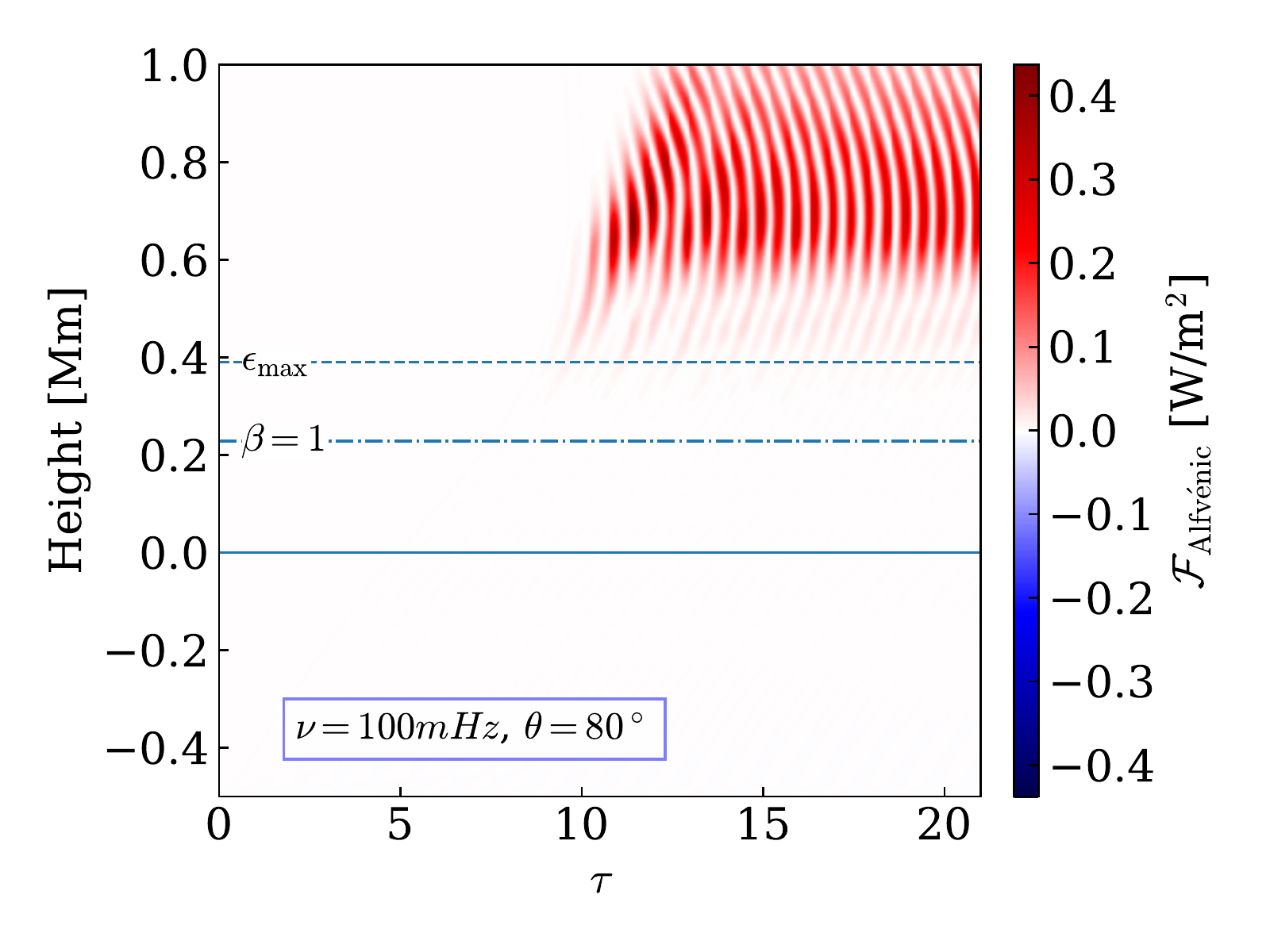}
 \includegraphics[width=0.32\textwidth]{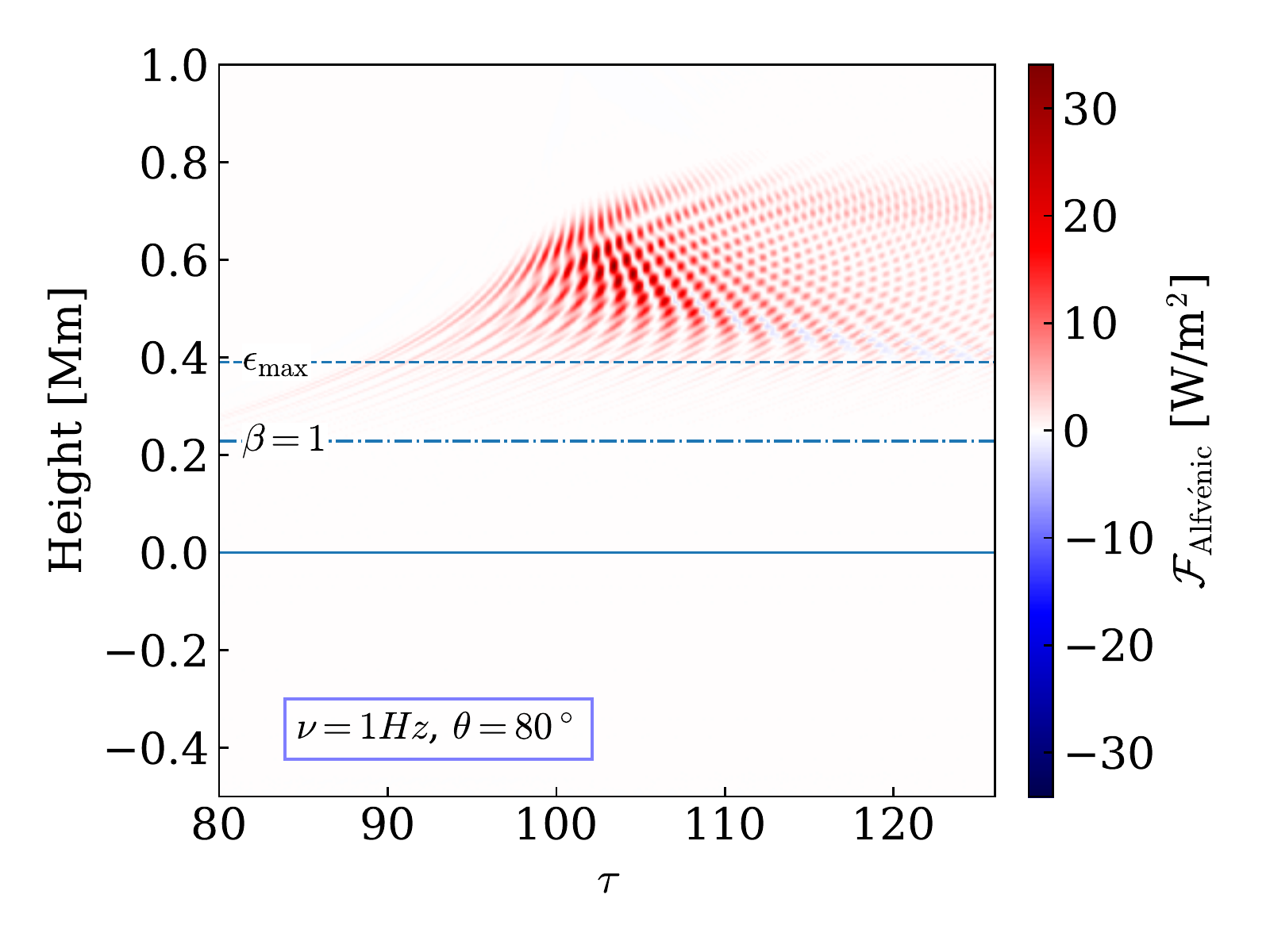}
 \caption{\footnotesize Time-height diagrams of $\mathcal{F}_\textrm{\alfvenic}$ in simulations with varying frequency (panels from left to right) and varying inclination angle of the magnetic field (panels from top to bottom).  The values of the frequencies and inclination angles are indicated in each panel. Positive value of the magnetic flux (red colour) mean upward propagation.}
 \label{fig:alfvenflux} 
\end{figure*}

Figure \ref{fig:vperp} shows the results of these experiments. It presents the time-height diagrams for the perpendicular component of the projected velocity ($v_\mathrm{perp}=v_y$) for three of the selected frequencies, $\nu= 0.01$, $0.1$, and $1$ Hz (columns) and five different inclination angles of the magnetic field, $\theta = 0$, $20$, $40$, $60$, and $80$ degrees (rows). All simulations share the same numerical setup including number of grid points per wavelength in the horizontal direction and the magnitude of the numerical diffusivity (necessary for the stability of the simulations). Nevertheless, it is unavoidable that higher frequency waves are affected more by numerical effects, and therefore their amplitude is inevitably lower than it should be. In order to account for these numerical effects and to standardize the experiments on the same scale, the velocities for a given frequency are scaled by setting the amplitude of the longitudinal component of the velocity $v_\mathrm{long}$ at the equipartition layer height in the simulation with $\theta=10^\circ$ to 500 m s$^{-1}$. The rest of amplitudes and perturbed quantities for this frequency are then scaled according to this factor. 

We observe in Figure \ref{fig:vperp} that the amplitudes of both upward and downward propagating \alfven waves depend on the inclination and frequency. In particular, the amplitude of the down-going wave becomes progressively smaller for larger magnetic field inclination angles. 
Similarly, simulations with higher frequencies show less down-going \alfven wave, as measured by the amplitude. In all cases, the amplitudes of the down-going waves are smaller than of the up-going waves. The amplitudes of the up-going wave significantly increase with increasing frequency. The region where these amplitudes are maximal are located at or immediately above the height with maximum Hall parameter.

In order to confirm the visual impression about the presence of the \alfven waves and their direction of propagation, and also in order to quantify our results in terms of energy supply to the upper layers, we computed the magnetic Poynting flux carried by waves:
\begin{equation}\label{eq:fmag}
    \mathbf{F}_\mathrm{mag}=\langle\mathbf{B}_1\times(\mathbf{v}_1\times\mathbf{B}_0/\mu_0)\rangle\ \coma
\end{equation}
and the acoustic flux
\begin{equation}\label{eq:facu}
    \mathbf{F}_\mathrm{acu}=\langle p_1 \mathbf{v}_1 \rangle \coma
\end{equation}
where the subscript ``$1$'' denotes a small perturbation in velocity ($\mathbf{v}_1$), magnetic field ($\mathbf{B}_1$), and pressure ($p_1$). The angled brackets denote phase averages.
Using Eqs.~(\ref{eq:pro}) together with Equation (\ref{eq:fmag}) we obtain the longitudinal component of the magnetic wave-energy flux. Considering only the perpendicular component of the velocity, we obtain a longitudinal magnetic flux associated exclusively with \alfven waves:
\begin{equation}
    \mathcal{F}_\textrm{\alfvenic}=-B_1^\textrm{perp}v_1^\textrm{perp}B_0^\textrm{long}/\mu_0 \punto
\end{equation}

Figure \ref{fig:alfvenflux} shows the \alfvenic flux in simulations with varying magnetic field inclination angle and frequency in the same format as Fig.~\ref{fig:vperp}. The amplitudes of the velocity and magnetic field oscillations were normalized for each frequency setting $v_{\rm long} = 500$ m s$^{-1}$ at $\beta=1$ height, in the simulation with $\theta=10$ degrees, as before. The results for the flux confirm that, indeed the ridges with the opposite inclination in Fig.~\ref{fig:vperp} correspond to downward magnetic flux (blue colour). They indeed vanish (in comparison to the upward fluxes) when the inclination angle and wave frequency are increased, though notice that the magnitude of the fluxes is orders of magnitude higher at 1 mHz than at the lower frequencies. Also, we observe how this flux increases up to a maximum value at a certain inclination angle, and then starts to decrease; see the column for $\nu=100$ mHz for example. The magnetic flux for a given inclination angle increases with increasing frequency.

\begin{figure}
 \includegraphics[width=1\columnwidth]{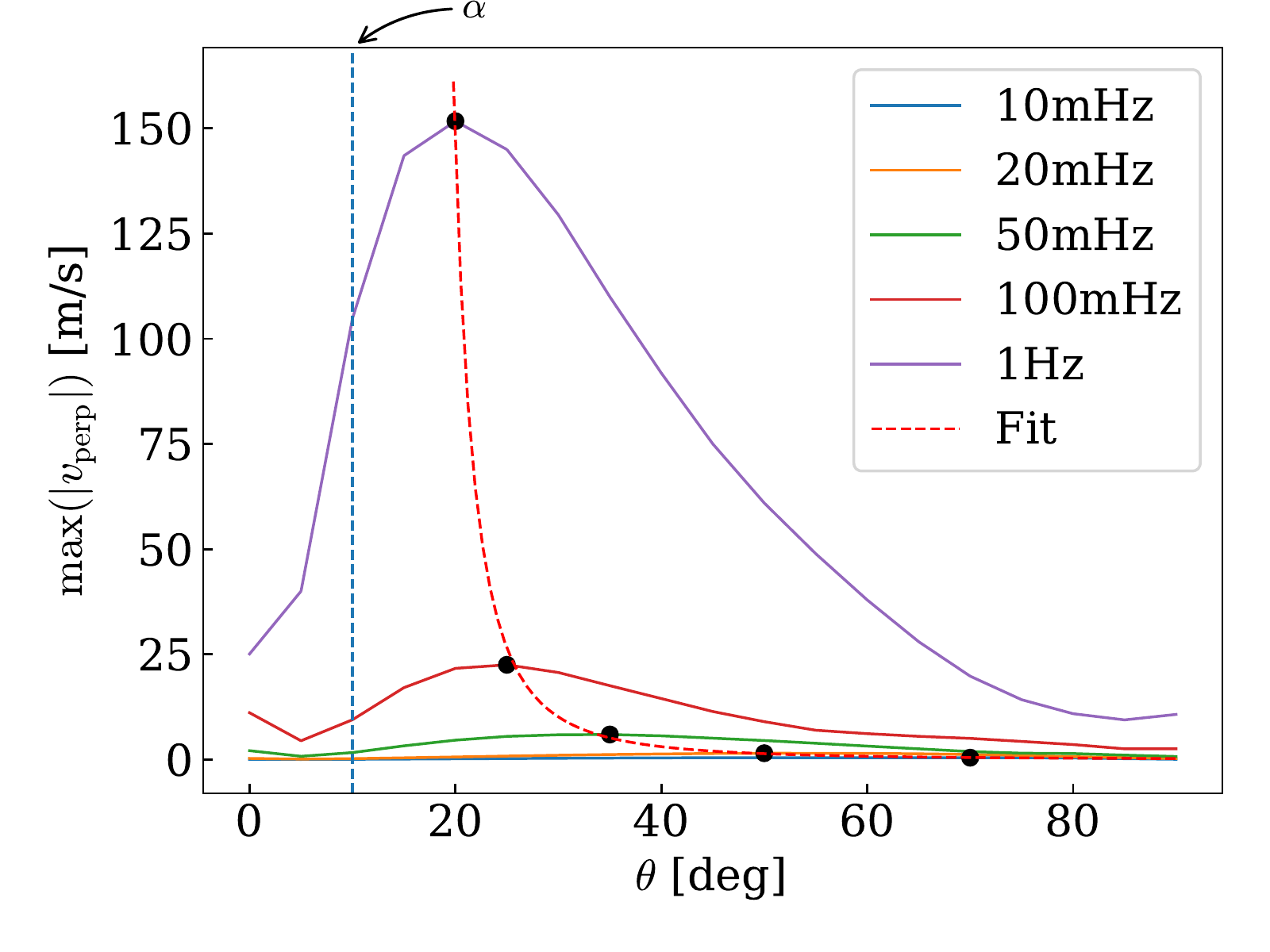}
 \caption{\footnotesize Velocity $v_\mathrm{perp}=v_y$ of the \alfven component as a function of  magnetic field inclination angle. Different colour lines present the results of the simulations with different frequencies, indicated in the figure.  The maximum of each curve is marked with a black dot, the dotted vertical line corresponds to the wave vector angle ($\alpha$). We observe how this maximum approaches to $\alpha$ for increasing frequencies.}
 \label{fig:maxvperp}
\end{figure}

\begin{figure}
 \includegraphics[width=1\columnwidth]{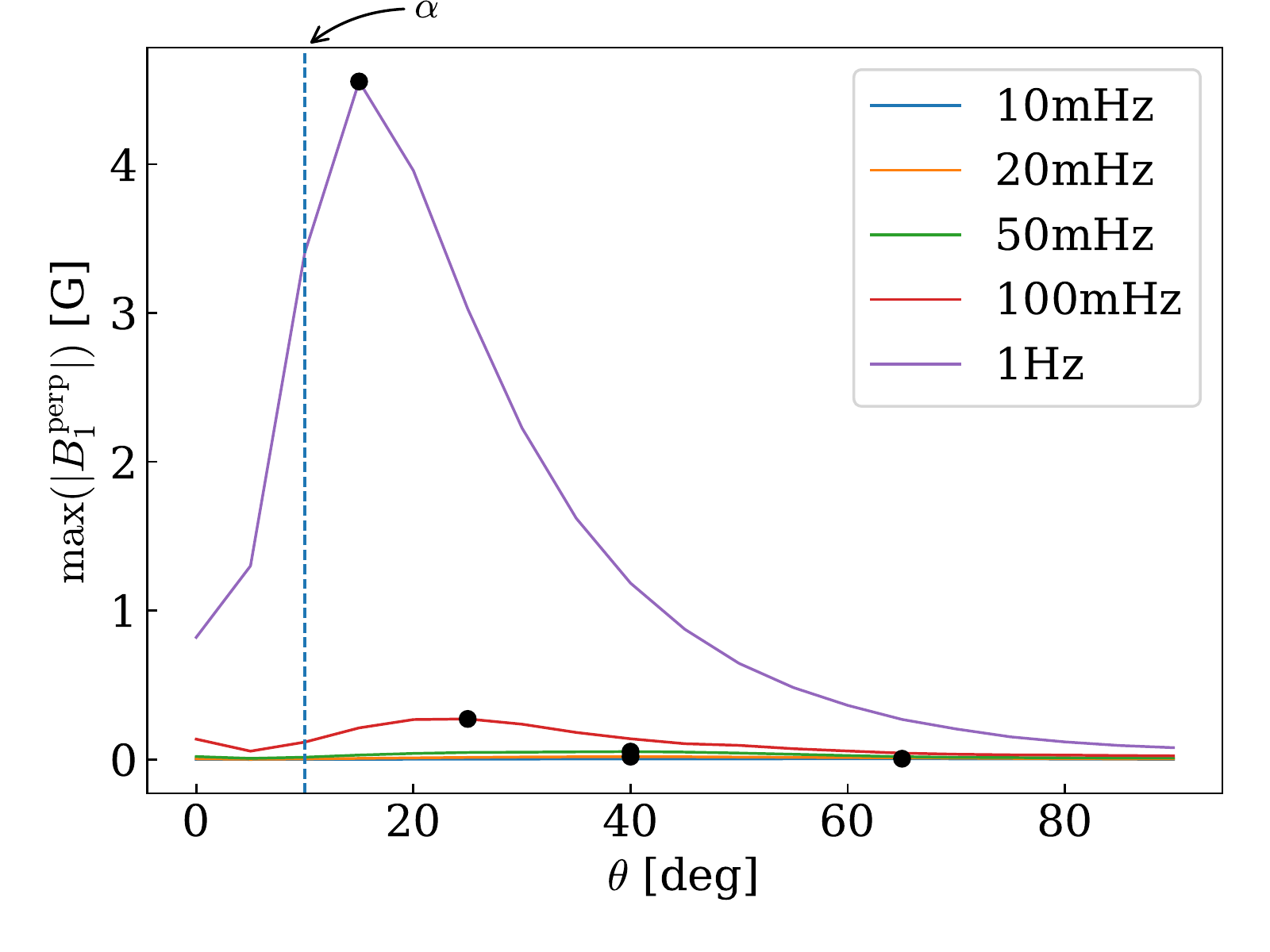}
 \caption{\footnotesize Perpendicular projection of the perturbed magnetic field as a function of the magnetic field inclination angle. Colour lines correspond to results of the simulations with different frequencies. The maximum of each curve is marked with a black dot, the dotted vertical line corresponds to the wave vector angle ($\alpha$). As in Fig. \ref{fig:maxvperp}, we observe how this maximum approaches to $\alpha$ for increasing frequencies.}
 \label{fig:maxb1perp}
\end{figure}

Finally, gathering together the results of all simulations, we have computed the amplitude of the perpendicular velocity once the system reaches a stationary regime. Figure \ref{fig:maxvperp} shows the results of this calculation as a function of inclination angle for all considered frequencies. This figure nicely summarizes the behaviour mentioned above. Firstly, we see that the amplitude of \alfven waves increases with increasing wave frequency. The maximum amplitude reached for waves of 1 Hz constitutes about 30\% of the amplitude of the longitudinal wave component, which is significant. Next, for a given frequency, there is magnetic field inclination where the amplitude of the generated \alfven waves reaches a maximum value. For low frequency waves this maximum falls at large inclination angles. But for progressively larger frequencies, the inclination with maximum amplitude approaches asymptotically the inclination of 10 degrees, i.e. $\theta=\alpha$, the inclination of the wave vector. In Figure \ref{fig:maxb1perp} we can see similar behaviour in the perpendicular projection of the perturbed magnetic field.

As mentioned previously, Hall-mediated transformation acts everywhere in the numerical domain and its efficiency increases with wave frequency. Because our numerical scheme contains numerical noise, it is important to carefully choose parameters such as the filtering cadence and the artificial diffusivity to keep this noise as low as possible to avoid its growth and the artificial Hall-mediated transformation that can be introduced into the simulations. Our chosen setup parameters made the simulations computationally very costly at progressively higher frequencies, so we had to truncate our numerical analysis at 1 Hz. Nevertheless, Figure \ref{fig:maxvperp} clearly suggests that for higher frequencies, the amplitudes would increase further. In order to quantify this increase we performed fitting to the maximum values of velocity versus angle, using the following power law:
\begin{equation}
\max\left(\left| v_\mathrm{perp}\right|\right)=A\left(\theta-\theta_0\right)^m
\end{equation}
The bst fit is for $A= 4475.88$ deg m s$^{-1}$, $m=-2.28$, $\theta_0 = 15.5^\circ$. According to these parameters, an asymptotic angle with maximum velocity lies around $15.5$ degrees. Nevertheless, this particular number should only be taken as an indication. Possibly, simulations with better sampling in $\theta$ would allow us to make a more precise fit. In addition, the amplitude of waves in numerical simulations is affected by numerical effects (such as numerical diffusivity). Although we kept the numerical parameters as similar as possible between all simulations, it is still possible that the numerical diffusivity affects the higher frequency waves (especially those at 1 Hz) more severely. Nevertheless, given all the uncertainties, we conclude that the amount of energy transferred from fast to \alfven  mode can be considerable. For a reference, Figure \ref{fig:maxb1perp} provides the corresponding amplitudes of the magnetic field perturbation. For all the frequencies except 1 Hz, the amplitudes of $B_1$ do not reach above 1 G. This means that detecting such perturbations in observations would only be possible with the highest sensitivity polarimeters on the largest-aperture telescopes such as DKIST or future EST.

\begin{figure}
 \includegraphics[width=1\columnwidth]{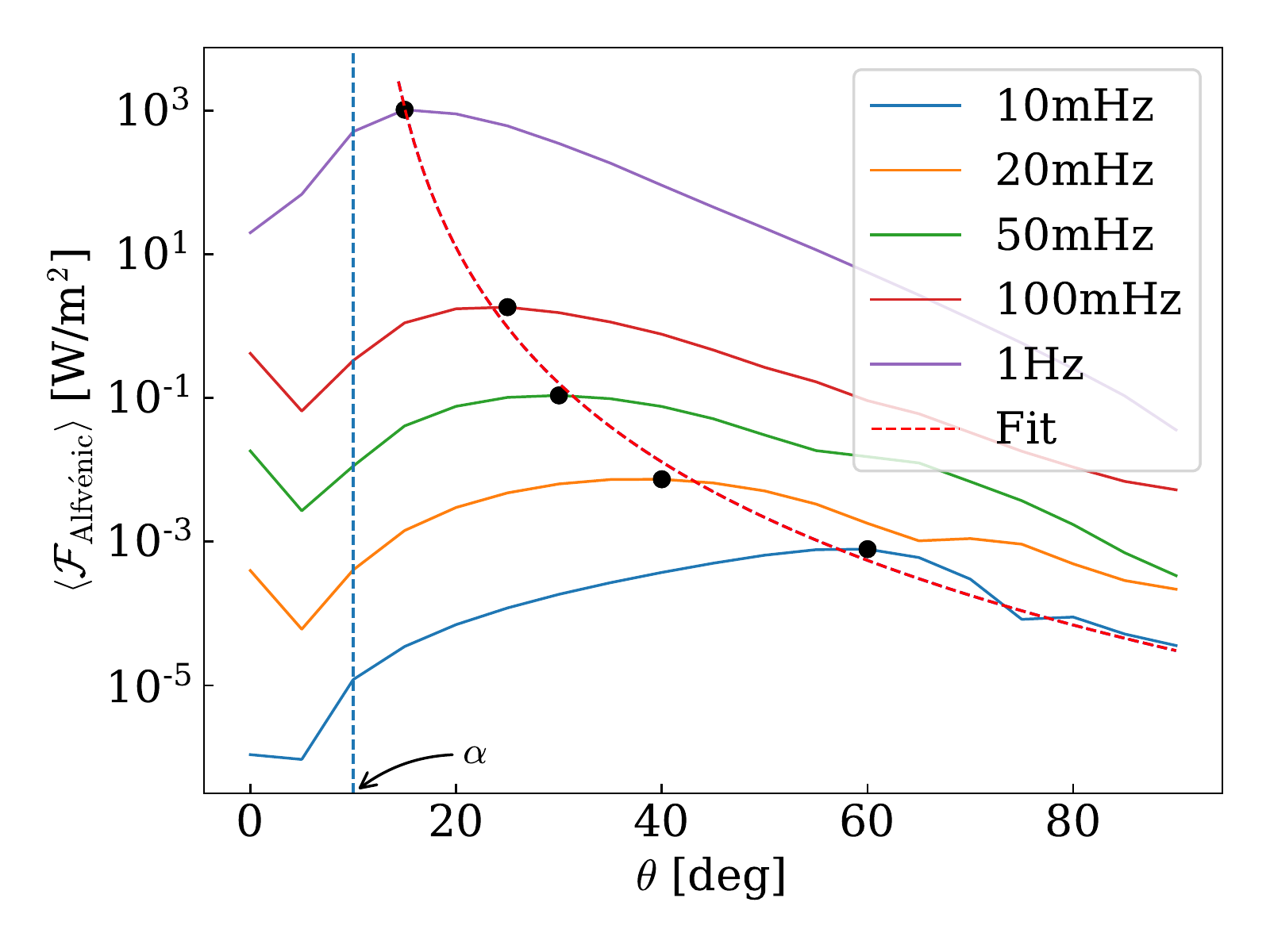}
 \caption{Temporal average of the \alfvenic flux at $z=450$ km. Solid coloured lines represents different frequencies, the dotted vertical line marks the wave vector angle $\alpha$. The maximum of each curve is marked with a black dot. We observe a displacement toward the left of those maximum for increasing frequencies. }
 \label{fig:meanfalf_logy}
\end{figure}

In Figure \ref{fig:meanfalf_logy} we collect the results for the average \alfvenic flux for all the inclinations and frequencies. This flux is calculated at $z=450$ km, above the height with the maximum Hall parameter. This figure allows us to quantify the energy input by \alfven waves into the higher layers. The \alfvenic flux reaches a maximum around $10^3$ W m$^{-2}$ at 15 degrees for the 1 Hz wave. The spatio-temporal root mean square (RMS) of the \alfven flux, calculated for this frequency inside the green box marked in Fig.~\ref{fig:maxvperp}, is about 780 W m$^{-2}$.  Making a similar fit  to the \alfvenic flux as before, the asymptotic angle is $\theta_0 = 10.36^\circ$, very close to the inclination angle of $\vec{k}$ with respect to the background magnetic field. 

\begin{figure}
 \includegraphics[width=1\columnwidth]{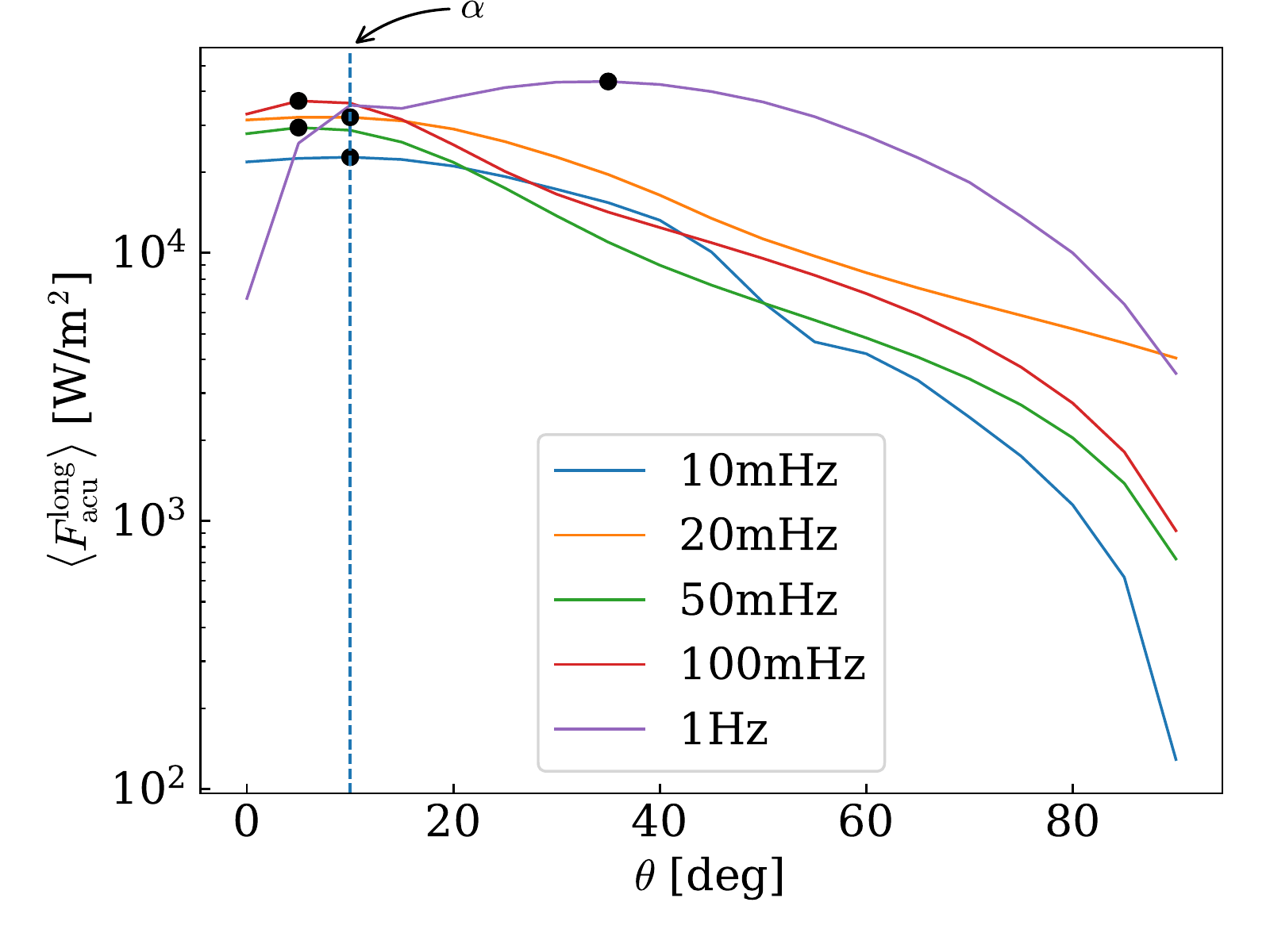}
 \caption{Temporal average of the longitudinal projection of the acoustic flux at the equipartition level. The dotted vertical line indicates the wave vector angle $\alpha$. Solid curves correspond to different frequencies, and its maximum is marked with a black dot. We observe how the acoustic flux, for higher inclination angles of the magnetic field, increases with the wave frequency.}
 \label{fig:meanfacul_logy}
\end{figure}

The values of the magnetic flux of \alfven waves are to be compared to the available acoustic flux at some reference layer. We take for the reference the equipartition layer as the layer where the acoustic waves start to transform. Figure \ref{fig:meanfacul_logy} shows the mean value of the longitudinal acoustic flux at the equipartition level. We observe that, in general, the acoustic flux pumped into the corona and then available to be transformed into \alfvenic flux slightly increases with frequency. The behaviour of the curve for 1 Hz is slightly different from the others, probably due to numerical effects on this simulation. The values of the acoustic flux vary in the range of  $10^2-10^4$ W m$^{-2}$ for all the frequencies. One can conclude that only for the highest frequency of 1 Hz does the \alfven flux shown in Figure \ref{fig:meanfalf_logy} contribute a significant fraction of the available acoustic flux.

On the other hand, in the full non-linear regime, which is not explored here, acoustic waves are much more subject to attenuation (by shocking for example) in the chromosphere than are Alfv\'en waves. Our results might therefore be expected to overestimate the acoustic wave flux reaching higher levels.


\section{Discussion and Conclusions}\label{sec:sum}

In this paper simulations are performed of Hall current mediated mode conversion to \alfven waves for plasma parameters approximating the solar atmosphere. Numerical solution allow us to relax the cold plasma approximation assumed in the initial analytic study of Paper I. We consider an acoustic-gravity wave with various frequencies propagating upwards from the lower boundary of our simulation domain located in the high plasma $\beta$ region below the photosphere. Our simulations indeed show the presence of \alfven waves when the Hall effect is acting. Therefore we confirm that this effect can be responsible for coupling fast magneto-acoustic and \alfven waves even when there is no cross-field wave propagation (the usual coupling mechanism). 

In Paper I, \citet{2015ApJ...814..106C} concluded that the transformation is more efficient for vertical fields and wave propagation aligned with the field. The efficiency is also a sensitive function of the Hall parameter, and therefore it increases for increasing wave frequency $\nu$ and decreasing ionization fraction $f$. Our numerical experiments in warm plasma have partially confirmed this picture, but also shown a more nuanced behaviour. We conclude that the efficiency of the transformation for low-frequency waves is maximal for strongly inclined fields ($50$-$70$ degrees). However, for waves at higher frequencies, the maximum becomes progressively aligned with the field. The asymptotic fit for the perpendicular velocity shows that the alignment between the directions of $\mathbf{k}$ and $\mathbf{B}$ is within 5.5 degrees for waves with frequencies above 1 Hz. A similar fit to the \alfvenic flux curves results in a difference between $\mathbf{k}$ and $\mathbf{B}$ of just 0.36 degrees. 

A further conclusion concerns the absolute value of the effect. As discussed in the Introduction, in warm plasmas the transformation is a two-step process. First acoustic fast waves are partially transformed into magnetic fast waves at the $v_A=c_s$ equipartition layer, and then the latter are transformed into \alfven waves progressively where Hall coupling operates. For that to happen there should be a specific relation between the location of the equipartition layer and the level with maximum Hall parameter $\epsilon_{\rm Hall}$. While the latter depends exclusively on the temperature structure of the atmosphere, the former is a function of the magnetic field. For the Hall-mediated transformation to be efficient one needs to satisfy simultaneously both conditions: (1) that the layer with $v_A=c_s$ is located below the layer with maximum $\epsilon_{\rm Hall}$, therefore $B_0$ has to be sufficiently large to have $v_A=c_s$ located deep enough; (2) that $B$ is sufficiently small to maximize the value of $\epsilon_{\rm Hall}$. In our simulations we set $B_0=500$ G to satisfy both conditions. We reach the maximum amplitudes of Hall-excited \alfven waves, about 30\% of $v_{\rm long}$, for waves of the highest studied frequency, i.e., 1 Hz. 

In practice,the process of Hall-mediated transformation acts in addition to the geometrical mode transformation to \alfven waves suggested by \citet{2008SoPh..251..251C}. It does not need any particular relation between the wave vector and the orientation of the magnetic field, and, our simulations suggest that the maximum transformation can occur for a broad range of magnetic field inclinations, depending on the wave frequency. Also, the considerations above suggest that the process would be efficient for intermediate field strengths of the order of hG, comparable to those existing in solar network and quiet areas. Therefore, this process could provide a constant energy supply be means of \alfven waves to the solar corona. 

The 3D geometric fast/Alfv\'en coupling mechanism occurs near the fast wave reflection height, where the horizontal phase speed $\omega/k_h$ equals the Alfv\'en speed (in a low $\beta$ plasma), and $k_h = (k_x^2+k_y^2)^{1/2}$ is the horizontal wavenumber. At frequencies of a few mHz this may occur somewhere in the low-to-mid chromosphere, depending on magnetic field strength and $k_h$. However, at the high frequencies considered above (1 Hz etc), reflection may not occur till the transition region (TR) is reached. In that case, the two coupling regions (Hall and geometric) are spatially separated and would operate independently: Hall coupling would operate in the weakly ionized low chromosphere, and geometric coupling would set in once the TR is reached.

In our modelling, for waves of 1 Hz, the average \alfvenic energy flux we obtain at $450$ km height (above the height of peak Hall coupling) is about $10^3$ W m$^{-2}$ with a RMS of $780$ W m$^{-2}$. These numbers are bigger than what is required for heating the corona above quiet Sun regions, which is about 100-300 W m$^{-2}$, but are of the order of what is needed for heating the corona above active regions \citep{1977ARA&A..15..363W}. The numbers we provide were obtained after normalizing the wave amplitude of $v_\mathrm{long}$ to be $500$ m s$^{-1}$ at $\beta=1$ which, using Eq (\ref{eq:facu}), corresponds to an average acoustic wave energy flux of $\sim 3.8\times10^4$ W m$^{-2}$ and a spatio-temporal RMS value of about 368 W m$^{-2}$. These values depend of the adopted base amplitude, which is probably overestimated at this frequency.

Theoretical estimates and measurements of the wave energy flux for such high frequency waves are uncertain \citep{Fossum:2006kj}. These authors provide measurements of the acoustic energy flux from TRACE observations for frequencies up to a few hundred mHz, showing an exponential decrease of flux with frequency. The flux measured at largest frequencies by these authors makes 0.1 - 1 W m$^{-2}$. Simulations of acoustic wave generation by turbulence show a maximum wave energy at frequencies around 0.1 Hz with a strong decrease toward the higher frequencies \citep{1994ApJ...423..474M}. The acoustic energy flux at 1 Hz reported by \citet{1994ApJ...423..474M} makes 10$^3$ W m$^{-2}$. Assuming we can convert 1\% of this acoustic flux into \alfven waves and bring it to the solar corona, a good fraction of the energy needed to compensate its losses would be provided by Hall coupling. Although this flux is small compared to the peak acoustic flux at the same height, the \alfven flux is far more able to penetrate to the corona, and so is more relevant to coronal heating.

Whilst fast modes refract and reflect in low-$\beta$ regions and slow modes rapidly shock and damp, \alfven waves can reach upper regions of the solar atmosphere due to their incompressible nature ($\nabla\cdot\mathbf{v} = 0$), making them an attractive mechanism to transport energy up to the corona. It is then a challenge to find mechanisms to dissipate \alfven energy there.

Recently, simulations from \citet{Santamaria:2017ko} have shown that slow magneto-acoustic shock waves coming from the chromosphere can trigger a jet-like structure of slow magneto-acoustic waves with frequencies up to $80$ mHz around null points. These shock waves could be converted into fast modes around $\beta=1$ regions with complex topology, and then they can be converted into \alfven waves via the Hall term. Because the presence of null points is theoretically predicted almost everywhere, this double transformation could be an important source of production of \alfven waves in complex magnetic topologies.

Possibly, Hall mediated conversion can be important for other astrophysical scenarios, e.g. in star formation regions or reconnection events as well as in other cooler stars. Further investigations of 3D effects via the azimuth ($\phi$) and stratification dependence, and taking into account other scenarios may also be interesting.\\

{\footnotesize \emph{Acknowledgments}: This work is supported by the Ministerio de Econom\'ia y Competitividad through project  AYA2014-55078-P and the predoctoral training grants in Centres/Units of Excellence `Severo Ochoa' with reference SEV-2011-0187. It contributes to the deliverable identified in FP7 European Research Council grant agreement 277829, `Magnetic connectivity through the Solar Partially Ionized Atmosphere'. We also wish to acknowledge the contribution of Teide High-Performance Computing facilities to the results of this research. TeideHPC facilities are provided by the Instituto Tecnol\'ogico y de Energ\'ias Renovables (ITER, SA). URL: http://teidehpc.iter.es. We thankfully acknowledges the technical expertise and assistance provided by the Spanish Supercomputing Network (Red Espa\~nola de Supercomputaci\'on), as well as the computer resource LaPalma Supercomputer, located at the Instituto de Astrof\'isica de Canarias.}

\begin{appendix}
\section{Bottom boundary condition}\label{sec:source}

For our experiments we impose a bottom boundary condition as a source. This source introduces an acoustic-gravity wave of a given frequency $\omega$ and wavenumber $k$. The analytical solution for the perturbations in pressure, density, temperature, and velocity are calculated according to \cite{1984oup..book.....M}, providing a self-consistent solution. The temperature gradient and magnetic field are neglected in this formulation since the bottom boundary is located in the region where $v_A \ll c_s$ and they are unimportant at this depth. The analytical solutions applied are identical to those used by \citet{Khomenko:2012gw}:
\begin{subequations}
\begin{alignat}{1}
    \delta V_z = {} & V_0\exp{\left(\frac{z}{2H}+k_{zi}z\right)}\sin{\left(\omega t - k_{zr}z - k_{x} x\right)}\coma \\
    \frac{\delta p}{p_0} = {} & V_0|P|\exp{\left(\frac{z}{2H}+k_{zi}z\right)}\sin{\left(\omega t - k_{zr}z - k_{x} x + \phi_P\right)}\coma \\
    \frac{\delta \rho}{\rho_0} = {} & V_0|R|\exp{\left(\frac{z}{2H}+k_{zi}z\right)}\sin{\left(\omega t - k_{zr}z - k_{x} x + \phi_R\right)}\coma\\
    \delta V_x = {} & V_0|U|\exp{\left(\frac{z}{2H}+k_{zi}z\right)}\sin{\left(\omega t - k_{zr}z - k_{x} x + \phi_U\right)}\coma
\end{alignat}
\end{subequations}
where with the $\delta$ symbol we are indicating a perturbed quantity, $p_0$ and $\rho_0$ are the pressure and density at the bottom, $V_0$ is the initial amplitude for the imposed perturbation in velocity, $H=c_s^2/2g$ is the pressure scale height, $k_x$ is the horizontal wavenumber, $k_z$ is the vertical wavenumber made up of a real part $k_{zr}$ and an imaginary part $k_{zi}$, and $\omega = 2\pi\nu$ is the angular frequency of the perturbation. $|P|$, $|R|$, and $|U|$ are the  amplitudes given by
\begin{subequations}
\begin{alignat}{1}
    |P| = {} & \frac{\gamma \omega}{\omega^2-c_s^2 k_x^2}\sqrt{k_{zr}^2 + \left(k_{zi} + \frac{1}{2H}\frac{\gamma - 2}{\gamma}\right)^2} \coma \\
    |R| = {} & \frac{\omega}{\omega^2-c_s^2 k_x^2}\sqrt{k_{zr}^2+\left(k_{zi}+\frac{\gamma-1}{\gamma H}\frac{c_s^2 k_x^2}{\omega^2}-\frac{1}{2H}\right)^2} \coma \\
    |U| = {} & \frac{c_s^2 k_x}{\gamma\omega}|P| \coma
\end{alignat}
\end{subequations}
where $\gamma$ is the adiabatic coefficient.
The phases $\phi_P$, $\phi_U$, and $\phi_R$ are given by
\begin{subequations}
\begin{alignat}{1}
    \phi_P = {} & \phi_U=\arctan{\left(\frac{k_{zi}}{k_{zr}}+\frac{1}{2Hk_{zr}}\frac{\gamma-2}{\gamma}\right)} \coma\\
    \phi_R = {} & \arctan{\left(\frac{k_{zi}}{k_{zr}}+\frac{\gamma-1}{\gamma H k_{zr}}\frac{c_S^2 k_x^2}{\omega^2}-\frac{1}{2Hk_{zr}}\right)}.
\end{alignat}
\end{subequations}
The vertical wave number is
\begin{equation}
    k_z=k_{zr}+ik_{zi}=\sqrt{\frac{\omega^2-\omega_c^2}{c_s^2}-k_x^2\frac{\omega^2-\omega_g^2}{\omega^2}} \coma
\end{equation}
where the cut-off frequency is
\begin{equation}
    \omega_c = \frac{\gamma g}{2c_s}\coma
\end{equation}
and the Brunt-V\"ais\"al\"a frequency is
\begin{equation}
    \omega_g = \frac{2\omega_c}{\gamma}\sqrt{\gamma-1} \punto
\end{equation}
The dispersion relation can be written in terms of the wavenumber modulus $k$ and the propagation angle $\alpha$ as
\begin{equation}\label{eq:kmodule_app}
k=\frac{\omega}{c_s}\sqrt{\frac{\omega_c^2-\omega^2}{\omega_g^2\sin^2{\alpha}-\omega^2}}\punto
\end{equation}

\section{A note on the velocity projections}
In a semi empirical solar stratification, as the one we had considered for our experiments, the identification of the different modes is a tough task because they are physically mixed. In this paper, in order to carry out the mode selection we have used the properties of MHD waves and made the change of base using the triad $(\mathbf{e}_\mathrm{long}, \mathbf{e}_\mathrm{perp}, \mathbf{e}_\mathrm{trans})$ as used by \citet{2008SoPh..251..251C}, also mention as the \emph{mixed field line/magnetic surface triad} by \citet{Goedbloed:2010tq},
\begin{eqnarray}\label{eq:pro_apendix}
 \mathbf{e}_\mathrm{long} &=& \cos{\phi}\sin{\theta}\,\mathbf{e}_x + \sin{\phi}\sin{\theta}\,\mathbf{e}_y + \cos{\theta}\,\mathbf{e}_z \coma \\ \nonumber
\mathbf{e}_\mathrm{perp} &=& -\cos{\phi}\sin^2{\theta}\sin{\phi}\,\mathbf{e}_x + (1-\sin^2{\phi}\sin^2{\theta})\,\mathbf{e}_y - \cos{\theta}\sin{\theta}\sin{\phi}\,\mathbf{e}_z \coma \\ \nonumber
 \mathbf{e}_\mathrm{tran} &=& -\cos{\theta}\,\mathbf{e}_x, + \cos{\phi}\sin{\theta}\,\mathbf{e}_z \punto
\end{eqnarray}
This new basis allow us to select the \alfven mode (for any plasma $\beta$). However, $\mathbf{e}_\mathrm{long}$ and $\mathbf{e}_\mathrm{tran}$ generally select a mixture of slow/fast magneto-acoustic modes depending on plasma $\beta$. 

To verify the behavior of the projections, we can use the ideal MHD formulation and consider a simple case with an infinite uniform plasma with a constant vertical magnetic field $\mathbf{B}=B_0\mathbf{e}_z$ and $k_y=0$. In this case, the transformation between the bases is simplified to
\begin{equation}\label{eq:pro3}
 \mathbf{e}_\mathrm{long} = \mathbf{e}_z, \hspace{0.2cm} \mathbf{e}_\mathrm{perp} = \mathbf{e}_y, \hspace{0.2cm} \mathbf{e}_\mathrm{tran} = \mathbf{e}_x \punto
\end{equation}
We can define the displacement associated to the \alfven wave and the slow/fast magneto-acoustic waves as 
 \begin{equation}\label{eq:disp2}
\mathbf{X}_F =(\mathbold{\xi} \cdot \mathbf{e}_\mathrm{tran})\,\mathbf{e}_\mathrm{tran} =  \xi_x\mathbf{e}_\mathrm{x} , 
\hspace{0.2cm} 
\mathbf{X}_A = (\mathbold{\xi} \cdot \mathbf{e}_\mathrm{perp})\,\mathbf{e}_\mathrm{perp}= \xi_y \mathbf{e}_\mathrm{y} , 
\hspace{0.2cm} 
\mathbf{X}_S = (\mathbold{\xi} \cdot \mathbf{e}_\mathrm{long})\,\mathbf{e}_\mathrm{long}=  \xi_z \mathbf{e}_\mathrm{z}  \punto
\end{equation}

The eigenfunctions for the incompressible \alfven wave  $\mathbold{\xi}_A$, and both the slow and fast magneto-acoustic waves, $\mathbold{\xi}_{ \{S,F\} }$, in this case are given by
 \begin{equation}\label{eq:A}
 \mathbold{\xi}_A=\xi_y\mathbf{e}_y\coma
 \end{equation}
 \begin{equation}\label{eq:S}
 \mathbold{\xi}_S=\xi_z\left(\frac{\omega_S^2}{\omega_S^2-k^2v_A^2}\frac{k_x}{k_z}\mathbf{e}_x + \mathbf{e}_z \right) = \xi_x\left(\mathbf{e}_x+\frac{\omega_S^2-k^2v_A^2}{\omega_S^2}\frac{k_z}{k_x}\mathbf{e}_z\right)\coma
 \end{equation}
  \begin{equation}\label{eq:F}
 \mathbold{\xi}_F=\xi_z\left(\frac{\omega_F^2}{\omega_F^2-k^2v_A^2}\frac{k_x}{k_z}\mathbf{e}_x + \mathbf{e}_z \right) = \xi_x\left(\mathbf{e}_x+\frac{\omega_F^2-k^2v_A^2}{\omega_F^2}\frac{k_z}{k_x}\mathbf{e}_z\right)\coma
 \end{equation}
where the slow/fast magneto-acoustic frequencies, $\omega_{ \{S,F\} }$, are
\begin{equation}
\omega_{ \{S,F\} } = \frac{k^2(c_s^2+v_A^2)}{2}\left\{1\pm\left(1-\frac{4\omega_C^2}{k^2(c_s^2+v_A^2}\right)^{1/2}\right\}\coma
\end{equation}
with $\omega_C=k_zv_C$ being the cusp frequency, and $v_C$ the cusp velocity $v_C^2=c_s^2v_A^2/(c_s^2+v_A^2)$. It is clear from Eq. (\ref{eq:A}) that $\mathbold{\xi}_A =  \mathbf{X}_A$ holds in general.  While from Eq. (\ref{eq:S}) and (\ref{eq:F}) it can be seen that the slow and fast mode displacements are formed by a linear combination of eigen displacements (\ref{eq:disp2}). In the limit of low plasma $\beta$,
\begin{equation}
\mathbold{\xi}_S \approx  \mathbf{X}_S, \,\,\,  \mathbold{\xi}_F \approx  \mathbf{X}_F.
\end{equation}
while in the limit of high plasma $\beta$,
\begin{equation}
\mathbold{\xi}_S \approx  \xi_z(\mathbf{e}_z - \frac{k_z}{k_x}\mathbf{e}_x), \,\,\,  \mathbold{\xi}_F =  \xi_x(\mathbf{e}_x + \frac{k_z}{k_x}\mathbf{e}_z).
\end{equation}
So the above projection only allows us to separate fast and slow magneto-acoustic modes in the limit of low plasma $\beta$. More details about this can be found in \cite{2003ASSL..294.....G}, chapter 5.
\end{appendix}


%
\bibliographystyle{aasjournal} 

\begin{thebibliography}{}
\expandafter\ifx\csname natexlab\endcsname\relax\def\natexlab#1{#1}\fi
\providecommand{\url}[1]{\href{#1}{#1}}
\providecommand{\dodoi}[1]{doi:~\href{http://doi.org/#1}{\nolinkurl{#1}}}
\providecommand{\doeprint}[1]{\href{http://ascl.net/#1}{\nolinkurl{http://ascl.net/#1}}}
\providecommand{\doarXiv}[1]{\href{https://arxiv.org/abs/#1}{\nolinkurl{https://arxiv.org/abs/#1}}}

\bibitem[{Alfv{\'e}n(1942)}]{1942Natur.150..405A}
Alfv{\'e}n, H. 1942, Nature, 150, 405, \dodoi{10.1038/150405d0}

\bibitem[{Ballester {et~al.}(2018)Ballester, Alexeev, Collados, Downes, Pfaff,
  Gilbert, Khodachenko, Khomenko, Shaikhislamov, Soler, Vazquez-Semadeni, \&
  Zaqarashvili}]{2018SSRv..214...58B}
Ballester, J.~L., Alexeev, I., Collados, M., {et~al.} 2018, Space Science
  Reviews, 214, {\#}58, \dodoi{10.1007/s11214-018-0485-6}

\bibitem[{Berenger(1994)}]{1994JCoPh.114..185B}
Berenger, J.-P. 1994, Journal of Computational Physics, 114, 185,
  \dodoi{10.1006/jcph.1994.1159}

\bibitem[{Berenger(1996)}]{Berenger1996363}
---. 1996, Journal of Computational Physics, 127, 363, \dodoi{DOI:
  10.1006/jcph.1996.0181}

\bibitem[{Cally(2006)}]{2006RSPTA.364..333C}
Cally, P.~S. 2006, Royal Society of London Philosophical Transactions Series A,
  364, 333, \dodoi{10.1098/rsta.2005.1702}

\bibitem[{Cally(2007)}]{Cally:2007cn}
---. 2007, Astronomische Nachrichten, 328, 286, \dodoi{10.1002/asna.200610731}

\bibitem[{Cally(2009)}]{2009MNRAS.395.1309C}
---. 2009, Monthly Notices of the Royal Astronomical Society, 395, 1309,
  \dodoi{10.1111/j.1365-2966.2009.14708.x}

\bibitem[{Cally \& Goossens(2008)}]{2008SoPh..251..251C}
Cally, P.~S., \& Goossens, M. 2008, Solar Physics, 251, 251,
  \dodoi{10.1007/s11207-007-9086-3}

\bibitem[{Cally \& Hansen(2011)}]{2011ApJ...738..119C}
Cally, P.~S., \& Hansen, S.~C. 2011, The Astrophysical Journal, 738, 119,
  \dodoi{10.1088/0004-637X/738/2/119}

\bibitem[{Cally \& Khomenko(2015)}]{2015ApJ...814..106C}
Cally, P.~S., \& Khomenko, E. 2015, The Astrophysical Journal, 814, 106,
  \dodoi{10.1088/0004-637X/814/2/106}

\bibitem[{Cheung \& Cameron(2012)}]{2012ApJ...750....6C}
Cheung, M. C.~M., \& Cameron, R.~H. 2012, The Astrophysical Journal, 750, 6,
  \dodoi{10.1088/0004-637X/750/1/6}

\bibitem[{Christensen-Dalsgaard {et~al.}(1996)Christensen-Dalsgaard, Dappen,
  Ajukov, Anderson, Antia, Basu, Baturin, Berthomieu, Chaboyer, Chitre, Cox,
  Demarque, Donatowicz, Dziembowski, Gabriel, Gough, Guenther, Guzik, Harvey,
  Hill, Houdek, Iglesias, Kosovichev, Leibacher, Morel, Proffitt, Provost,
  Reiter, Rhodes~Jr, Rogers, Roxburgh, Thompson, \&
  Ulrich}]{1996Sci...272.1286C}
Christensen-Dalsgaard, J., Dappen, W., Ajukov, S.~V., {et~al.} 1996, Science,
  272, 1286, \dodoi{10.1126/science.272.5266.1286}

\bibitem[{De~Pontieu {et~al.}(2007)De~Pontieu, McIntosh, Carlsson, Hansteen,
  Tarbell, Schrijver, Title, Shine, Tsuneta, Katsukawa, Ichimoto, Suematsu,
  Shimizu, \& Nagata}]{DePontieu:2007eh}
De~Pontieu, B., McIntosh, S.~W., Carlsson, M., {et~al.} 2007, Science, 318,
  1574, \dodoi{10.1126/science.1151747}

\bibitem[{Felipe(2012)}]{Felipe:2012kq}
Felipe, T. 2012, The Astrophysical Journal, 758, 96,
  \dodoi{10.1088/0004-637X/758/2/96}

\bibitem[{Felipe {et~al.}(2010)Felipe, Khomenko, \&
  Collados}]{2010ApJ...719..357F}
Felipe, T., Khomenko, E., \& Collados, M. 2010, The Astrophysical Journal, 719,
  357, \dodoi{10.1088/0004-637X/719/1/357}

\bibitem[{Fossum \& Carlsson(2006)}]{Fossum:2006kj}
Fossum, A., \& Carlsson, M. 2006, The Astrophysical Journal, 646, 579,
  \dodoi{10.1086/504887}

\bibitem[{Goedbloed {et~al.}(2010)Goedbloed, Keppens, \&
  Poedts}]{Goedbloed:2010tq}
Goedbloed, J.~P., Keppens, R., \& Poedts, S. 2010, {Advanced
  Magnetohydrodynamics}, With Applications to Laboratory and Astrophysical
  Plasmas (Cambridge University Press).
\newblock \url{www.cambridge.org/9780521879576}

\bibitem[{Gonzalez-Morales {et~al.}(2018)Gonzalez-Morales, Khomenko, Downes, \&
  de~Vicente}]{2018arXiv180304891G}
Gonzalez-Morales, P.~A., Khomenko, E., Downes, T.~P., \& de~Vicente, A. 2018,
  arXiv.org, arXiv:1803.04891

\bibitem[{Goossens(2003)}]{2003ASSL..294.....G}
Goossens, M. 2003, An introduction to plasma astrophysics and
  magnetohydrodynamics, 294, \dodoi{10.1007/978-94-007-1076-4}

\bibitem[{Hesthaven(1998)}]{Hesthaven1998129}
Hesthaven, J.~S. 1998, Journal of Computational Physics, 142, 129,
  \dodoi{10.1006/jcph.1998.5938}

\bibitem[{Hu(1996)}]{1996JCoPh.129..201H}
Hu, F.~Q. 1996, Journal of Computational Physics, 129, 201,
  \dodoi{10.1006/jcph.1996.0244}

\bibitem[{Hu(2001)}]{2001JCoPh.173..455H}
---. 2001, Journal of Computational Physics, 173, 455,
  \dodoi{10.1006/jcph.2001.6887}

\bibitem[{Jess {et~al.}(2009)Jess, Mathioudakis, Erd{\'e}lyi, Crockett, Keenan,
  \& Christian}]{Jess-2009}
Jess, D.~B., Mathioudakis, M., Erd{\'e}lyi, R., {et~al.} 2009, Science, 323,
  1582, \dodoi{10.1126/science.1168680}

\bibitem[{Jess {et~al.}(2012)Jess, Pascoe, Christian, Mathioudakis, Keys, \&
  Keenan}]{2012ApJ...744L...5J}
Jess, D.~B., Pascoe, D.~J., Christian, D.~J., {et~al.} 2012, The Astrophysical
  Journal Letters, 744, L5, \dodoi{10.1088/2041-8205/744/1/L5}

\bibitem[{Khomenko \& Cally(2011)}]{Khomenko:2011fx}
Khomenko, E., \& Cally, P.~S. 2011, Journal of Physics Conference Series, 271,
  012042, \dodoi{10.1088/1742-6596/271/1/012042}

\bibitem[{Khomenko \& Cally(2012)}]{Khomenko:2012gw}
---. 2012, The Astrophysical Journal, 746, 68,
  \dodoi{10.1088/0004-637X/746/1/68}

\bibitem[{Khomenko \& Collados(2006)}]{2006ApJ...653..739K}
Khomenko, E., \& Collados, M. 2006, The Astrophysical Journal, 653, 739,
  \dodoi{10.1086/507760}

\bibitem[{Khomenko \& Collados(2009)}]{Khomenko:2009cx}
---. 2009, Astronomy and Astrophysics, 506, L5,
  \dodoi{10.1051/0004-6361/200913030}

\bibitem[{Khomenko {et~al.}(2014)Khomenko, Collados, D{\'\i}az, \&
  Vitas}]{2014PhPl...21i2901K}
Khomenko, E., Collados, M., D{\'\i}az, A., \& Vitas, N. 2014, Physics of
  Plasmas, 21, 092901, \dodoi{10.1063/1.4894106}

\bibitem[{McIntosh {et~al.}(2011)McIntosh, De~Pontieu, Carlsson, Hansteen,
  Boerner, \& Goossens}]{McIntosh:2011cy}
McIntosh, S.~W., De~Pontieu, B., Carlsson, M., {et~al.} 2011, Nature, 475, 477,
  \dodoi{10.1038/nature10235}

\bibitem[{Mihalas \& Mihalas(1984)}]{1984oup..book.....M}
Mihalas, D., \& Mihalas, B.~W. 1984, {Foundations of radiation hydrodynamics}
  (New York: Oxford University Press).
\newblock \url{http://adsabs.harvard.edu/abs/1984oup..book.....M}

\bibitem[{Musielak {et~al.}(1994)Musielak, Rosner, Stein, \&
  Ulmschneider}]{1994ApJ...423..474M}
Musielak, Z.~E., Rosner, R., Stein, R.~F., \& Ulmschneider, P. 1994, The
  Astrophysical Journal, 423, 474, \dodoi{10.1086/173825}

\bibitem[{Parchevsky \& Kosovichev(2009)}]{2009ApJ...694..573P}
Parchevsky, K.~V., \& Kosovichev, A.~G. 2009, The Astrophysical Journal, 694,
  573, \dodoi{10.1088/0004-637X/694/1/573}

\bibitem[{Santamaria {et~al.}(2015)Santamaria, Khomenko, \&
  Collados}]{Santamaria:2015boa}
Santamaria, I.~C., Khomenko, E., \& Collados, M. 2015, Astronomy and
  Astrophysics, 577, A70, \dodoi{10.1051/0004-6361/201424701}

\bibitem[{Santamaria {et~al.}(2017)Santamaria, Khomenko, Collados, \&
  de~Vicente}]{Santamaria:2017ko}
Santamaria, I.~C., Khomenko, E., Collados, M., \& de~Vicente, A. 2017,
  Astronomy and Astrophysics, 602, A43, \dodoi{10.1051/0004-6361/201629729}

\bibitem[{Schunker \& Cally(2006)}]{2006MNRAS.372..551S}
Schunker, H., \& Cally, P.~S. 2006, Monthly Notices of the Royal Astronomical
  Society, 372, 551, \dodoi{10.1111/j.1365-2966.2006.10855.x}

\bibitem[{Schunker {et~al.}(2011)Schunker, Cameron, Gizon, \&
  Moradi}]{2011SoPh..271....1S}
Schunker, H., Cameron, R.~H., Gizon, L., \& Moradi, H. 2011, Solar Physics,
  271, 1, \dodoi{10.1007/s11207-011-9790-x}

\bibitem[{Srivastava {et~al.}(2017)Srivastava, Shetye, Murawski, Doyle,
  Stangalini, Scullion, Ray, W{\'o}jcik, \& Dwivedi}]{Srivastava:2017kw}
Srivastava, A.~K., Shetye, J., Murawski, K., {et~al.} 2017, Scientific Reports,
  7, 43147, \dodoi{10.1038/srep43147}

\bibitem[{Tomczyk {et~al.}(2007)Tomczyk, McIntosh, Keil, Judge, Schad, Seeley,
  \& Edmondson}]{2007Sci...317.1192T}
Tomczyk, S., McIntosh, S.~W., Keil, S.~L., {et~al.} 2007, Science, 317, 1192,
  \dodoi{10.1126/science.1143304}

\bibitem[{Tracy {et~al.}(2003)Tracy, Kaufman, \& Brizard}]{2003PhPl...10.2147T}
Tracy, E.~R., Kaufman, A.~N., \& Brizard, A.~J. 2003, Physics of Plasmas, 10,
  2147, \dodoi{10.1063/1.1543579}

\bibitem[{Vernazza {et~al.}(1981)Vernazza, Avrett, \&
  Loeser}]{1981ApJS...45..635V}
Vernazza, J.~E., Avrett, E.~H., \& Loeser, R. 1981, The Astrophysical Journal
  Supplement Series, 45, 635, \dodoi{10.1086/190731}

\bibitem[{Withbroe \& Noyes(1977)}]{1977ARA&A..15..363W}
Withbroe, G.~L., \& Noyes, R.~W. 1977, In: Annual review of astronomy and
  astrophysics. Volume 15. (A78-16576 04-90) Palo Alto, 15, 363,
  \dodoi{10.1146/annurev.aa.15.090177.002051}

\end{thebibliography}

\end{document}